\documentclass[aps,pra,10pt,twocolumn,floatfix,nofootinbib,longbibliography,nobibnotes]{revtex4-2}

\usepackage{amsmath}
\usepackage{amssymb}
\usepackage{mathtools}
\usepackage{wasysym}
\usepackage{newtxmath}
\usepackage{csquotes}

\allowdisplaybreaks

\usepackage{array} 
\usepackage[T1]{fontenc}
\usepackage{newtxtext}
\usepackage{dsfont}                 
\usepackage{bbold}                  
\usepackage[normalem]{ulem}         

\usepackage{enumerate}
\usepackage[shortlabels]{enumitem}

\usepackage[dvipsnames,table]{xcolor}
\usepackage{graphicx}
\graphicspath{{figures/}}              

\usepackage[flushleft]{threeparttable}
\usepackage{multirow}

\usepackage{physics}                
\usepackage{csquotes}               
\usepackage{comment}
\usepackage[dvipsnames]{xcolor}

\usepackage[caption=false]{subfig}
\usepackage[colorlinks=true,citecolor=cyan,linkcolor=magenta,filecolor=magenta]{hyperref}
\usepackage[capitalize]{cleveref}

\usepackage[globalcitecopy]{bibunits}
\defaultbibliography{bib}
\defaultbibliographystyle{apsrev4-1}
\usepackage{orcidlink}

\DeclareMathAlphabet{\mathbbold}{U}{bbold}{m}{n}

\def\bra#1{\langle{#1}|}				
	\def\ket#1{|{#1}\rangle}		

\definecolor{TB}{rgb}{0.93,0.47,0.2}

\definecolor{WD}{rgb}{0,0,0} 

\usepackage{bm}

\begin{document}


\title{Topological states and flat bands in exactly solvable decorated Cayley trees}

\author{Wanda P.~Duss\,\orcidlink{0009-0000-3785-9423}}
\email{wanda.duss@physik.uzh.ch}
\author{Askar Iliasov\,\orcidlink{0000-0003-2409-7292}}
\email{askar.iliasov@uzh.ch}
\author{Tom\'{a}\v{s} Bzdu\v{s}ek\,\orcidlink{0000-0001-6904-5264}}
\email{tomas.bzdusek@uzh.ch}

\affiliation{Department of Physics, University of Zurich, Winterthurerstrasse 190, 8057 Zurich, Switzerland}


\begin{abstract}
We derive the full spectrum of decorated Cayley trees that constitute tree analogs of selected two-dimensional Euclidean lattices; namely of the Lieb, the double Lieb, the kagome, and the star lattice.
The common feature of these Euclidean lattices is that their nearest-neighbor models give rise to flat energy bands interpretable through compact localized states.
We find that the tree analogs exhibit similar flat or nearly flat energy bands at the corresponding energies.
Interestingly, such flat bands in the decorated Cayley trees acquire an interpretation that is absent in their Euclidean counterparts: as edge states localized to the inner or the outer boundary of the tree branches.
In particular, we establish an exact correspondence between the Lieb-Cayley tree and an ensemble of one-dimensional Su-Schrieffer-Heeger chains, which maps topological edge states on one side of the chains to flat-band states localized in the bulk of the tree, furnishing the flat energy band with a topological stability.
Similar mapping to topological edge states or to states bound to edge defects in one-dimensional chains is shown for flat-band states in all the considered tree decorations.
We finally show that the persistence of exact flat bands on infinite decorated trees (i.e., Bethe lattices) arises naturally from a covering interpretation of tree graphs.
Our findings reveal a rich landscape of flat-band and topological phenomena in non-Euclidean systems, where geometry alone can generate and stabilize unconventional quantum~states.
\end{abstract}

\maketitle


\section{Introduction} 

Tree structures represent perhaps the simplest examples of non-Euclidean systems, characterized by their loopless connectivity and hierarchical growth. A peculiar aspect of trees is that they can be viewed simultaneously as one-dimensional and infinitely-dimensional: locally, each branch supports linear propagation, yet the number of sites grows exponentially with distance from the root. Their recursive geometry enables exact analytic treatments of single-particle problems and significant simplifications of many-body ones, making finite (Cayley) and infinite (Bethe) trees a versatile theoretical laboratory across physics.

In the field of many-body quantum physics, infinite trees (also known as Bethe lattices) are primarily known for their applications in dynamical mean-field theory (DMFT), where, in the limit of infinite coordination number, the mean-field approximation becomes exact~\cite{Georges:1992,Jarrell:1992}.
This idea motivated studies of Bethe lattices in the context of DMFT and related models~\cite{eckstein_hopping_2005,Kollar:2005,Peters:2009}. 
Bose-Hubbard models were studied both analytically~\cite{Semerjian:2009} and numerically with the help of tensor networks~\cite{Lunts:2021}.
Density matrix renormalization group and tensor network approaches were employed for the Fermi-Hubbard model at half-filling in Refs.~\citenum{Lepetit2000} and~\citenum{Chen_etal:2025}. 
Recent studies of correlated states on trees include a mean-field treatment of $s$-wave
superconductivity in the attractive Hubbard model \cite{Bashmakov:2025, pavliuk_work_2025} and the formation of a quantum spin liquid in the Kitaev model~\cite{Vidal:2025}.

In the realm of single-particle and (semi-)classical models, the analytical tractability allows for a fruitful investigation of glassiness~\cite{Thouless:1986,Laumann:2008,Mezard:2001,Savitz:2019}, effects of disorder~\cite{Abou_Chacra:1973,Aizenman2006,Tommaso:2024}, and critical phenomena~\cite{Eggarter:1974,Bing:2000,Kopec:2006,Hu:1998}. Another line of research is the study of (multi)fractal properties of the spectrum and eigenstates of Hermitian and non-Hermitian models defined on tree graphs~\cite{Tikhonov:2016,Tikhonov:2017,Junsong:2024,hamanaka_multifractal_2024}. 
Non-Hermitian models on trees were also investigated recently in Refs.~\citenum{Junsong:2025} and~\citenum{Hatano:2024}.
Several intriguing field theoretic constructions were studied on Bethe lattices as well, including AdS/CFT duality with $p$-adic boundary conformal field theory~\cite{Gubser2017}, and fracton phases~\cite{Shenoy:2023}.

In most of the discussed research on trees, the focus is on bulk properties, and the difference between finite and infinite trees can be neglected. 
However, in contrast to Euclidean systems, there is an important distinction between large finite trees and infinite Bethe lattices. 
This distinction arises because trees constitute an example of so-called expander graphs~\cite{Laumann:2009,Breach:2025,Placke:2025}.
This means that, irrespective of the system size, one always finds that a finite fraction of all the sites reside at the system's boundary~\cite{Basteiro:2023}.
The distinction between finite and infinite trees becomes apparent when analyzing the spectral properties of tree graphs~\cite{mahan_energy_2001,aryal_complete_2020,ostilli_spectrum_2022,Yorikawa:2018}, and phase transitions on large finite trees~\cite{Eggarter:1974,Wang:2025}. 
To ensure convergence to a genuinely infinite tree, one should impose proper boundary conditions~\cite{lux_converging_2023}.

Despite this vast body of work on tree structures, the topological and flat-band aspects of quantum models on Cayley trees have remained largely unexplored. 
While topological and flat-band models have been extensively studied in the context of quasicrystals~\cite{Verbin:2013,Fan:2021,Nielsen:2020,DongHui:2020,LihKing:2024,Akhmerov:2019,Ouyang:2024,Johnstone:2022,Nielsen:2025}, fractals~\cite{Pai:2019,Canyellas:2024,Neupert:2018,Ivaki:2022,Osseweijer:2024,Pal:2018,Manna:2022,Manna:2024,Roy:2024,Eek:2025}, and hyperbolic lattices~\cite{Yu:2020,Urwyler:2022,Liu:2022,Tao:2023,Chen:2023,Tummuru:2024,Bzdusek:2022}, only a few recent works have focused on topological models on decorated Cayley trees ~\cite{Weststrom:2023, singh_arboreal_2024}.  
In Ref.~\citenum{Weststrom:2023}, the authors have revealed a connection between topological modes and the topology of effective one-dimensional chains, and they have identified bulk localization of topological modes. 
In addition, the work of Ref.~\citenum{Eek:2025} considered topology of a tree-like system describes as a Vicsek fractal.
However, these works considered only trees without loop decorations 
and did not delve into the specifics,
such as the calculation of the density of states (DOS). 
Our manuscript is devoted to filling this niche and carefully investigating the differences and similarities between several Euclidean and non-Euclidean geometries, making connections between the topological in-gap bands on tree graphs and flat bands on Euclidean lattices.

We focus on tree graphs with decorations corresponding to the Lieb, double-Lieb, kagome, and star lattices, each of which hosts flat bands in its Euclidean realization. 
We demonstrate that these flat bands persist in their tree analogs, but with profoundly altered characteristics: 
in certain cases, they become topological or symmetry-protected states that are localized in a large region of the bulk rather than at the outer boundary.
The resulting states can be interpreted as topological modes of effective one-dimensional SSH-like chains, establishing a correspondence between the flat-band physics of Euclidean lattices and the topological structure of decorated trees.
It is important to point out, however, that due to innate aspects of the tree geometry one has to be careful about a meaningful definition of flat bands in the studied models.
We elaborate on these subtleties and on our adopted definition of flat bands at the end of Sec.~\ref{Sec:Methods_B} while discussing the analytical solution of the elementary (i.e., non-decorated) Cayley-tree.

The remainder of this article is organized as follows.
In Sec.~\ref{sec:methods} we introduce our general methodology, construct the symmetry-adapted basis for Cayley trees with nearest and next-nearest neighbor hopping, and show how it enables 
exact diagonalization of large trees.
In Sec.~\ref {Sec:LiebDecoration}, we analyze the Lieb-decorated Cayley tree as the simplest model constituting an analog of a Euclidean lattice with flat energy bands,
and we reveal the topological origin of the zero-energy flat band in such a decorated tree. 
We also show that the presence of zero-energy states can be explained by the rank-nullity theorem, similarly to the Euclidean case.
We continue the analysis with the double Lieb decoration in Sec.~\ref{Sec:DoubleLiebDecoration}, where several flat bands acquire the topological index of the corresponding trimer SSH model. 
In the next Secs.~\ref{Sec:HusimiDecoration} and ~\ref{Sec:CliqueDecoration}, we turn to the looped decorations corresponding to the kagome and the star lattice, respectively.
In these cases, no topological protection can be found; however, the flat bands persist and can be interpreted through 
localization at a crystalline 
defect on the boundary of effective one-dimensional chains. 
In Sec.~\ref{Sec:CoveringGraphs}, we consider a broader picture and argue that the persistence of flat energy bands when passing from the Euclidean lattice to the infinite tree geometry is related to
covering properties of the decorated trees. 
Finally, in Sec.~\ref{Sec:Conclusion}, we summarize our work and discuss potential avenues for further endeavors.
Several technical aspects of the analytical constructions appear in the Appendix.

\section{Methods}\label{sec:methods}

To study topological states on decorated Cayley trees, we need a technique to determine their energy spectrum. 
To that end, we adapt in this work the method for exactly solving the density of states on a Bethe lattice initially proposed by Mahan in Ref.~\citenum{mahan_energy_2001}, which was later extended by Refs.~\citenum{aryal_complete_2020}, \citenum{ostilli_spectrum_2022} and \citenum{hamanaka_multifractal_2024} (a similar symmetry-adapted basis was developed and employed also in works
\cite{Aizenman2006,Petrova:2016,Weststrom:2023}) to find the complete spectrum on Cayley trees with finite radius.
In this section, we review how the technique works for graphs with a general branching factor $K$ while explicitly showcasing its action for the specific choice $K=2$. 
We use this as an opportunity to explicitly demonstrate that the technique can be applied without modifications to also study isotropic trees with next-to-nearest hopping amplitudes.
In subsequent sections, we will introduce the suitably adapted versions of this exact approach as applicable for the various presented decorations of the Cayley trees. 
To streamline the discussion, only the key ideas of the generalization will be presented in the main text, with the technical details (such as the precise form of the symmetry-adapted basis states) postponed to the respective appendices. Readers that are familiar with Refs.~\citenum{aryal_complete_2020,ostilli_spectrum_2022,hamanaka_multifractal_2024} should be able to move directly to Sec.~\ref{Sec:LiebDecoration} without impeding the clarity of the presentation.

Our discussion of the computational method is structured as follows.
In Sec.~\ref{Sec:Methods_A} we introduce the set of 
shell-symmetric basis states and use them to compute the bulk density of states. 
The result of this analysis applies in the absence of a boundary (i.e., for the infinite Bethe lattice) or at the center of a finite Cayley tree in the limit of large radial size.
Subsequently, in Sec.~\ref{Sec:Methods_B} we complement the shell-symmetric basis states with shell-non-symmetric basis states, which allow us to calculate the exact spectrum of finite Cayley trees. 
We will use the term ``symmetry-adapted basis states'' for the union of shell-symmetric and shell-non-symmetric basis states. 
They will feature prominently throughout our analysis of decorated Cayley trees as they constitute a convenient choice of orthonormal basis which results in block-diagonalization of the investigated tight-binding models.

\subsection{Shell-Symmetric Basis States}\label{Sec:Methods_A}
A Cayley tree is a finite rooted tree graph where each interior node has a fixed number of neighbors (called the \emph{coordination number} or \emph{degree}) $q\geq 3$. 
Starting from a central (or \emph{root}) node $|0\rangle$, the tree branches out layer by layer without forming loops. 
Each interior node, except the central node $0$, then has $K \geq 2$ \emph{children} 
in the next layer, where $K = q-1$
is called the \emph{connectivity} or the \emph{branching factor} of the tree. 
We let $M$ be the number of layers beyond the root and let $l$ represent the layer index. 
We refer to the nodes in the outermost layer (i.e., with $l=M$), which do not branch any further, as the \emph{leaf nodes}.
A single layer therefore has $N_\textrm{L} = (K+1)K^{\,l-1}$ nodes, and the total number of nodes is
\begin{equation}
\label{eqn:Cayley-number-of-sites}
    N_{\text{total}} = 1 + \sum_{l=1}^{M} (K+1)\,K^{\,l-1} = 1 + \frac{(K+1)}{(K-1)}(K^M  - 1). 
\end{equation}
We consider a tight-binding model on this tree given by the Hamiltonian
\begin{equation}
\label{eqn:Ham-NN+NNN}
    H = \sum_{\langle i,j\rangle } t_1\big( \ket{i}\bra{j} + \text{h.c.} \big) + \sum_{\langle \langle m,n\rangle \rangle} t_2 \big( \ket{m}\bra{n} + \text{h.c.} \big),
\end{equation}
where $\langle\cdot, \cdot \rangle$ denotes nearest-neighbor (NN) and $\langle \langle \cdot , \cdot \rangle \rangle$ denotes next-nearest-neighbor (NNN) pairs of sites, and $t_{1,2}$ are the corresponding hopping amplitudes. 

\begin{figure}[t!]
\centering
\includegraphics[width=\linewidth]{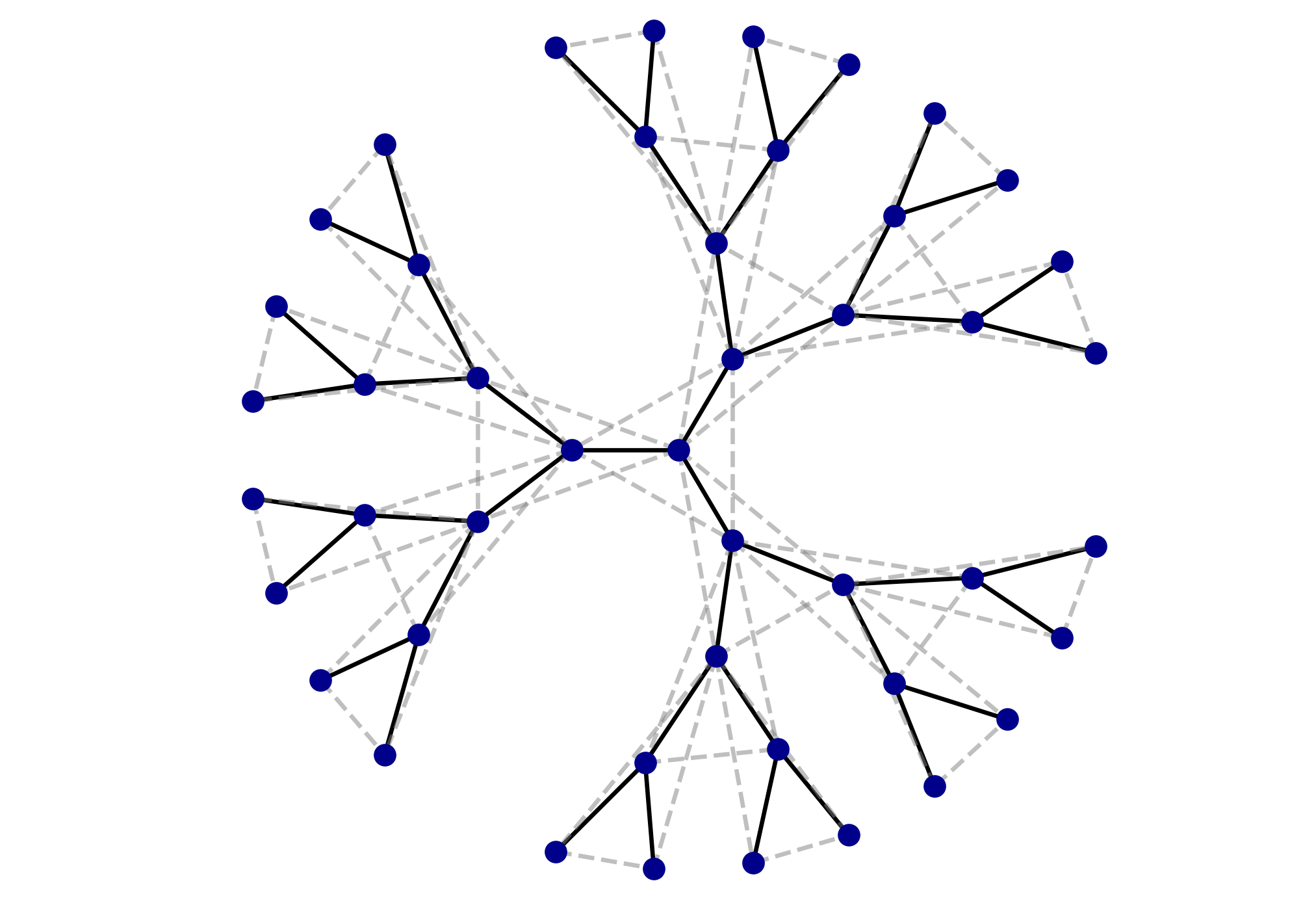}
\caption{A Cayley tree with nearest and next-nearest neighbor hopping. The solid-black bonds represent a nearest-neighbor hopping amplitude of $t_1$, while the gray-dashed bonds represent a next-nearest neighbor hopping amplitude of $t_2$. 
Here we set the number of layers to $M=4$ and the branching factor to $K=2$. 
The tree is invariant under permutation of individual sub-branches and under cyclic rotations of the three main branches. 
}
\label{Fig:NNN_CayleyTree_Sketch}
\end{figure}

We emphasize that we consider Hamiltonians with NNN hoppings in this section only, for the purpose of illustrating how the underlying method can be extended beyond the elementary NN models (all models on decorated Cayley trees considered in Sec.~\ref{Sec:LiebDecoration} and later assume strictly NN Hamiltonians).
An example of a Cayley tree with branching factor $K=2$ that supports NN hopping (solid black lines) and NNN hopping (dashed gray lines) is shown in Fig.~\ref{Fig:NNN_CayleyTree_Sketch}.  

Cayley trees possess a high degree of permutation symmetry. 
An intuitive approach to understand this is to visually rotate branches around the tree. Because the orbitals at all vertices are identical and the hoppings are homogenous, such a transformation will leave the tree invariant. More formally, there is no intrinsic ordering of the branches around any given node: the system will always look the same under cyclic rotation of the main branches or permutations of the subbranches.

We can exploit this symmetry by constructing a basis of symmetry-adapted states that block-diagonalizes the tight-binding Hamiltonian.
Observe that a Cayley tree with a central node (layer $l=0$) and $M$ layers of branching can be divided into $K+1$ equivalent branches emanating from the center. 
The Hamiltonian in Eq.~(\ref{eqn:Ham-NN+NNN}) 
is invariant under any cyclic permutation of these $K+1$ branches. 
For example, in a Cayley tree with branching factor  $K=2$ (coordination number $q=3$) 
this amounts to a $C_{K+1} \equiv C_3$ rotation cycling the three main branches.
Furthermore, a large set of additional symmetries $C_{K,\alpha}$ can be introduced where $\alpha$ is any node in layers $l\in\{1,\ldots,M-1\}$ (i.e., it is neither the center nor a leaf node).
Specifically, $C_{K,\alpha}$ corresponds to cycling the $K$ sub-branches rooted at node $\alpha$. 
These symmetry operations commute with $H$ (i.e., $[H, C_{K+1}] = [H, C_{K,\alpha}] = 0$)  which implies that the Hamiltonian and any one individual symmetry can be diagonalized simultaneously. 
While this does not imply that the individual symmetries can be simultaneously diagonalized with each other, we do find that the symmetry-adapted basis states achieve a block-diagonalization of the Hamiltonian, where each block is associated with \emph{one} of these symmetries.
Specifically, we will obtain a single block transforming in the trivial representation of $C_{K+1}$, which we call the ``shell-symmetric sector'' and denote $\mathcal{H}_{\textrm{sym.}}$.
The remaining blocks, denoted $\mathcal{H}_{\textrm{nonsym.}}^{\alpha}$ 
are termed ``shell-non-symmetric sectors'', with each such sector corresponding to states transforming in a non-trivial representation of $C_{K,\alpha}$.

\begin{figure}[t!]
\centering
  \centering
  \includegraphics[width=\linewidth]{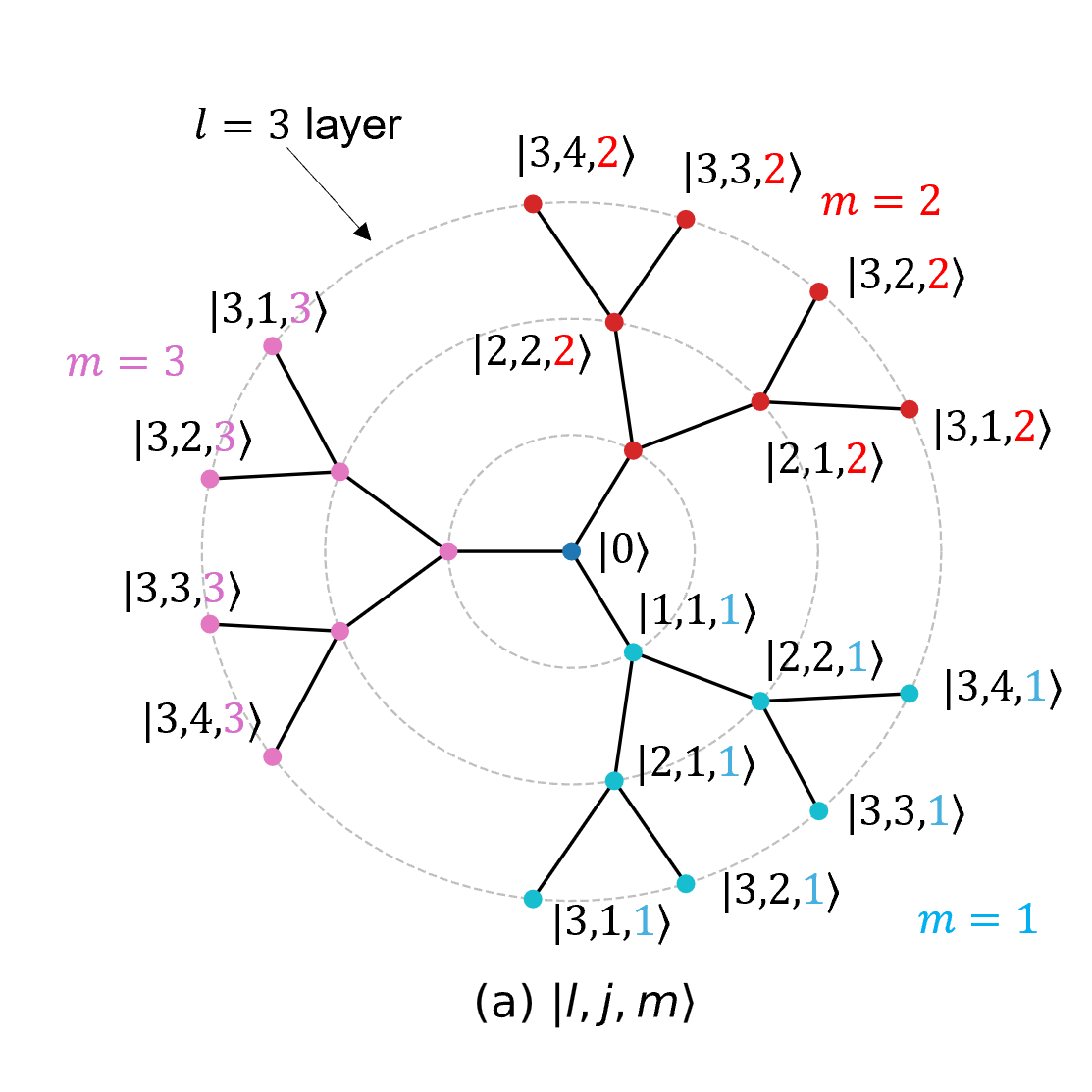} \\
  \includegraphics[width=\linewidth]{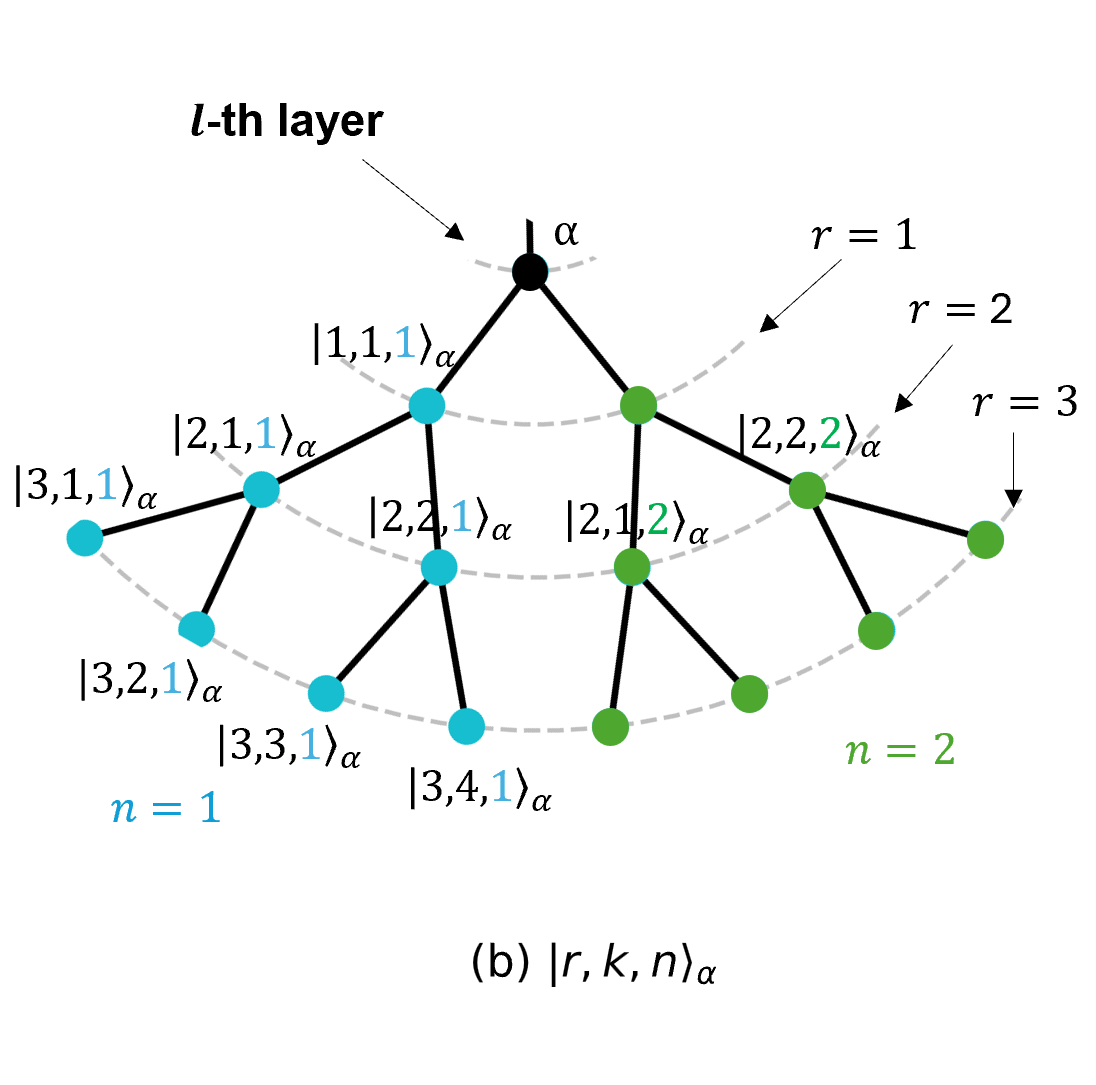}
\caption{A sketch demonstrating the notation of the position basis in the 
construction of
(a) the shell-symmetric states and of 
(b) the shell-non-symmetric states. (Similar schematics appear in Ref.~\citenum{hamanaka_multifractal_2024})}
\label{Fig:Labeling}
\end{figure}

We proceed to explicitly construct the basis states that achieve the outlined block-diagonalization. 
We first consider the the shell-symmetric sector, and express its
symmetry-adapted basis that consists of \emph{shell-symmetric basis states}. 
We will express the symmetry-adapted basis as a linear combination of the original \emph{position basis} (equivalently called the \emph{site basis}) employed in Eq.~(\ref{eqn:Ham-NN+NNN}). 
Specifically, a shell-symmetric basis state is constructed by taking an equal-amplitude superposition of all site states on a given \emph{layer} (i.e., collection of sites with the same distance from the center) of the tree. 
First, we choose the central node as one of the shell-symmetric basis states: 
\begin{equation}
    |0\rangle = |0).
\end{equation}
Here $|\cdots\rangle$ denotes
the position basis states, and $|\cdots)$ indicates the symmetric and non-symmetric basis states. 
The remaining shell-symmetric basis states $|l)$ are constructed by symmetrizing the position basis $|l,j,m\rangle$ [where $m\in\{1,\ldots,K+1\}$ labels the main branches emanating from the root node, $l\in\{1,\ldots,M\}$ labels the layers of the tree, and $j \in\{ 1,\dots,K^{l-1}\}$ labels all sites in a specified branch $m$ and layer $l$; see Fig.~\ref{Fig:Labeling}(a)] within the layer $l$, i.e., 
\begin{equation}
\label{eqn:Lieb-sym-basis-states}
    |l) = \frac{1}{\sqrt{(K+1)K^{l-1}}} \sum_{m=1}^{K+1}\sum_{j=1}^{K^{l-1}} |l,j,m\rangle
\end{equation}
where the prefactor ensures normalization to $(l|l) = 1$.
There are a total of $N_S = M+1$ symmetric basis states which are orthogonal to each other. 

Observe that the action of the Hamiltonian on the symmetric basis state $|l)$ generates symmetric basis states in the adjacent layers $l \pm 1$, effectively turning the shell-symmetric sector into a linear 1D chain.
To illustrate the emergence of the effective 1D description, we briefly consider the action of $H$ in Eq.~(\ref{eqn:Ham-NN+NNN}) on a few shell-symmetric basis states. Setting $K=2$, we find
\begin{subequations}
\begin{equation}
    \begin{aligned}
        H|0) & =  t_1 \big(|1,1,1\rangle + |1,1,2\rangle + |1,1,3\rangle \big) \\
        & +  t_2 \big(|2,1,1\rangle + |2,2,1\rangle + |2,1,2\rangle \\
        & + |2,2,2\rangle + |2,1,3\rangle + |2,2,3\rangle \big) = \ldots
        \end{aligned}
        \end{equation}
which further simplifies to 
        \begin{equation}
        \begin{aligned}
        \ldots & = \sqrt{3} \; t_1 \; |1) + \sqrt{6} \; t_2 \; |2) \\
        & = \sqrt{K+1} \; t_1 \; |1) + \sqrt{(K+1)K} \;t_2 \; |2).
    \end{aligned}
\end{equation}
\end{subequations}
Similarly, applying the Hamiltonian to $|1)$ (where we drop the expansion in position states for the sake of brevity), we obtain
\begin{equation}
    \begin{aligned}
        H |1) & = \sqrt{3} \; t_1 \; |0) + 2t_2 \; |1) + \sqrt{2} \; t_1 \; |2) + 2t_2 \; |3) \\
        & = \sqrt{K+1} \; t_1 \; |0) + Kt_2 \; |1) + \sqrt{K} \; t_1 \; |2) + Kt_2 \; |3).
    \end{aligned}
\end{equation}
Generalizing to the full eigenvalue equation $H|\Psi \rangle = E \ket{\Psi}$ while expanding the shell-symmetric eigenstates using the orthonormal basis as $\ket{\Psi} = \sum_l \psi_l |l)$, we find the following system of equations:
\begin{subequations}
\begin{flalign}
    & E \psi_0 = \sqrt{K+1} \; t_1 \; \psi_1 + \sqrt{(K+1) K} \; t_2 \; \psi_2, \\
    & E \psi_1 = \sqrt{K+1} \; t_1\; \psi_0 + Kt_2 \; \psi_1 + \sqrt{K} \; t_1\; \psi_2 + Kt_2 \; \psi_3, \\
    & \begin{aligned}
            E \psi_l &= \sqrt{K} \; t_1 \; \psi_{l+1} + \sqrt{K} \; t_1 \; \psi_{l-1} \\ &+ K t_2 \; \psi_{l+2} + K t_2 \; \psi_{l-2} + (K-1)t_2\; \psi_{l},
    \end{aligned} \label{eqn:last-from-recurrence}
\end{flalign}
\end{subequations}
where Eq.~(\ref{eqn:last-from-recurrence}) applies for $l \geq 2$. 
These equations can be expressed in terms of the Hamiltonian of the shell-symmetric sector:
\begin{widetext}
\begin{equation}
\label{Eq:Cayley_Symmetic_Sector}
\hspace{-2cm}
    \mathcal{H}_{\textrm{sym.}} = 
    \begin{pmatrix}
    0 & \sqrt{K+1} \;t_1 & \sqrt{K(K+1)} t_2 \; & 0 & 0 & \cdots\\
    \sqrt{K+1} \; t_1 & Kt_2 & \sqrt{K} \; t_1 & K t_2 & 0&  \cdots\\
    \sqrt{K(K+1)} \; t_2 & \sqrt{K} \; t_1 & (K-1)t_2 & \sqrt{K} \; t_1 & Kt_2 & \cdots \\
    0 & Kt_2 & \sqrt{K} \; t_1 & (K-1) t_2 & \sqrt{K} \; t_1 & \cdots\\
    0 & 0 & Kt_2 & \sqrt{K} \; t_1 & (K-1)t_2 & \cdots\\
    \vdots & \vdots  & \vdots & \vdots & \vdots & \ddots \\ 
\end{pmatrix}\!
\end{equation}
\end{widetext}
with dimension $\dim(\mathcal{H}_{\textrm{sym.}})=(M+1)\times(M+1)$.

The central (root) node $\ket{0}$ has nonzero amplitude only for shell-symmetric basis states, because any state transforming non-trivially under rotations of the $K+1$ branches must have zero amplitude at the invariant center. This is because only the zero amplitude remains invariant under picking up a phase that is characteristic of all non-trivial representation of $C_{K+1}$.
Therefore, the local density of states (LDOS) at the central site receives contributions only from shell-symmetric eigenstates, which are also the only eigenstates extended over the whole tree. Hence, calculating LDOS at the central node of the Cayley tree becomes equivalent to calculating LDOS at the end of a one-dimensional chain, which converges to a smooth function in the limit of an infinite chain. Furthermore, one can anticipate that in the limit of an infinite tree, the LDOS at the center should converge to the DOS of the Bethe lattice. In accordance with this idea, in Ref.~\citenum{mahan_energy_2001}, Mahan focused on this symmetric sector to derive the energy band and density of states for the Bethe lattice ($M \to \infty$) with NN hopping. We should also remark that it is important to distinguish the local density of states at the root node and the total density of the states of the finite trees. In contrast with LDOS at the center, the total density of states receive contributions from all symmetry-adapted sectors and becomes a singular distribution as we discuss in Sec.~\ref{Sec:Methods_B}.

\begin{figure}[t!]
\centering
\includegraphics[width=\linewidth]{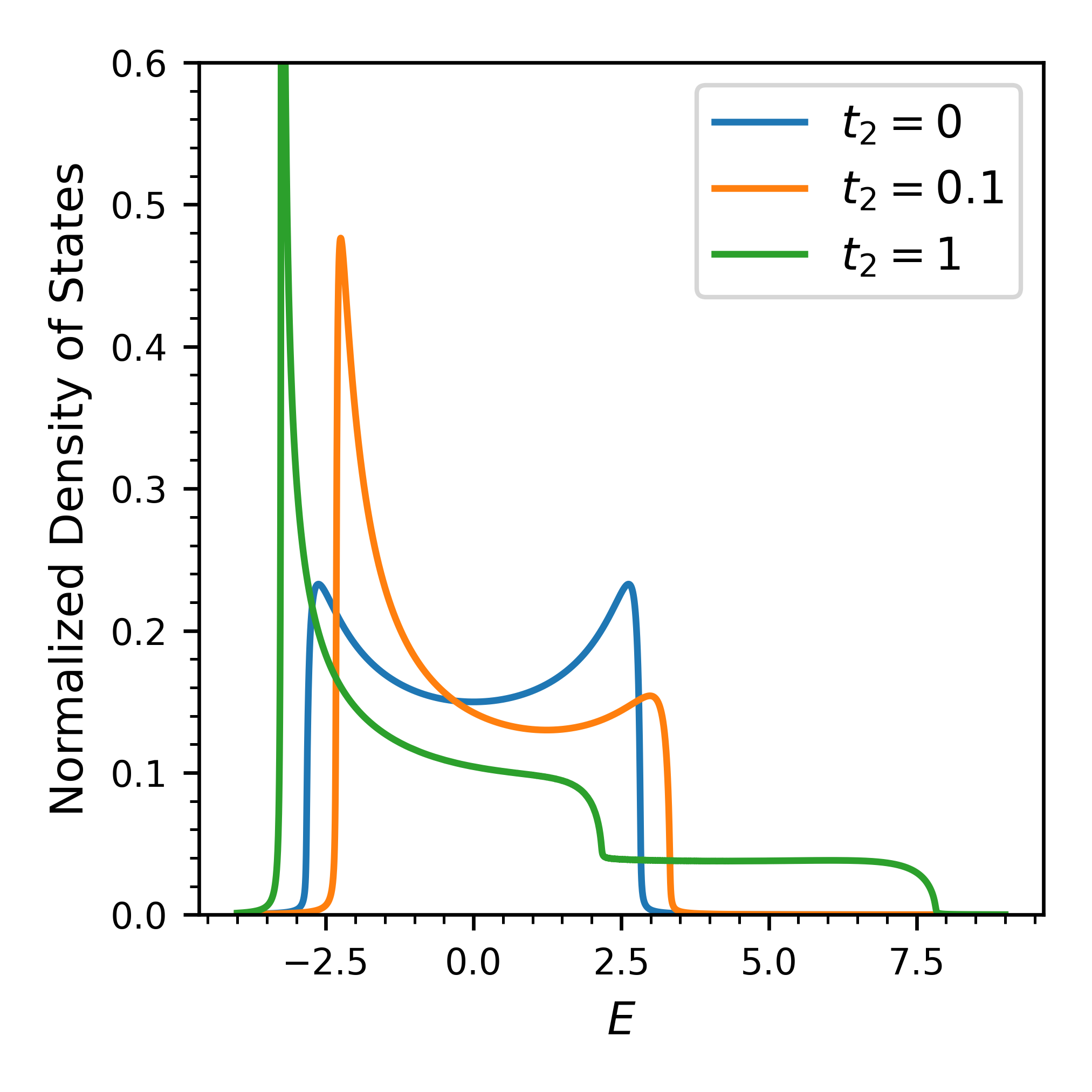}
\caption{The bulk density of states of the NNN-hopping Calyey tree with connectivity $K=2$ with nearest-neighbor hopping $t_1 = 1$.
For $t_2=0$ we recover the density of states of the simple Bethe lattice~\cite{mahan_energy_2001}. As we increase $t_2$ the spectrum becomes asymmetric. 
For $t_2 = 1$ we find a square-root singularity at the lower band-edge as well as a cusp (i.e., infinite slope of the density of states) inside the energy band~\cite{eckstein_hopping_2005}.
The data shown was computed from Eqs.~(\ref{eqn:Green-00}) and~(\ref{Eq:LDOS_Formula}) with $M= 3000$ and $\delta=0.01$.}
\label{Fig:NNN_CayleyTree_DensityOfStates}
\end{figure}

By the same principle, we can compute the LDOS at the central node on Cayley trees involving NNN hopping, which converges to the DOS of the Bethe lattice with NNN hoppings for large system sizes. 
Let $G_{00}$ be the local Green's function, which we can write using the spectral representation of the matrix elements of the full Green's function as
\begin{equation}
\label{eqn:Green-00}
    G_{00}(z) = \sum_n \frac{\langle 0|v_n\rangle \langle v_n | 0\rangle}{z-E_n} = \sum_{n=1
    }^{M+1} \frac{|\langle 0 | v_n \rangle |^2}{z - E_n}.
\end{equation}
Here $E_n$ are the eigenvalues associated with the eigenstate $v_n$, $z\in \mathbb{C}$, and the sum `$\sum_n$' runs over all eigenstates of the Hamiltonian. 
However, only the $M+1$ eigenstates of the shell-symmetric sector have a nonzero overlap with the central site $|0\rangle$, allowing us to reduce at the second step of Eq.~(\ref{eqn:Green-00}) the summation over all eigenstates to a summation over the $M+1$ shell-symmetric eigenstates.  
Taking sufficiently large $M$ to ensure convergence, the bulk density of states can be obtained~as
\begin{equation} \label{Eq:LDOS_Formula}
    \rho_0 (\omega) = \lim_{\delta\to 0^+}\frac{1}{\pi}\text{Im} \,G_{00}(\omega + i \delta).
\end{equation}
We found that $M=3000$ was necessary to ensure a smooth convergence around the cusp of the $t_2=1$ tree. Using the outlined procedure, we numerically find the LDOS at the center for the NNN-hopping Cayley tree shown in Fig.~\ref{Fig:NNN_CayleyTree_DensityOfStates}.
Let us point out that in the absence of NNN hoppings, the eigensystem of the symmetric sector~(\ref{Eq:Cayley_Symmetic_Sector}) can be obtained analytically~\cite{mahan_energy_2001}, which in turn enables an analytical expression for $\rho_0(\omega)$.

In the next subsection, we discuss the shell-non-symmetric basis states, which complete our symmetry-adapted basis and that enable us to find the complete and exact spectrum of the NNN-hopping Cayley tree. 

\subsection{Shell-Non-Symmetric Basis States}\label{Sec:Methods_B}

To complete the orthonormal symmetry-adapted basis, we next turn to the shell-non-symmetric basis states.
Their construction proceeds in two steps. 
First, we choose a non-leaf node $\alpha$ located on a chosen layer $l\in\{1,\ldots,M-1\}$ as the \emph{originator} or \emph{seed} of the specific non-symmetric sector $\mathcal{H}_{\textrm{non-sym}}^{\alpha}$. (The case when the seed coincides with the root in layer $l=0$ is discussed separately below.) 
Beyond the seed, there are $M-l$ layers remaining which we label with $r \in \{1,2,\dots , M-l\}$. 
We characterize the sites that trace back to the seed using the position basis state $\ket{r,k,n}_\alpha$ where $n\in\{1,\ldots,K\}$ labels the specific sub-branch emanating from the seed and the number $k\in\{1,\ldots,K^{r-1}\}$ enumerates all sites in a specified sub-branch $n$ and layer $r$; see Fig.~\ref{Fig:Labeling}(b). 
To proceed, let further $\omega$ be a nontrivial $K$-th root of unity. 
The shell-non-symmetric basis states $|l,r,\omega)_\alpha$ are generated by weighing the position basis $|r,k,n\rangle_\alpha$ by powers of $\omega$, where all states in a branch $n$ have the same weight,
\begin{equation}
\label{eqn:shell-non-sym-state-def-nonroot}
    |l,r,\omega)_\alpha = \frac{1}{\sqrt{K^r}}\sum_{n=1}^{K} \omega^n \sum_{k=1}^{K^{r-1}} |r,k,n\rangle_\alpha.
\end{equation}
The weighting with the powers of $\omega$ implies destructive interference at the seed $\alpha$ when acting on $|l,r,\omega)_\alpha$ with the Hamiltonian, meaning the the dynamics of shell-non-symmetric states $|l,r,\omega)_\alpha$ is bound to the sub-branches emanating from~$\alpha$.
(Let us also point out a redundancy of our notation, since the layer $l = l(\alpha)$ is implicitly specified by the choice of the~seed~$\alpha$.) 

We illustrate the action of the Hamiltonian in Eq.~(\ref{eqn:Ham-NN+NNN}) on a few shell-non-symmetric states for the simplest case $K=2$, when the only non-trivial root of unity is $\omega = -1$. We find
\begin{equation}
\begin{aligned}
        H |l,1,\omega)_\alpha & = H \frac{1}{\sqrt{2}}\big(|1,1,1\rangle_\alpha - \ket{1,1,2}_\alpha  \big) \\
        & = \frac{1}{\sqrt{2}} \big( t_1(1-1)\ket{\alpha} + t_2(1-1)\ket{\alpha-1} \\
        &+ t_2( \ket{1,1,2}_\alpha - \ket{1,1,1}_\alpha)\\
        & + t_1(\ket{2,1,1}_\alpha + \ket{2,2,1}_\alpha -  
        \ket{2,1,2}_\alpha - 
        \ket{2,2,2}_\alpha)  \\
        &+ t_2(\ket{3,1,1}_\alpha + \ket{3,2,1}_\alpha + \ket{3,3,1}_\alpha + \ket{3,4,1}_\alpha \\
        & - 
        \ket{3,1,2}_\alpha - 
        \ket{3,2,2}_\alpha - 
        \ket{3,3,2}_\alpha - 
        \ket{3,4,2}_\alpha)  \\
        &= -t_2 \; |l,1,\omega)_\alpha + \sqrt{2}\;t_1 \; |l,2,\omega)_\alpha + 2t_2 \; |l,3,\omega)_\alpha, 
\end{aligned}
\end{equation}
where $\ket{\alpha-1}$
refers to the parent node of $\alpha$. 
Inspecting the next layer but omitting the expansion in position the basis, we find
\begin{equation}
\begin{aligned}
        H |l,2,\omega)_\alpha & = 0 \; \ket{\alpha} 
        + \sqrt{2}\;t_1 \; |l,1,\omega)_\alpha + t_2 \; |l,2,\omega)_\alpha \\ 
    & + \sqrt{2} \; t_1 \; |l,3,\omega)_\alpha + 2t_2 \; |l,4,\omega)_\alpha. 
\end{aligned}
\end{equation}
Observe that the ``quantum numbers'' $l$, $\omega$ and $\alpha$ are preserved under the action of the Hamiltonian, with only the layer index $r$ altered. 
Therefore, a shell-non-symmetric eigenstate in the sector specified by seed $\alpha$ and root $\omega$ can be expanded as $\ket{\Phi} = \sum_r \phi_r |l,r,\omega )_\alpha$. 
The eigenvalue problem $H|\Phi \rangle = E \ket{\Phi}$ then reduces to the system of equations
\begin{subequations}
\begin{eqnarray}
     E\phi_{1} &=& -t_2 \; \phi_{1} + \sqrt{K}\;t_1 \; \phi_{2} + Kt_2 \; \phi_{3} \\
     E\phi_{2} &=& \sqrt{K} t_1\; \phi_{1} + (K-1)t_2\phi_{2} \; \nonumber \\
     &\phantom{=}&  + \sqrt{K}t_1\; \phi_{3} + Kt_2 \; \phi_{4} \\
    E\phi_{r} 
    & =& \sqrt{K} \; t_1 \, \phi_{r-1} \!+\! \sqrt{K} \; t_1 \; \phi_{r+1} \!+\! (K\!-\!1)t_2 \; \phi_{r} \nonumber \\ 
    &\phantom{=}& +Kt_2 \; \phi_{r-2} + Kt_2\; \phi_{r+2} \\
     0  
     &=& \phi_{0} = \phi_{M-l+1}.
\end{eqnarray}
\end{subequations}
This equations can be written in terms of the Hamiltonian of the shell-non-symmetric sector
\begin{equation}
    \label{Eq:Cayley_NonSymmetric_Sector}
    \mathcal{H}_{\textrm{nonsym.}}^\alpha \!=\! \begin{pmatrix}
    -t_2 & \sqrt{K} \;t_1 & Kt_2 & 0 &\cdots\\
    \!\sqrt{K} \;t_1\! & \!(K-1)t_2\!\! & \sqrt{K} \; t_1 & Kt_2 &\cdots\\
    Kt_2 & \sqrt{K} \; t_1 & \!\!(K-1)t_2\!\! & \sqrt{K} \;t_1 &\cdots \\
    0 & Kt_2 & \sqrt{K} \; t_1 & \!\!(K-1)t_2\!\! & \cdots\\
    \vdots & \vdots  & \vdots & \vdots & \ddots \\ 
\end{pmatrix}\! ,
\end{equation} 
which does not depend on the choice of non-trivial root of unity $\omega$, and that only depends on the choice of seed $\alpha$ through its layer $l$; namely, the dimension of the Hamiltonian is $\dim(\mathcal{H}_{\textrm{nonsym.}}^\alpha)=(M-l)\times(M-l)$.
 
We count how many shell-non-symmetric basis states are defined by Eq.~(\ref{eqn:shell-non-sym-state-def-nonroot}). 
We can choose $\alpha$ from $M-1$ layers, where each layer has $N_\textrm{L} = (K+1)K^{l-1}$ nodes. 
There are $K-1$ nontrivial roots of unity $\omega$ and each non-symmetric sector consists of $M-l$ states. 
The total number of states described by these non-symmetric sectors is therefore
\begin{eqnarray}
    N_{B_{{\geq}1}} &=& \sum^{M-1}_{l=1} (K-1)(K+1) \times K^{l-1}(M-l) \nonumber \\
    &=& N_{\text{total}} - (K+1)M - 1.
\end{eqnarray}
Here the `$B$' subscript stands for ``Branch'', as these states live on sub-branches. 
The subsubscript `${\geq}1$' denotes that this only counts the non-symmetric states up to layer $l=1$ but does not account for the non-symmetric states that originate from the central node $0$. 

Finally, we turn to the shell-non-symmetric sector originating from the center node $0$. 
This sector is sometimes referred to as the ``shell-symmetric sector with $\psi_0=0$''; however, we here find it more convenient to interpret it as a shell-non-symmetric 
sector, because it is usually described (up to minor modifications) by the Hamiltonian $\mathcal{H}_{\textrm{nonsym.}}^\alpha$. 
Specifically, there are differences with respect to normalization and degeneracies, because for a tree with a branching factor $K$, we have to weigh over $K+1$ (rather than $K$) branches that originate from the central node. 
To be precise, these additional shell-non-symmetric basis states are written as 
\begin{equation}
\label{eqn:nonsym-at-root}
    |r,\varpi)_{0} = \frac{1}{\sqrt{(K+1)K^{r-1}}}\sum_{m=1}^{K+1} \varpi^n \sum_{j=1}^{K^{r-1}} |r,j,m\rangle
\end{equation}
where $\varpi$ are nontrivial $(K\,{+}\,1)$-th roots of unity.
Observe also that, compared to Eq.~(\ref{eqn:shell-non-sym-state-def-nonroot}) for shell-non-symmetric basis states with seed $\alpha \neq \ket{0}$, we have droped the layer $l \equiv l(\ket{0}) = 0$ from the notation.
The block Hamiltonian $\mathcal{H}_\textrm{nonsym.}^0$ of this sector looks like Eq.~(\ref{Eq:Cayley_NonSymmetric_Sector}) and has dimension $\dim(\mathcal{H}_{\textrm{nonsym.}}^0)=M \times M$ and a multiplicity $K$, because there are $K$ non-trivial roots of unity. 
The total number of such states is thus
\begin{equation}
N_{B_0} = M\times K.
\end{equation}

For the sake of brevity, we will not derive the energy recursion relations of this sector explicitly but instead just highlight how the first layer behaves under application of the Hamiltonian for $K=2$.
\begin{equation}
\begin{aligned}
    H|1,\varpi)_{0} & = H \big( \varpi^1 \ket{1,1,1} + \varpi^2 |1,1,2\rangle + \varpi^3 \ket{1,1,3}\big) \\
    & = (\varpi^2 + \varpi^3) \ket{1,1,1} + (\varpi^1 + \varpi^3) \ket{1,1,2} \\
    & + (\varpi^1 + \varpi^2) \ket{1,1,3} +\Big(\sum_i \varpi^i\Big)|0\rangle \\ & + \sqrt{2} \; |2,\varpi)_{0} + 2 \; |3,\varpi)_{0} 
    \\
    & = -\; |1,\varpi) +\sqrt{2} \; |2,\varpi)_{0} + 2 \; |3,\varpi)_{0} 
\end{aligned}
\end{equation}
Because the transformation of the $|2,\varpi)_{0}$ and $|3,\varpi)_{0}$ states is straightforward, we skipped the expansion in the position basis. 
When simplifying the sums over $\varpi$, we have used that
\begin{equation}\label{Eq:Sum_of_Roots_of_Unity}
    \sum_i^K \varpi^i = 0, \quad \text{therefore} \quad \sum_{i, i \neq j}^K \varpi^i = - \varpi^j. 
\end{equation}
Adding $N_{B_0}$ to the total state count, we can verify that we have found a complete set of basis states: $N_{B_{\geq1}} + N_S + N_{B_{0}} = N_{\text{total}}$.

We now  
address the question of computing
the full spectrum of a finite Cayley tree. 
Using the presented symmetry-adapted basis states,
we block-diagonalize the Hamiltonian of our system, which results in
an effective 1D chain for each symmetry sector. 
In total, there are $M+1$  
unique symmetry sectors, most of which appear with a large multiplicity. 
Only a single sector  
corresponds to shell-symmetric states [Eq.~(\ref{Eq:Cayley_Symmetic_Sector})], whose multiplicity is $N_\textrm{deg(sym.)}=1$. 
The remaining $M$ 
unique sectors are non-symmetric ones [Eq.~(\ref{Eq:Cayley_NonSymmetric_Sector}), plus the additional $\mathcal{H}_\textrm{nonsym.}^0$]. 
The multiplicity of the non-symmetric sector with seed node in layer $l$ is expressed as 
$N_{\text{deg}(l)} = N_\textrm{L} \times N_{\textrm{roots}(l)}$, where $N_\textrm{L}$ is the number of nodes in layer $l$ and $N_{\textrm{roots}(l)}$ is the number of non-trivial roots of unity at that depth $l$ of the graph. 
This multiplicity evaluates to
\begin{equation}
N_{\text{deg}(l)} = \left\{ 
\begin{array}{ll}
K   & \textrm{for $l = 0$} \\
 (K^2 - 1)K^{l-1}   & \textrm{for $l \in \{1,\ldots,M-1\}$} 
\end{array}\right.
\end{equation}
It follows that we only need to diagonalize $M+1$ unique Hamiltonians and then count their eigenvalues with multiplicity 
$N_{\text{deg}(l)}$. 
Collecting all of these eigenvalues will give us the full spectrum of the Cayley tree. 
In this way, it is possible to find the spectra for extremely large trees, which would not be possible with exact diagonalization (ED) of the full Hamiltonian. 
Specifically, the largest matrix we diagonalize has dimension $M+1$. 
At the same time, the number of nodes in the tree scales with $N \propto K^{M}$. 
The time complexity for ED is thus reduced from $O(N^3)$ to $O(\log(N)^4)$. 

\begin{figure}[t!]
\centering
\includegraphics[width=\linewidth]{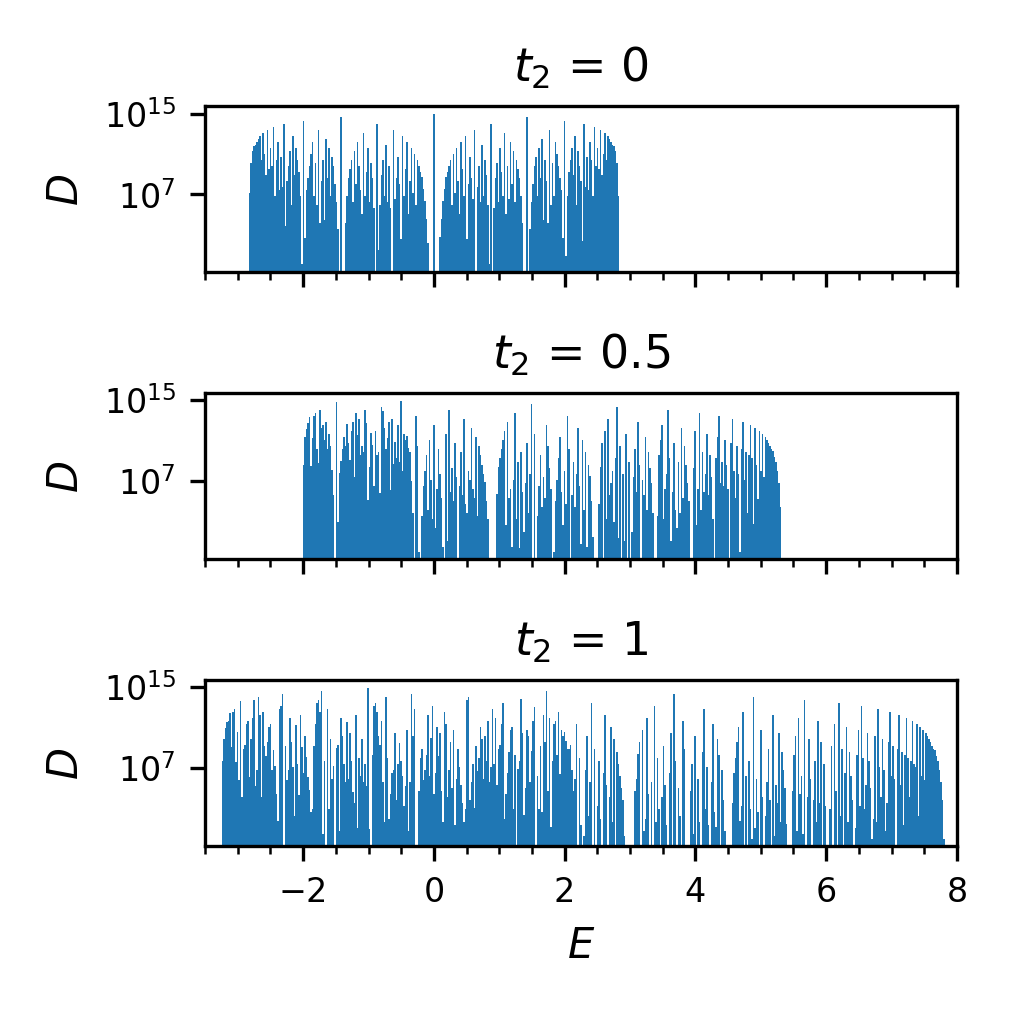}
\caption{Spectrum of a Cayley tree with $M=50$ layers and branching factor $K=2$. 
Nearest-neighbor hopping is fixed at $t_1=1$ and next-nearest-neighbor (NNN) hopping $t_2$ is tuned from $0$ to $1$. The symbol $D$ at the vertical axis denotes the degeneracy of the energy level; this notation, is also adopted in subsequent figures. 
For $t_2=0$ we recover the spectrum of a simple Cayley tree with only NN hopping \cite{aryal_complete_2020}.
As we turn on NNN hopping $t_2 > 0$, the spectrum shifts to larger energies and loses its symmetrical arrangement around~$E=0$. 
}
\label{Fig:NNN_CayleyTree_Spectra}
\end{figure}

We present in Fig.~\ref{Fig:NNN_CayleyTree_Spectra} the full spectrum of a Cayley tree with branching factor $K=2$ and $M = 50$ layers, containing ${\sim}10^{15}$ sites, 
computed using the symmetry-adapted sectors as described  
above. We plot the energy ($E$) of the eigenstates against their degeneracy ($D$).
The coefficient 
$t_2$ tunes the NNN-hopping. 
Let us point out several aspects of this spectrum.
First, observe that the density of states is plotted on a logarithmic scale, meaning that there is an exponentially large separation between the heights of the histogram peaks.
This is understood upon realizing that the non-symmetric sector with seed nodes in layer $l=M-1$, whose Hamiltonian is a single number that directly encodes the eigenvalue of a length-$1$ chain, appears with multiplicity $N_{\text{deg}(M-1)} = (K^2 - 1)K^{M-2}$. 
For large $M$, this single eigenvalue constitutes a fraction \begin{equation}
N_{\text{deg}(M-1)}/N_\textrm{total}=\left(1-\frac{1}{K}\right)^2
\end{equation} 
(i.e., $25\,\%$ for $K=2$, and more for larger $K$) of the entire spectrum. 
Consecutively smaller peaks in the density of states are generated by the non-symmetric sectors with smaller $l$, with every decrease of $l$ by $1$ resulting in a suppression of the corresponding peak heights by $1/K$.

It further follows from the above considerations that the full spectrum does not become continuous even in the limit $M\to \infty$. 
Instead, the density of states 
is a distribution that consists of a large collection of Dirac delta functions.
The set of energies that support a Dirac peak becomes dense in the limit $M\to\infty$. 
While these peaks are distributed within certain well-specified windows of energies $[E_\textrm{min},E_\textrm{max}]$---the energy bands of
the tight-binding Hamiltonian on the tree---its  
\emph{integrated} density of states is a function that is discontinuous for all energies within the energy band \cite{Deford:2020}.

In light of the above, we also need to sharpen our understanding of `flat bands' in Cayley trees and in their decorations.
Clearly, in a spectrum that consists solely of a discrete set of Dirac functions, any eigenenergy may be claimed to constitute an exact flat band.
While we occasionally use the term `flat band' for the \emph{largest} peak in the spectrum (i.e., for the energy of the shortest and most abundant non-symmetric sector), we mostly reserve the name for peaks in the density of state that appear \emph{outside of the energy bands}, i.e., in energy intervals where eigenvalues do not form a dense set over real numbers. 
We will find for several of the studied decorations of the Cayley trees that states forming flat bands---understood in this concrete and more narrow sense---can be interpreted as topological boundary modes of the one-dimensional chains captured by Hamiltonians~(\ref{Eq:Cayley_Symmetic_Sector}) and~(\ref{Eq:Cayley_NonSymmetric_Sector}) (with large-$l$ non-symmetric sectors giving the dominant contribution to the flat band).

Finally, let us comment on the notion of \emph{compact localized states} (CLS), which are a convenient tool for explaining the origin of flat bands from real-space topology in Euclidean lattices~\cite{bergman_band_2008}.
Conventionally, CLS is an eigenstate of a lattice Hamiltonian that is exactly confined to a finite set of lattice sites due to destructive interference of the (next-)nearest-neighbor hoppings and that has exactly zero amplitude on the remaining sites of the lattice.
From this perspective, almost all eigenstates on a Cayley tree (except eigenstates of sectors $\mathcal{H}_\textrm{sym.}$ and $\mathcal{H}_\textrm{nonsym.}^0$) are CLSs. 
Specifically, eigenstates of the sector $\mathcal{H}_\textrm{nonsym.}^\alpha$ with a seed node in layer $l$ are localized to the $K+K^2+\ldots K^{M-l} = K(K^{M-l}-1)/(K-1)$ sites on the branches emanating from $\alpha$, while having exactly zero amplitude everywhere else.
This confinement is ensured by destructive interference of the hopping processes at the seed node. 
Since essentially all eigenstates can be interpreted as CLSs, this notion clearly ceases to be a useful concept. 
For this reason, we will largely refrain from adopting the notion of CLSs in the context of Cayley trees, using it solely for the eigenstates of the shortest non-symmetric sectors whose support extends over ${\sim}{K}$ sites.

\section{Lieb decoration}\label{Sec:LiebDecoration}

In this section we introduce the Lieb decoration on the Cayley tree (the `Lieb-Cayley tree' for short), which corresponds to including an additional node in the middle of every edge of the lattice.
By adapting the technique of symmetry-adapted basis states, we find the exact spectrum, which exhibits 
a flat band at $E=0$. 
We show that this flat band cannot be explained through the standard geometric consideration of CLSs
localized over a small set (${\sim}{K}$) of sites; rather, 
its origin is 
traced to 
topological properties of the decorated tree. 

Our discussion of Lieb-decorated lattices is structured as follows.
In Sec.~\ref{Sec:Lieb_1_euclidean} we introduce the Lieb decoration of the Euclidean square lattice, 
and we show how the emergence of a flat band in this lattice is explained with CLSs. 
In Sec.~\ref{Sec:Lieb_2_exact} we outline how to generalize the approach of symmetry-adapted basis states to the Lieb-decorated Cayley tree, and we present exact results for its spectrum, including an analysis of the flat-band degeneracy. 
Finally, in Sec.~\ref{Sec:Lieb_3_topological} we show how a mapping of the individual symmetry sectors to the one-dimensional Su-Schriefer-Heeger model gives us an understanding for the \emph{topological} origin of the flat band. 
We also present here an alternative explanation for the formation of the zero-energy flat band that is rooted in 
the rank-nullity theorem.

\subsection{Euclidean Lieb Lattice}\label{Sec:Lieb_1_euclidean}
\begin{figure}[t!]
\centering
\includegraphics[width=\linewidth]{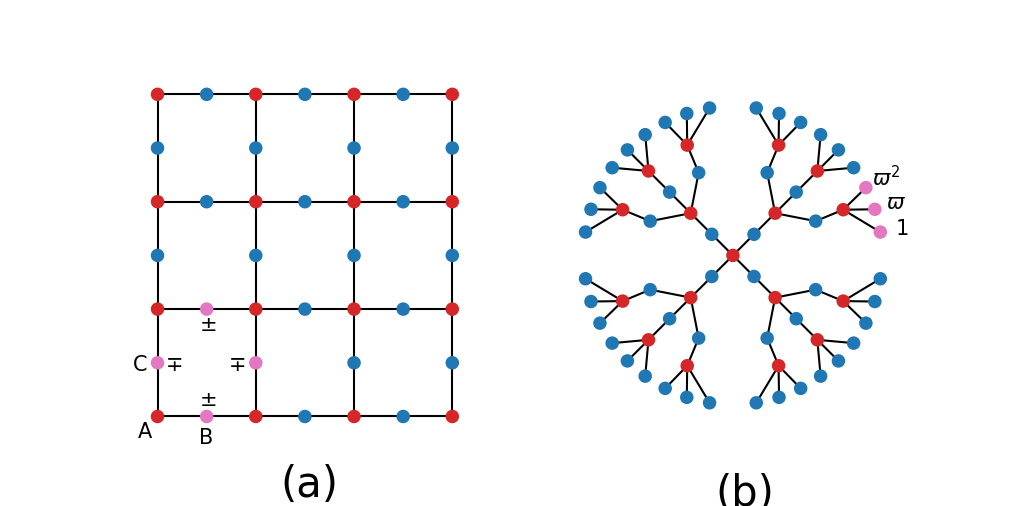}
\caption{(a) The Euclidean Lieb lattice: an additional site is placed at the center of each edge of the square lattice. 
The unit cell contains three sites, labeled $A$, $B$, $C$. 
A typical compact localized state has support on the four sites highlighted in pink, 
with positive (negative) amplitude on the $B$ sites (on the $C$ sites) as indicates with the $\pm$ signs. (b) The Lieb decoration of a Cayley tree with 
branching factor $K=3$. 
The Cayley nodes of the decorated tree are colored in red, while the Lieb nodes are colored in blue. A compact localized state has support on three leaves of the same parent on the outermost layer, with the nodes amplitude corresponding to powers of a non-trivial third root of unity, $\varpi$ [cf.~discussion below Eq.~(\ref{eqn:nonsym-at-root})].}
\label{Fig:LiebLattice_LiebCayleyTree}
\end{figure}

In this subsection, we introduce the Euclidean 
Lieb lattice 
which serves as the motivation for introducing the Lieb decoration of Cayley trees.
It is formed by adding sites to the midpoints of each edge in a square lattice unit cell, yielding a three-site basis inside the unit cell: one central site ($A$) and two edge sites ($B$ and $C$), as shown in Fig.~\ref{Fig:LiebLattice_LiebCayleyTree}. 
In the tight-binding approximation with nearest-neighbor hopping $t$ on a lattice with periodic boundary conditions, the momentum-space Hamiltonian can be written in the basis $(A, B, C)$ as
\begin{equation}
H(\mathbf{k}) = -t \begin{pmatrix}
0 & 1 + e^{-ik_x} & 1 + e^{-ik_y} \\
1 + e^{ik_x} & 0 & 0 \\
1 + e^{ik_y} & 0 & 0 \\
\end{pmatrix}
\end{equation}
where $\mathbf{k}=(k_x,k_y)$ is the two-dimensional momentum.
For simplicity,
we assumed the absence of on-site potentials. 
Solving the characteristic polynomial $\det(H - E I) = 0$ gives the energy spectrum,
\begin{equation}
E(\mathbf{k}) = 0,\; \pm t \sqrt{4 + 2\cos(k_x) + 2\cos(k_y)},
\end{equation}
where the zero-energy flat band appears between two symmetric dispersive bands \cite{lieb_two_1989,nita_spectral_2013,Mukherjee:2015}. 
Note that at $\mathbf{k} = (\pi,\pi)$ all three bands touch.

The existence of a flat band calls for its interpretation in terms of compact localized states (CLSs)~\cite{bergman_band_2008}, 
i.e., 
eigenstates with strictly finite spatial extent which emerge 
due to destructive interference patterns unique to the lattice geometry. 
In the Lieb lattice, a typical CLS can be constructed by placing non-zero amplitudes on a cross formed by four edge sites (two of each $B$ and $C$
). 
Specifically, as shown in Fig.~\ref{Fig:LiebLattice_LiebCayleyTree}, setting positive amplitude $+\psi$ on the two $B$ sites and negative $-\psi$ on the two $C$ sites ensures a destructive interference of the amplitudes at the adjacent $A$ sites, i.e., a formation of a zero-energy CLS: $H \ket{\psi_{\text{CLS}}}= 0$.
The possibility to construct such zero-energy CLS in each unit cell results in the formation of the flat band.

Alternatively, one can understand the appearance of the flat band through the rank-nullity theorem~\cite{bzdusek_flat_2022}. 
The Lieb lattice is bipartite with sublattice $\textrm{A}$ containing all sites with label $A$, and with sublattice $\textrm{L}$ (for `Lieb') containing all sites located on square-lattice edges ($B$ and $C$). 
Denote by $N_\textrm{A}$ ($N_\textrm{L}$) the number of sites on sublattice $\textrm{A}$ ($\textrm{L}$). Then the NN Hamiltonian of the Lieb 
lattice can be written in a block form~as
\begin{equation}
\label{Eq:RankNullity_BlockMatrix}
    \mathcal{H} = \begin{pmatrix}
        0_\textrm{L} & M \\
        M^{\dagger} & 0_\textrm{A}
    \end{pmatrix}
\end{equation}
where $0_\textrm{L}$ denotes the $N_\textrm{L} \times N_\textrm{L}$ zero matrix, $0_\textrm{A}$ the $N_\textrm{A} \times N_\textrm{A}$ zero matrix, $M$ is an $N_\textrm{L} \times N_\textrm{A}$ matrix which contains the hopping amplitudes, and $M^{\dagger}$ is its $N_\textrm{A} \times N_\textrm{L}$ Hermitian conjugate.

The rank-nullity theorem then applies as follows. Let $\widetilde{\psi}_\textrm{L} \in \text{null} (M^{\dagger} )$ be a right zero eigenvector of $M^\dagger$, and $\widetilde{\psi}_\textrm{A} \in \text{null} (M)$ a right zero eigenvector of $M$. 
The dimensions of these null spaces are $\widetilde{N}_L = \text{dim null} (M^\dagger)$ and $\widetilde{N}_A = \text{dim null} (M)$. 
The rank-nullity theorem ensures that 
\begin{equation}
    N_\textrm{L} = \text{rank}(M^\dagger) + \widetilde{N}_\textrm{L} \quad \text{and} \quad N_\textrm{A} = \text{rank} (M) + \widetilde{N}_\textrm{A}
\end{equation}
Combined with the fact that the ranks of the rectangular matrices $M$ and $M^\dagger$ must match, we find 
$N_\textrm{L} - \widetilde{N}_\textrm{L} = N_\textrm{A} -\widetilde{N}_\textrm{A}$, which 
implies $\widetilde{N}_\textrm{L} \geq \abs{N_\textrm{L} -N_\textrm{A}}$. 
The absolute value follows from replacing the role of the two sublattices in the derivation.
It follows 
that the dimension $\text{dim null} (M^\dagger)$ 
has a lower bound given by the imbalance of the two sublattices~\cite{sutherland_localization_1986}.
Finally,
for any $\widetilde{\psi}_L \in \text{null} (M^{\dagger} )$, the vector 
\begin{equation}
    \widetilde{\Psi}_L = \begin{pmatrix}
        \widetilde{\psi}_L \\
        0
    \end{pmatrix}
\end{equation}
is a zero eigenvector of $\mathcal{H}$ with support on the sublattice L only. 
Therefore, 
the Hamiltonian has a flat band at $E=0$ with degeneracy $N_\textrm{FBS}$, which obeys a lower bound given by the sublattice imbalance: $N_{\text{FBS}} \geq \abs{N_\textrm{L} - N_\textrm{A}}.$
For a system with $N$ unit cells and periodic boundary condition, we find that $N_\textrm{FBS} \geq N$ and the flat-band fraction $f = N_\textrm{FBS}/N_\textrm{sites} \geq 1/3$, where $N_\textrm{sites}=3N$ is the total number of sites in the lattice. 
A more careful analysis~\cite{bergman_band_2008} reveals that $N_\textrm{FBS}=N+2$, and that the flat-band fraction saturates the bound, $f = 1/3$, in the limit $N\to\infty$.
With open boundary condition, the extraction of $N_\textrm{FBS}$ requires more care, but the same flat-band fraction $f=1/3$ is recovered in the thermodynamic limit $N\to\infty$.

\subsection{Exact Solution of Lieb-Cayley tree}\label{Sec:Lieb_2_exact}

We now apply the Lieb decoration to the Cayley tree. Consider a Cayley tree with connectivity $K_\textrm{C}$, i.e., each node is connected to its $K_\textrm{C}$ children via an edge. 
We now add an additional node, called a ``Lieb node'', in the middle of each edge [see Fig.~\ref{Fig:LiebLattice_LiebCayleyTree}(b)].  
The symmetrical placement ensures that the decorated tree exhibits the same 
hopping amplitude $t$ from the original parent node to the Lieb node and from the Lieb node to the original child node. 
In this process, we 
effectively double the number of layers. 
Each Lieb node has a connectivity $K_\textrm{L} = 1$, i.e., only one single child node; therefore, we classify Lieb-Cayley trees according to the connectivity $ K_\textrm{C} \equiv K$ of the Cayley nodes. 
In graph theory, the process of including an additional site at the middle of each edge corresponds to the construction of 
a \emph{subdivision graph}; in the present case, the Lieb-Cayley tree `$S(\textrm{Cay})$' is the subdivision graph of the Cayley tree `$\textrm{Cay}$'. 

Consider a Lieb-Cayley tree with connectivity $K$ and the total number of layers $M = M_\textrm{C} + M_\textrm{L}$, where $M_\textrm{C}$ is the number of ``Cayley-layers''and $M_\textrm{L}$ the number of ``Lieb-layers'' (the root site $0$ counts to neither of these quantities). 
In practice, depending on the boundary termination, we have either $M_\textrm{L} = M_\textrm{C}$ or $M_\textrm{L} = M_\textrm{C}+1$.
The total number of states on this tree is
\begin{equation}
\label{eq:total_number_of_states_lieb_lattice}
    \begin{aligned}
    N_\textrm{total} 
    & = 1 + (K+1) \! \times \! \! \sum_{l_\textrm{C} = 1}^{M_\textrm{C}} K^{l_\textrm{C} - 1} + (K+1) \! \times \! \! \sum_{l_\textrm{L} = 1}^{M_\textrm{L}} K^{l_\textrm{L} - 1} \\
    & = 1 + \frac{(K+1)}{(K-1)} (K^{M_\textrm{C}} + K^{M_\textrm{L}} -2).
  \end{aligned}  
\end{equation}
Furthermore, let  $l$ be the layer index. 
We can introduce two sublattices on the Lieb-Cayley tree, with sublattice $\textrm{L}$ corresponding to $l$ odd (`Lieb layers') and sublattice $\textrm{C}$ corresponding to $l$ 
even (`Cayley layers'). 

\begin{figure}[t!]
\centering
\includegraphics[width=\linewidth]{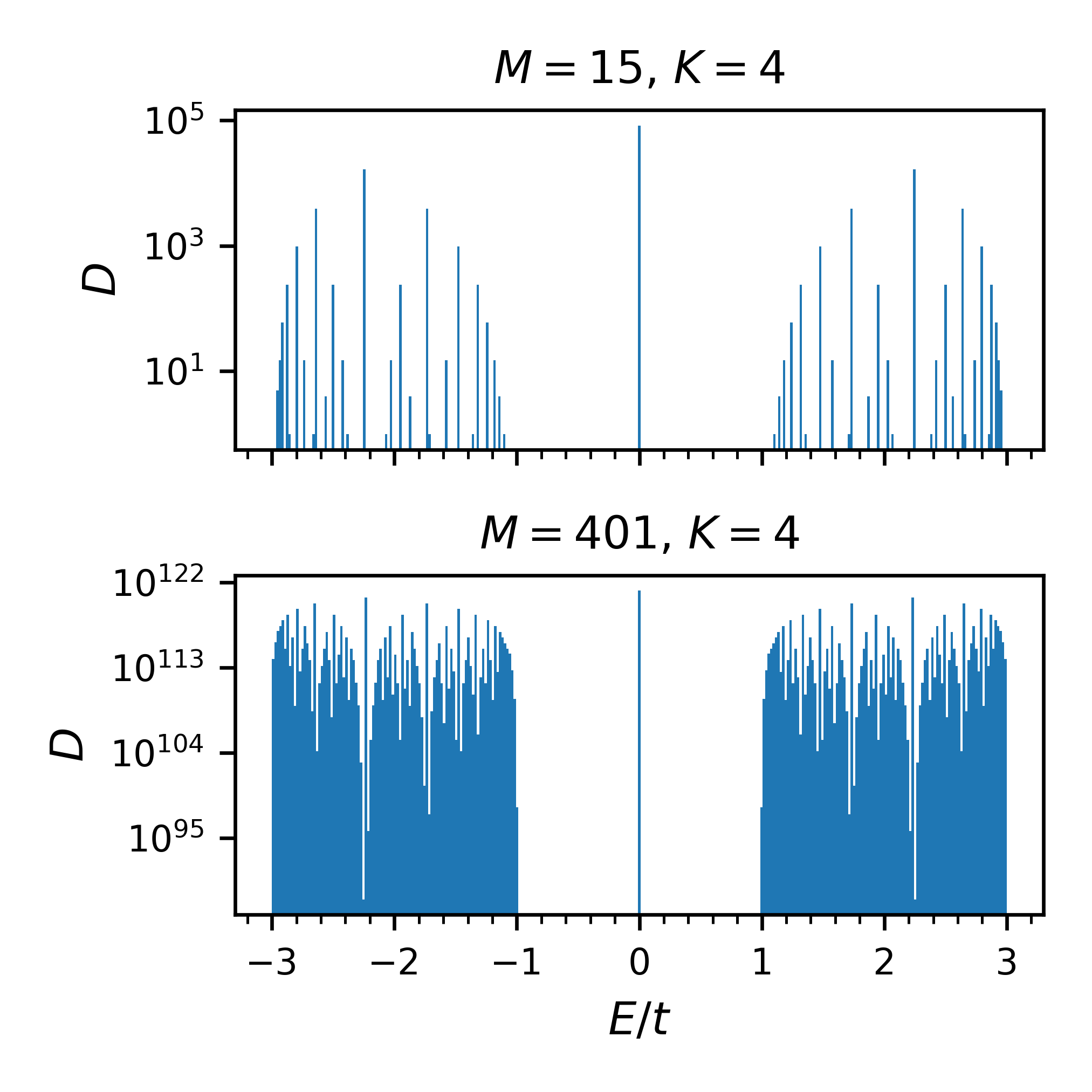}
\caption{Spectrum of 
a Lieb-Cayley tree for two choices of the number of shells $M$ and with connectivity $K=4$. 
Both spectra exhibit a flat band at $E=0$. 
}
\label{Fig:LiebCayley_Spectra}
\end{figure}

We aim to investigate the spectrum of the NN model on the Lieb-Cayley tree, with the hopping amplitude set to $t_1 = 1$.
To utilize the method of symmetry-adapted basis states,
one needs to introduce an additional set of basis states which constitute linear combinations of the Lieb nodes. 
Since the connectivity of the Lieb nodes is $K_\textrm{L}=1$, the hopping amplitude between a Lieb layer and its children in the subsequent Cayley layer is renormalized trivially with $\sqrt{K_\textrm{L}}=1$ inside the effective one-dimensional Hamiltonians of the symmetric and non-symmetric sectors (see Appendix~\ref{Sec:App_LiebCayley} for further details), resulting in
\begin{subequations}
\begin{equation}
\label{Eq:symm_LiebCayley_Hamiltonian}
    \mathcal{H}_{\textrm{sym.}} = \begin{pmatrix}
        0 & \sqrt{K+1} & 0 & 0 & 0 & 0 & \cdots \\
        \sqrt{K+1} & 0 & 1 & 0 & 0 & 0 & \cdots \\
        0 & 1 & 0 & \sqrt{K} & 0 & 0 &  \cdots \\
        0 & 0 & \sqrt{K} & 0 & 1 & 0 & \cdots \\
        0 & 0 & 0 & 1 & 0 & \sqrt{K} &  \cdots \\
        0 & 0 & 0 & 0 & \sqrt{K} & 0 & \cdots \\
        \vdots & \vdots & \vdots & \vdots & \vdots & \vdots & \ddots
        
    \end{pmatrix}
\end{equation}
and
\begin{equation}
\label{Eq:nonsymm_LiebCayley_Hamiltonian}
    \mathcal{H}_{\textrm{nonsym.}}^\alpha = \begin{pmatrix}
        0 & 1 & 0 & 0 & 0 & 0 & \cdots \\
        1 & 0 & \sqrt{K} & 0 & 0 & 0 & \cdots \\
        0 & \sqrt{K} & 0 & 1 & 0 & 0 &  \cdots \\
        0 & 0 & 1 & 0 & \sqrt{K} & 0 & \cdots \\
        0 & 0 & 0 & \sqrt{K} & 0 & 1 &  \cdots \\
        0 & 0 & 0 & 0 & 1 & 0 & \cdots \\
        \vdots & \vdots & \vdots & \vdots & \vdots & \vdots & \ddots
        
    \end{pmatrix}
\end{equation}
\end{subequations}
The simple form of the effective 1D Hamiltonians allows us to obtain the full spectrum of Lieb-Cayley trees using 
the procedure presented in Sec.~\ref{sec:methods}. 
In Fig.~\ref{Fig:LiebCayley_Spectra} we present the spectrum for the Lieb-Cayley tree with $K = 4$ for two different choices of the number of layers $M$. 
We find that the spectrum describes an insulator with a flat band at $E=0$. 

Drawing from an analogy with the Euclidean Lieb lattice, one may anticipate that the flat band of the Lieb-Cayley tree is underpinned by a CLS construction as well. 
Assuming a termination with a Lieb layer [corresponding to the case $M_\textrm{L} = M_\textrm{C} + 1$, shown in Fig.~\ref{Fig:LiebLattice_LiebCayleyTree}(b)], we can construct such a CLS by choosing the set of all leaf nodes $\{|i\rangle_{\alpha} \}_{i=1}^{K}$ (i.e., nodes on the outermost layer) originating from the same parent node $|\alpha \rangle$.
Specifically, we can assign these $K$ sites
wave function amplitudes that interfere destructively upon hopping to their parent node $\ket{\alpha}$. 
For a given connectivity $K$, such a CLS is written as
\begin{equation}\label{Eq:CLS_LiebCayley}
    \vert \psi_{\text{CLS}} \rangle = \frac{1}{\sqrt{K}} \sum_{i=1}^K \omega^i \ket{i}_\alpha
\end{equation}
where $\omega$ is a non-trivial $K$-th root of unity. 
There are $K\,{-}\,1$ possible choices for this root, giving us $K\,{-}\,1$ CLSs for each  
choice of a parent node. 
Inspecting Fig.~\ref{Fig:LiebLattice_LiebCayleyTree}, we recognize  that each node on the second outermost layer can be used as a parent node for constructing such CLSs. 
There are $N_\alpha = (K\,{+}\,1)K^{M_\textrm{C} -1}$ such nodes, leading to a total of 
\begin{equation}
\label{eqn:Lieb-Cayley-CLS-count}
    N_{\text{CLS}} = (K-1)(K+1)K^{M_\textrm{C}-1} = (K+1)(K^{M_\textrm{C}} - K^{M_\textrm{C}-1})
\end{equation}
such CLS. 
However, we numerically 
find that this value does not accurately predict the degeneracy of the flat band: 
the number of flat-band states $N_{\text{FBS}}$ is \emph{larger} than the derived number $N_{\text{CLS}}$, i.e., we are underestimating the flat-band fraction  
if we only consider the described CLSs.
The missing flat-band states correspond to eigenstates of shell-non-symmetric sectors with seeds at deeper layers. 
We show in the next subsection that these states are more conveniently interpreted as certain topological states rather than CLSs.

\subsection{Topological States}\label{Sec:Lieb_3_topological}

We here clarify the cause for underestimating the flat-band degeneracy by reconsidering the block Hamiltonians in Eqs.~(\ref{Eq:symm_LiebCayley_Hamiltonian}) and (\ref{Eq:nonsymm_LiebCayley_Hamiltonian}). 
Observe that these Hamiltonians map exactly to the Su-Schrieffer-Heeger (SSH) model --- a one-dimensional (1D) topological insulator originally introduced to describe the spectrum of polyacetylene molecules~\cite{su_solitons_1979}. 
This mapping implies that the spectrum of the Lieb-Cayley tree can be understood through an analogy with the SSH model. 

Recall that the SSH model consists of a bipartite tight-binding chain with alternating hopping amplitudes $t$ and $t'$ (assumed real and non-negative). 
The model exhibits distinct topological phases characterized by a winding number, and in the topological regime it hosts robust zero-energy edge states. 
The real-space tight-binding Hamiltonian in the presence of open boundary conditions is given by
\begin{eqnarray}
    H &=& t\sum_{m=1}^{N} \left( |m,B\rangle \langle m,A|  + \text{h.c.} \right) \nonumber \\
    &\phantom{=}& + \, t'\sum_{m=1}^{N-1} \left( |m+1,A\rangle \langle m,B|  + \text{h.c.} \right).
\end{eqnarray}
where $N$ is the number of unit cells, with each unit cell hosting two sites (corresponding to sublattices $A$ and $B$). 
Furthermore, 
$|m,A\rangle$ and $|m,B\rangle$, with $m\in\{1,2,\dots,N \}$, denote the basis states 
where the electron is in  
unit cell $m$  
on sublattice $A$ or $B$, respectively.
The Hamiltonian matrix in this basis reads
\begin{equation}\label{Eq:SSH_Hamiltonian_Matrix}
    \mathcal{H} 
    = \begin{pmatrix}
        0 & t & 0 & 0 & 0 & \cdots\\
        t & 0 & t' & 0 & 0 & \cdots\\
        0 & t' & 0 & t & 0 & \cdots\\
        0 & 0 & t & 0 & t' & \cdots\\
        0 & 0 & 0 & t' & 0 & \cdots \\
        \vdots & \vdots & \vdots & \vdots & \vdots & \ddots
    \end{pmatrix}
\end{equation}
Comparing this expression against 
Eqs.~(\ref{Eq:symm_LiebCayley_Hamiltonian}) and (\ref{Eq:nonsymm_LiebCayley_Hamiltonian}), we find that the shell-non-symmetric sectors map to the $t < t'$ instance of Eq.~(\ref{Eq:SSH_Hamiltonian_Matrix}), while  
the shell-symmetric sector maps to $t>t'$.

\begin{figure}[t!]
\centering
\includegraphics[width=\linewidth]{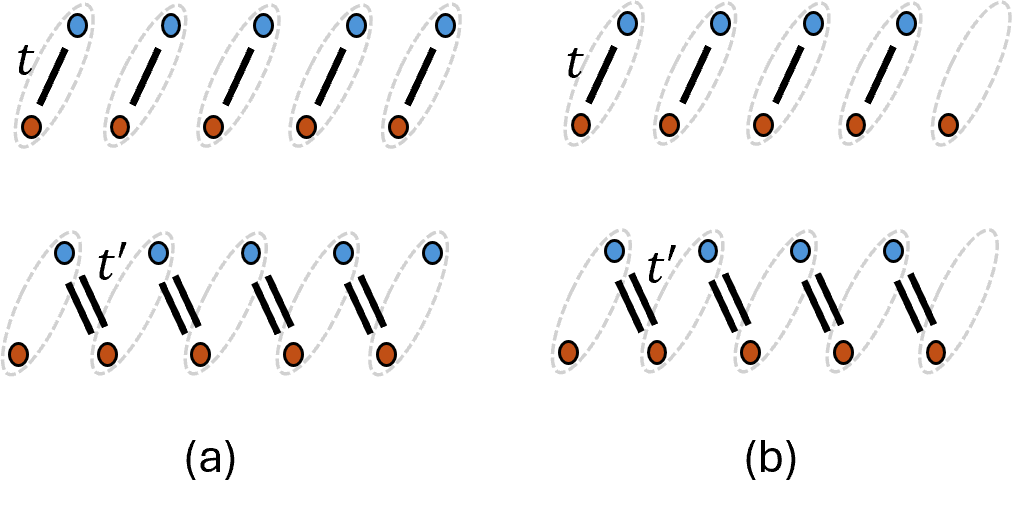}
\caption{Atomic limits of two SSH chains with different total number of sites. 
Sublattice $A$ is marked in red and sublattice $B$ in blue. 
The gray circles denote the unit cell. 
We use single (double) black lines for hopping amplitude $t$ ($t'$).
(a)~SSH chain with an even number of sites. In the topological phase of the model ($t'> t$) we find two isolated edge sites which are hosting the edge states. 
No edge states occur in the trivial phase ($t' < t$).
(b) SSH chain with an odd number of sites, which translates to a half-integer number of unit cells. 
We find that in both the trivial ($t' < t$) as well as in the topological phase ($t' > t$) there is an isolated edge site, but its location within 
the chain changes. 
In the topological phase (bottom panel), the edge state is 
localized at the beginning of the chain.}
\label{Fig:SSH_Chains}
\end{figure}

The momentum-space Hamiltonian of the SSH model reads
\begin{equation}
\label{Eq:SSH_bulk_momentumspace_hamiltonian}
    H (k) = \begin{pmatrix}
0 & t +t'e^{-ik}\\
t +t' e^{ik} & 0 
\end{pmatrix},
\end{equation}
with energy bands $E(k) = \pm \sqrt{t^2 + t'^2 + 2t t' \cos k}$.
The system is gapped for all $t \neq t'$, with a gap closing at $t = t'$. 
The winding number of the model is non-trivial for the $t < t'$ phase, which is associated with the formation of a zero-energy mode at each edge. 
The occurrence of the edge states can be understood from the real-space perspective by considering the atomic limits of the model (i.e., setting either $t=0$ or $t'=0$), when all inner sites form dimers with a gapped spectrum $E=\pm t$ (or $\pm t'$).
However, at the boundaries of the chain, and depending of the choice of the atomic limit and on the boundary termination, isolated sites may occur, each hosting a single eigenstate at 
energy $E=0$ (see Fig.~\ref{Fig:SSH_Chains}). 
As one moves 
away from the atomic limit, these states develop exponential tails seeping into the chain; however, the presence of sublattice symmetry fixes their energy at $E=0$.

\begin{figure}[t!]
\centering
\includegraphics[width=\linewidth]{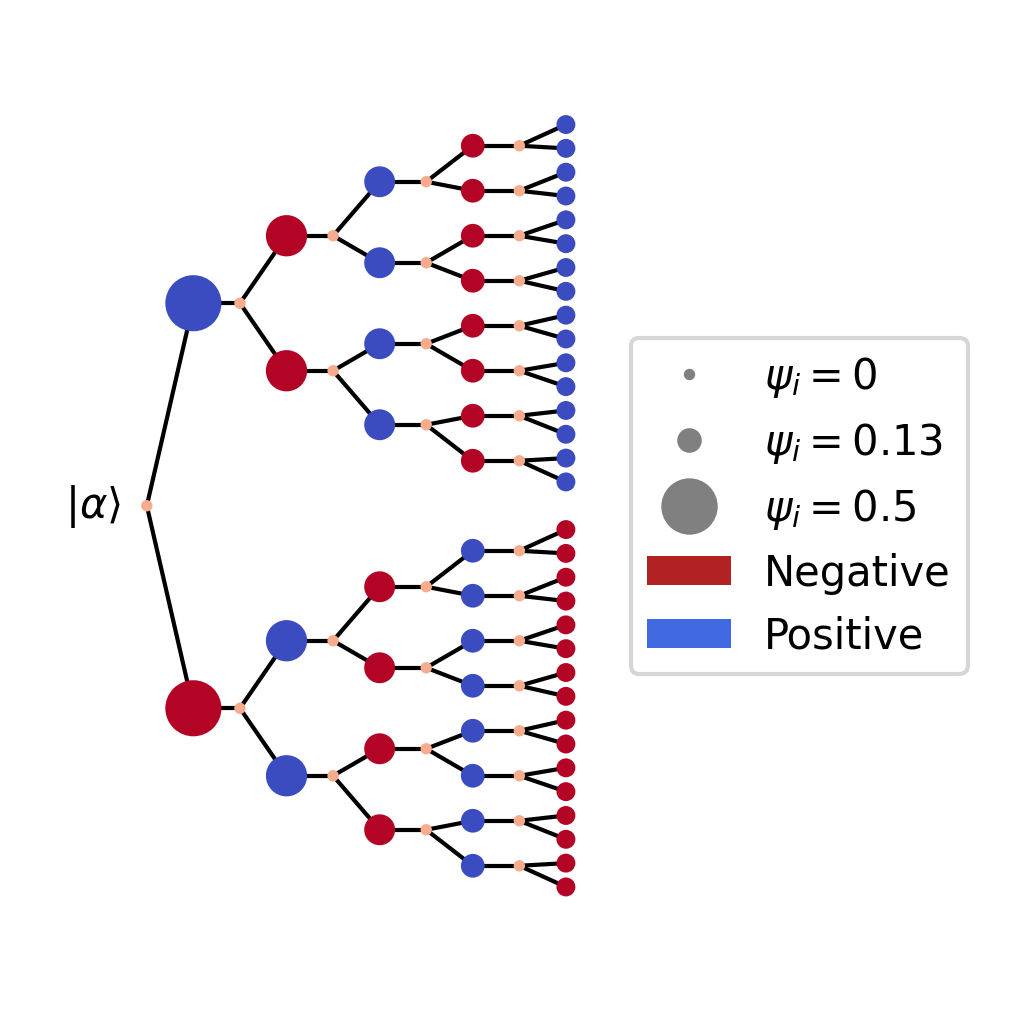}
\caption{An example of a zero-energy topological state 
hosted on a branch of the Lieb-Cayley tree with length $M-l = 9$ (not counting the seed node $\alpha$). 
The size (color) of the nodes corresponds to the absolute value (resp.~the sign) of the state's amplitude at the corresponding site 
according to the legend.  
Nodes with zero amplitude are 
colored in light red for better contrast. 
The amplitudes were found by exactly solving the Hamiltonian $\mathcal{H}_{\textrm{nonsym.}}^\alpha$ that describes the shell-non-symmetric sector hosted on this branch and then mapping the eigenstate with $E=0$ back to the branch via the definition of the shell-non-symmetric states (see Appendix~\ref{Sec:App_LiebCayley}). 
The topological state 
has support on the Lieb sublattice only and 
its amplitude decays exponentially  
as we move from the seed node to the outer boundary. }
\label{Fig:LiebCayley_TopologicalState}
\end{figure}

By mapping the block  
Hamiltonians in Eqs.~(\ref{Eq:symm_LiebCayley_Hamiltonian}) and~(\ref{Eq:nonsymm_LiebCayley_Hamiltonian}) 
to the SSH model of variable length, it is possible to explain the origin of the zero-energy flat band visible in Fig.~\ref{Fig:LiebCayley_Spectra}. 
There are two specific conclusions to be drawn for the Lieb-Cayley tree.  
First, the occurrence of topological edge states at the beginning of a SSH chain in the topological phase ($t' > t$) implies the existence of topological states \emph{in the bulk} of the Lieb-Cayley tree. 
This is an implication of our exact mapping from a shell-non-symmetric  
sector to the SSH model: the beginning of the SSH-chain maps directly to the first non-symmetric state of the respective sector. 
This first state is located \emph{inside} the tree for every non-symmetric symmetry sector except for the shortest  
one. [The shortest non-symmetric sector only lives on the leaf nodes and it has length $M-l = 1$, 
describing the CLSs in Eq.~(\ref{Eq:CLS_LiebCayley}).] 
In Fig.~\ref{Fig:LiebCayley_TopologicalState} we show an example of a branch of the Lieb-Cayley tree emanating from seed node $\ket{\alpha}$ together with 
the topological state hosted by this branch.  
The figure shows how the zero-energy states are exponentially localized at the inner boundary of the symmetry sector and thereby in the bulk of the Lieb-Cayley tree.
We find that for the Lieb-Cayley tree with the number of layers $M$ odd, all topological states are localized in the bulk through such a mechanism.

\begin{figure}[t!]
\centering
\includegraphics[width=\linewidth]{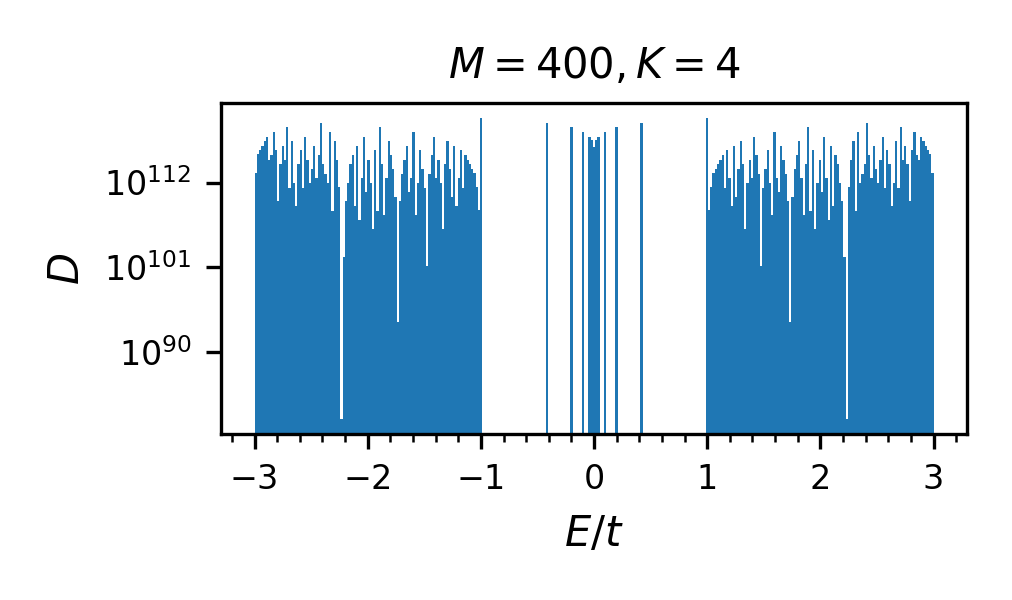}
\caption{Spectrum of a Lieb-Cayley tree with an even number of shells $M=400$ and connectivity $K=4$. We find a large number of states within the band gap which are symmetrically arranged around $E=0$. The origin of these states lies in the hybridization of edge states of short chains and in the fact that these states are not protected by the rank-nullity theorem.}
\label{Fig:LiebCayley_EvenVsOdd}
\end{figure}

The second conclusion is that the occurrence of the edge states depends on the length of the chain being even vs.~odd [recall Fig.~\ref{Fig:SSH_Chains}(a) vs.~(b)]. We therefore expect to see different patterns in the spectrum of the Lieb-Cayley tree depending on whether it terminates on a Lieb layer or on a Cayley layer. 
This prediction is confirmed by showing in Fig.~\ref{Fig:LiebCayley_EvenVsOdd} the spectrum of a Lieb-Cayley tree with an even number of layers. 
Specifically, while trees with an odd number of layers (exemplified by Fig.~\ref{Fig:LiebCayley_Spectra}) exhibit an exact flat band at $E=0$, trees with an even number of layers exhibit a symmetric distribution of near-zero (including only a few exact-zero) energy eigenstates spread over a finite energy interval.
The origin of this spreading lies in the hybridization of edge states located on opposite ends of a finite one-dimensional chain. 
Specifically, shell-non-symmetric sectors with $M$ even map to a SSH-chain with an even number of sites, which exhibits two edge states in the topological phase [bottom panel of Fig.~\ref{Fig:SSH_Chains}(a)]. 
When the chain is short (i.e., when the seed node is close to the outer boundary), the two edge states extend far enough into the chain to hybridize with each other, resulting in symmetrically arranged eigenvalues around $E=0$. This then also explains why we see a macroscopic fraction of states showing hybridization: the degeneracy of small sectors which map to short chains with a large overlap between the edge states, is much higher than the degeneracy of large sectors.
\subsection{Rank-Nullity Theorem and Disorder}
\begin{figure}[t!]
\centering
\includegraphics[width=\linewidth]{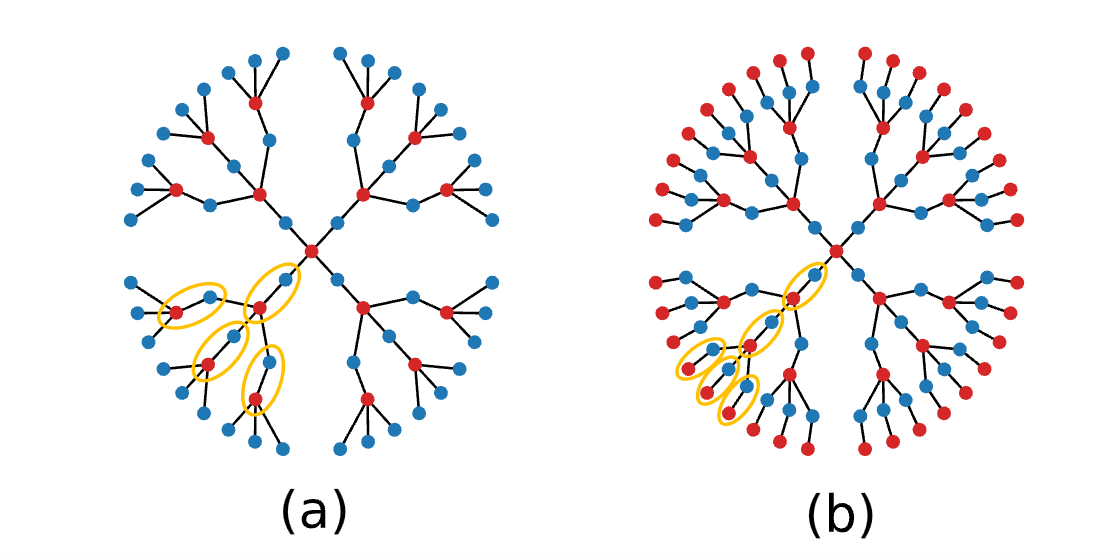}
\caption{We show two Lieb-Cayley trees with different numbers of layers. The Lieb sublattice is colored in blue, the Cayley sublattice is colored in red. Yellow ovals 
show how to pair elements of the sublattices 
in a way that allows for simple counting of the sublattice imbalance. 
(a)~For an odd number of layers it is not possible to match every node with a sublattice partner. 
Instead we find that all nodes at the boundary are lacking a partner and will therefore contribute to the sublattice imbalance. 
This results in a correspondingle large number of exact zero-energy eigenstates. 
(b)~For an even number of layers all nodes except the root can be matched with a sublattice partner. }
\label{Fig:LiebCayley_RankNullity}
\end{figure}

We can further elucidate the differences in the spectrum of Lieb-Cayley trees with odd and even number of shells by considering the consequences of the rank-nullity theorem, which dictates that a sublattice imbalance will lead to eigenstates with zero energy. Let $N_\textrm{L}$ be the number of sites on the Lieb sublattice and $N_\textrm{C}$ the number of sites on the Cayley sublattice. Then according to the rank-nullity theorem the number of flat band states is bounded by 
\begin{equation}
    N_{\text{FBS}} \geq \abs{N_\textrm{L} - N_\textrm{C}},
\end{equation}
which we can calculate exactly by using the considerations from Eq.~(\ref{eq:total_number_of_states_lieb_lattice}).
\begin{equation}
\begin{aligned}
    N_\textrm{L} - N_\textrm{C} &= (K+1) \times \sum_{l_\textrm{L} = 1}^{M_\textrm{L}} K^{l_\textrm{L} - 1} - (K+1) \times \sum_{l_\textrm{C} = 1}^{M_\textrm{C}} K^{l_\textrm{C} - 1} -1 \\
    & =\frac{K+1}{K-1}(K^{M_{L}} - K^{M_\textrm{C}}) - 1 
\end{aligned}
\end{equation}
There are two possible cases for the relation between $M_\textrm{L}$ and $M_\textrm{C}$. Either (1)~$M_\textrm{L} = M_\textrm{C}$ ($M$-even) in which case the above gives $-1$, translating to a single enforced state at $E=0$, or (2)~$M_\textrm{L} = M_\textrm{C} + 1$ ($M$-odd) which then gives us
\begin{equation}\label{Eq:LiebCayley_LowerBound_Flatband}
    N_\textrm{L} - N_\textrm{C} = \frac{K+1}{K-1}K^{M_\textrm{C}}(K - 1) - 1 = (K+1)\times K^{\lfloor M/2 \rfloor} - 1
\end{equation}
where we used that $M_\textrm{C} = \lfloor M/2\rfloor$. The above expression is exactly the number of nodes on the outermost layer minus the central node. That this must be the case can be proven visually. 
Consider Fig.~\ref{Fig:LiebCayley_RankNullity}: as we form Lieb-Cayley pairs on the $M$-odd tree, we find that the outermost layer and the central node cannot be matched with a partner. 

\begin{figure*}[t!]
\centering
\includegraphics[width=\textwidth]{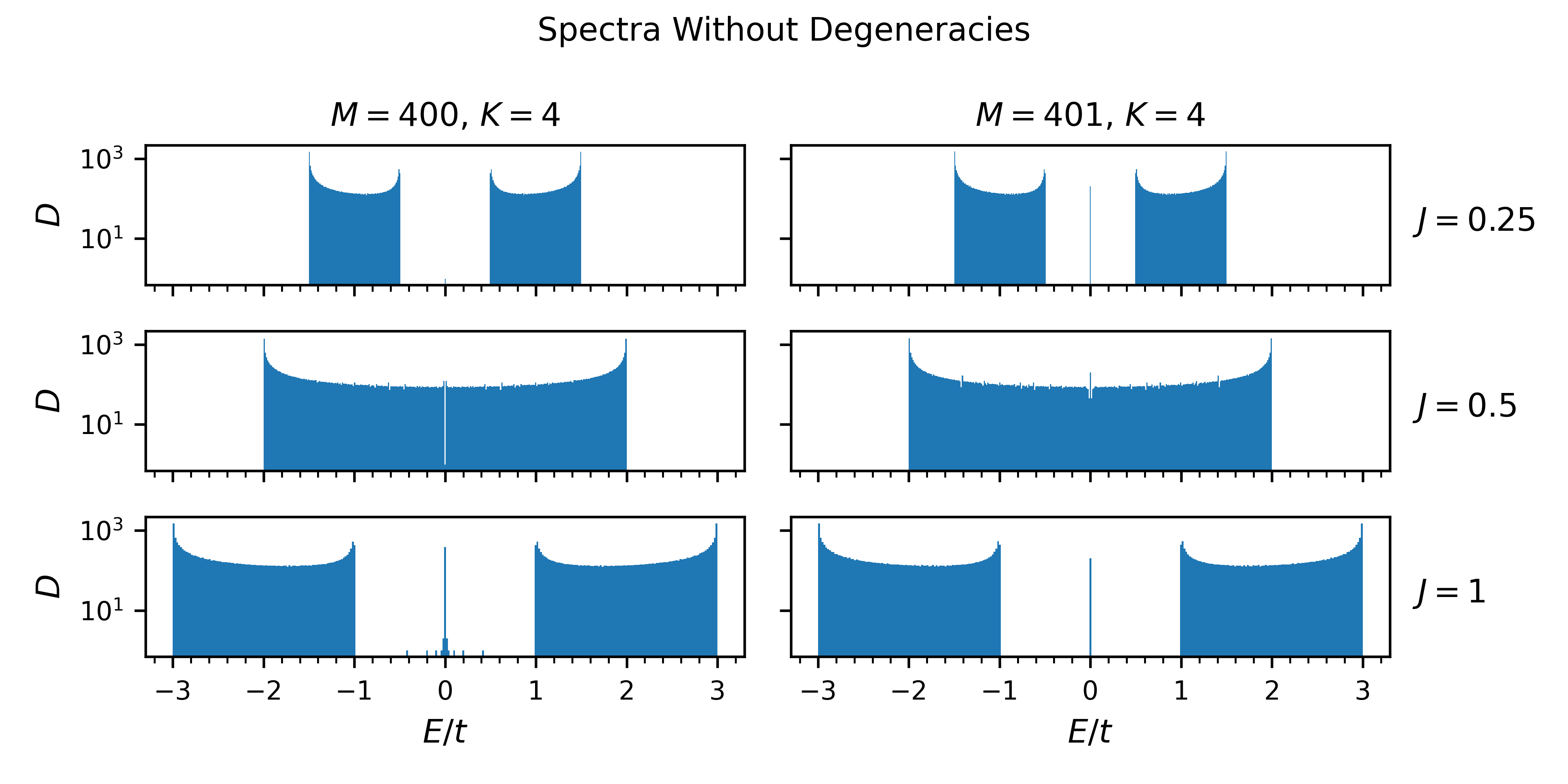}
\caption{Spectra of 
Lieb-Cayley trees with modulated hopping amplitudes $J=t_1/t_2$ and with an odd vs.~even number of layers.
For better readability, we here do not account for the (\emph{i})~the distinct roots of unity and for (\emph{ii})~the number of seed nodes in the Cayley layers, i.e., we solve the $M_\textrm{C} + 2$ unique symmetry sectors for their eigenvalues but do not count these eigenvalues with the sector's multiplicity. 
We find a gap closing for both trees at $J=0.5$, because at that point $\sqrt{K}\times J = 1$ such that the symmetry sectors are described by a simple 1D chain. 
In the trivial phase, $J=0.25$, we find that the tree with number of layers $M$ even has only a single in-gap state, as dictated by the rank-nullity theorem. 
In contrast, the $M$-odd tree exhibits the same number of in-gap states in both the topological and the trivial phase. 
The location of these in-gap states changes as explained from the SSH-model perspective in Fig.~\ref{Fig:SSH_Chains}.
}
\label{Fig:LiebCayley_PhaseTransition}
\end{figure*}

We find that the lower bound of Eq.~\ref{Eq:LiebCayley_LowerBound_Flatband}) directly corresponds to the number of zero-energy states that we find in the spectrum of our tree. 
This can be verified through analytical considerations.
Namely, we already explained that in the Lieb-Cayley tree with an odd $M$ each symmetry sector except the symmetric one contributes one topological zero-energy state at the beginning of the chain (recall Figs.~\ref{Fig:SSH_Chains} and \ref{Fig:LiebCayley_TopologicalState}). 
The number of seed nodes giving rise to non-symmetric sectors is 
\begin{equation}
\begin{aligned}
        N_\textrm{seeds}
                & = (K+1) \times \sum_{l_\textrm{C}=1}^{M_\textrm{C}} K^{l_\textrm{C}-1} + 1 \\
        & = \frac{(K+1)}{(K-1)}(K^{M_\textrm{C}}-1) + 1.
\end{aligned}
\end{equation}
The extra `$+1$' derives from taking the root node $0$ as the seed. 
For $K \geq 2$ each of these seeds (except the root node) results in $(K-1)$ non-symmetric sectors according to the possible choices of the $K$-th root of unity $\omega$, while the root node generates $K$ non-symmetric sectors. 
The total number of topological zero-energy states hosted by symmetry sectors is therefore
\begin{equation}
    N_{\text{Edge-States}} = (K+1)\times (K^{M_\textrm{C}}-1) + K = (K+1)\times K^{M_\textrm{C}} - 1
\end{equation}
which matches Eq.~(\ref{Eq:LiebCayley_LowerBound_Flatband}).
In summary, 
we find that the number of zero-energy eigenstates protected by the rank-nullity theorem for $M$-odd Lieb-Cayley tree corresponds exactly to the number of topological edge states hosted by the individual non-symmetric sectors. 
For the $M$-even tree there is only one exact-zero-energy state protected by the rank-nullity theorem; all the other topological states are allowed to hybridize due to the finite length of the tree branches. 

That this result coincides with the prediction of edge states as presented in Fig.~\ref{Fig:SSH_Chains} can be further exemplified by studying phase transitions in the Lieb-Cayley tree which are driven by introducing a variation of the hopping amplitudes.
Specifically, we distinguish between ``branching-hoppings'' (which carry a factor of $\sqrt{K}$ inside the block Hamiltonians) and ``mono-hoppings'' (which carry a factor of $1$). 
We modify the NN Hamiltonian on the Lieb-Cayley tree by setting the hopping amplitude on the branching-hoppings to $t_1$ and on the mono-hoppings to $t_2$.
The formation of the energy gaps depends only on the ratio 
$J=t_1/t_2$. 
Setting $J \neq 1$ breaks the homogeneity of the tree; however, 
such modification at the level of the symmetry sectors actually preserves inversion and chiral symmetry of the effective one-dimensional Bloch Hamiltonians. 
By tuning $J$ we can induce a phase transition, moving from the topological phase into the trivial phase of the Lieb-Cayley tree. 
In Fig.~\ref{Fig:LiebCayley_PhaseTransition} we show how the spectrum evolves as we tune $J$. 
For the $M$-even Lieb-Cayley tree we find that the in-gap states disappear as we enter the trivial phase. This corresponds to reaching the arrangement in the top panel of Fig.~\ref{Fig:SSH_Chains}(a).
In contrast, for the $M$-odd tree we see that the in-gap states persist into the trivial phase. 
This agrees with how the termination conditions of SSH-chains influence the occurrence of edge states, as shown in Fig.~\ref{Fig:SSH_Chains}. The in-gap states in the trivial phase are located at the outer boundary of the tree (i.e., they are no longer localized in the bulk). 

\begin{figure}[t]
\centering
\includegraphics[width=\linewidth]{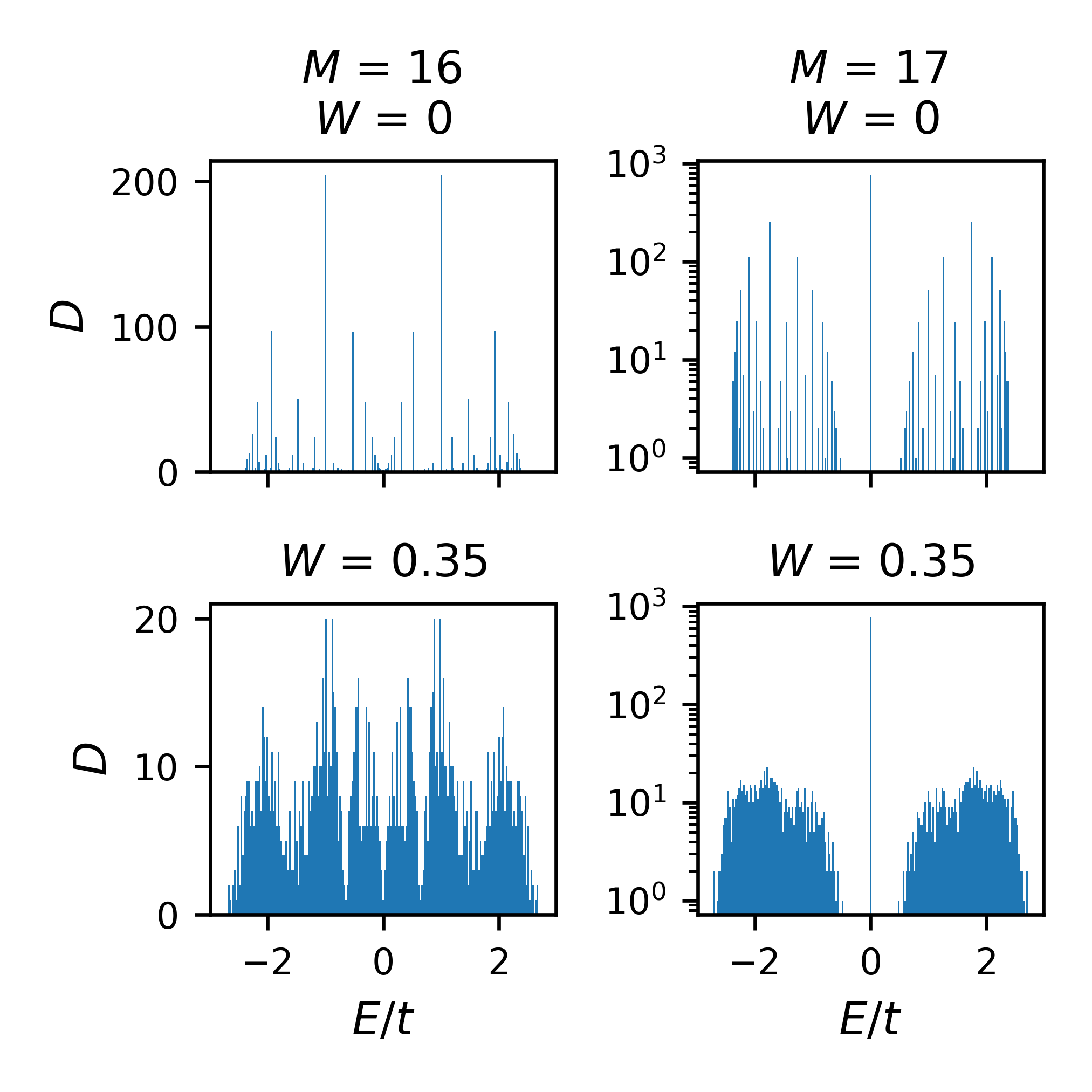}
\caption{The spectra of Lieb-Cayley trees with an even vs.~odd number of layers $M$ and with vs.~without bond disorder $W$. 
For the $M$-even case, the clean model ($W=0$) exhibits a single state at $E=0$ which cannot be resolved on the shown scale. Correspondingly, all sharp features of the spectrum disappear when bond disorder $W>0$ is introduced. 
In contrast, for the $M$-odd case, topological principles (captured either by the mapping onto SSH chains, or by virtue of the rank-nullity theorem) ensure the presence of a large number of exact $E=0$ states.
The bond disorder $W>0$ is compatible with the sublattice symmetry of the model, and therefore preserves the~$E=0$~peak.}
\label{Fig:LiebCayley_BondDisorder}
\end{figure}

Finally, we consider the impact of disorder on this system. 
Specifically, to preserve the block-odd-diagonal Hamiltonian with sublattice symmetry per Eq.~(\ref{Eq:RankNullity_BlockMatrix}), we consider \emph{bond disorder}, i.e., we introduce a random, uncorrelated noise on each hopping amplitude. 
The noise is drawn from a uniform distribution with $\epsilon_{ij} \in [-W,W]$ such that $t_{ij} \to t_{ij}(1+\epsilon_{ij})$ where $t_{ij} = t_{ji}$ is the hopping amplitude between NN sites $i$ and $j$. 
The spectrum is found using exact diagonalization. 
The bond disorder breaks the permutation symmetry of the tree; therefore, one can no longer rely on the symmetry-adapted basis. 
However, it preserves the sublattice symmetry and the assumptions for the rank-nullity theorem. 
In Fig.~\ref{Fig:LiebCayley_BondDisorder} we show the impact of bond disorder on the spectrum of a small Lieb-Cayley tree. 
We observe that the sharp features of the spectrum become smeared; with the sole exception of the exact-zero-energy states for the Lieb-Cayley tree with odd $M$. 
Due to continuity (i.e., a smooth turning on of the disorder), we anticipate that these zero-energy states remain localized in the bulk, which we confirmed with a numerical analysis.

In the subsequent sections, where we analyze other decorations of Cayley trees, we will not discuss the impact of disorder, as its effect on topological edge states and on its interplay with the rank-nullity is qualitatively similar.

We have shown that geometric considerations are not enough to explain the occurrence and behavior of zero-energy states but that through our mapping to 1D-models we can gain further insight into the behavior of the Lieb-Cayley tree, specifically showing that this system hosts topological edge modes in the bulk. 
This mapping further implies that the termination condition of the tree changes the number of edge states and thereby whether or not these edge states have a partner to hybridize with on a finite chain. 
We are then able to connect this with the real-space picture through the rank-nullity theorem, which dictates the stability of these states.

\section{Double Lieb decoration}\label{Sec:DoubleLiebDecoration}
In this section, we discuss a Cayley tree with a double Lieb decoration, i.e., with two additional nodes added on each edge. 
We dub this tree the ``Double Lieb-Cayley tree'' or ``2xLieb-Cayley tree''. 
Using the approach of symmetry-adapted basis states, 
we find the exact spectrum, which exhibits two flat bands, one at $E=1$ and the other at $E=-1$. 
We show that the exactness of these flat bands depends on the termination condition of the tree and that they can be understood to have a topological origin through a mapping to the trimer SSH model. 
Finally, we rewrite the eigenvalue equation of the system to reveal how the rank-nullity theorem underpins these flat bands.

Our discussion is structured as follows.
In Sec.~\ref{sec:2xLC_Euclidean} we introduce the Euclidean ``extended Lieb lattice'', which serves as a motivation for the 2xLieb-Cayley tree and which also exhibits two flat bands at $E=\pm1$. 
Next, in Sec.~\ref{sec:2xLC_ExactSolution} we extend the approach of symmetry-adapted basis states to the 2xLieb-Cayley tree, calculate the exact spectrum, and show that the rank-nullity theorem protects an additional flat band at $E=0$, which is hidden by the bulk energy bands. 
In Sec.~\ref{sec:2xLC_topology} we elaborate on how solutions of the 1D-chain models, to which the 2xLieb-Cayley tree maps within the individual symmetry sectors, depend on the termination condition of the chain. 
We further discuss the topological properties of the underlying 1D chain model, known as the ``trimer SSH'' model, to substantiate the topological origin of the $E=\pm1$ states. Finally, in Sec.~\ref{sec:2xLC_RankNullity} we demonstrate how one can rewrite the eigenvalue equation of the full 2xLieb-Cayley tree to prove that the flat bands at $E=\pm1$ are protected by the rank-nullity theorem.

\subsection{Extended Lieb Lattice}\label{sec:2xLC_Euclidean}
\begin{figure}[t!]
\centering
\includegraphics[width=\linewidth]{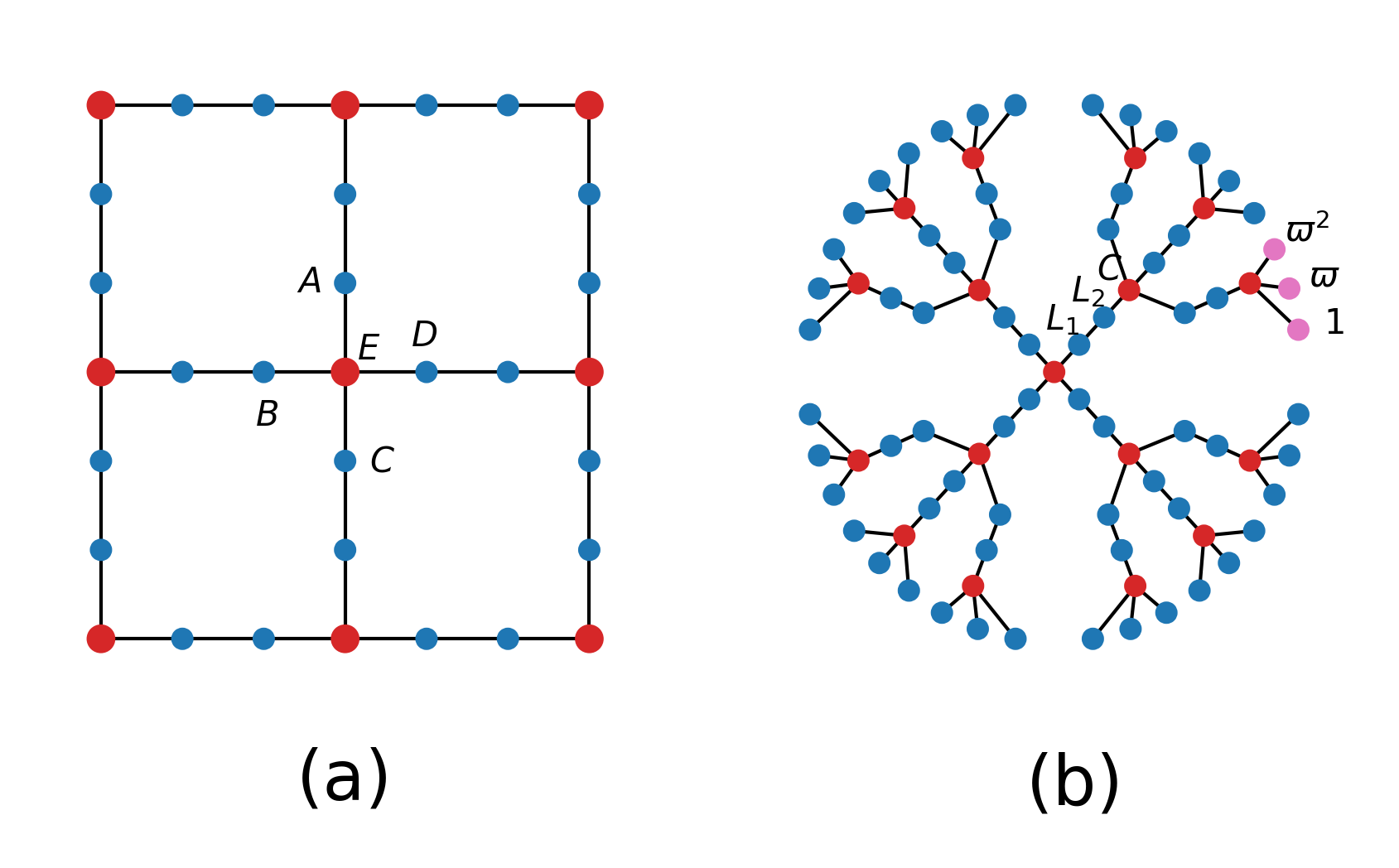}
\caption{(a) The extended Lieb-lattice with the atomic sites of the unit cell ($A,B,C,D,E$) marked. (b) A Cayley tree with double Lieb decoration. 
For reference we marked a Lieb node on a first Lieb-layer ($\textrm{L}_1$), a Lieb node on a second Lieb-layer ($\textrm{L}_2$) and a node on a Cayley-layer ($\textrm{C}$).
An example of a CLS on the tree has been added in pink, with $\varpi$ a non-trivial third root of unity. (Note that the exact form of the CLS depends on the termination of the tree; here, only one termination is shown and, therefore, only one instance of CLS~is~illustrated.)}
\label{Fig:2xLC_Sketch}
\end{figure}

The Euclidean analog of the 2xLieb-Cayley tree, which motivates this analysis, is the ``extended Lieb lattice'', sometimes called the ``Lieb-5 lattice'' \cite{zhang_new_2017,bhattacharya_flat_2019,centala_compact_2023,hanafi_localized_2022}. 
It is formed by adding two additional sites on each edge of the square lattice, resulting in five atomic sites ($A,B,C,D,E$) per unit cell as shown in Fig.~\ref{Fig:2xLC_Sketch}. 
In a tight-binding approximation with nearest-neighbor hopping $t$ and with periodic boundary conditions, the momentum-space Hamiltonian can be written in the basis ($A,B,C,D,E$) as
\begin{equation}
    H(\mathbf{k}) = t \begin{pmatrix}
        0 & 0 & e^{ik_y} & 0 & 1 \\
        0 & 0 & 0 & e^{-ik_x} & 1 \\
        e^{-ik_y} & 0 & 0 & 0 & 1 \\
        0 & e^{ik_x} & 0 & 0 & 1 \\
        1 & 1 & 1 & 1 & 0
    \end{pmatrix},
\end{equation}
where we assumed no on-site potentials. 
Solving for the dispersion relation one finds flat bands at $E=\pm 1$ with a dispersive band between them and two more dispersive bands, one below $E=-1$ and one above $E=+1$. For the exact form of the dispersion relation, we refer readers to Ref.~\citenum{zhang_new_2017}. 

\begin{figure}[t!]
\centering
\includegraphics[width=\linewidth]{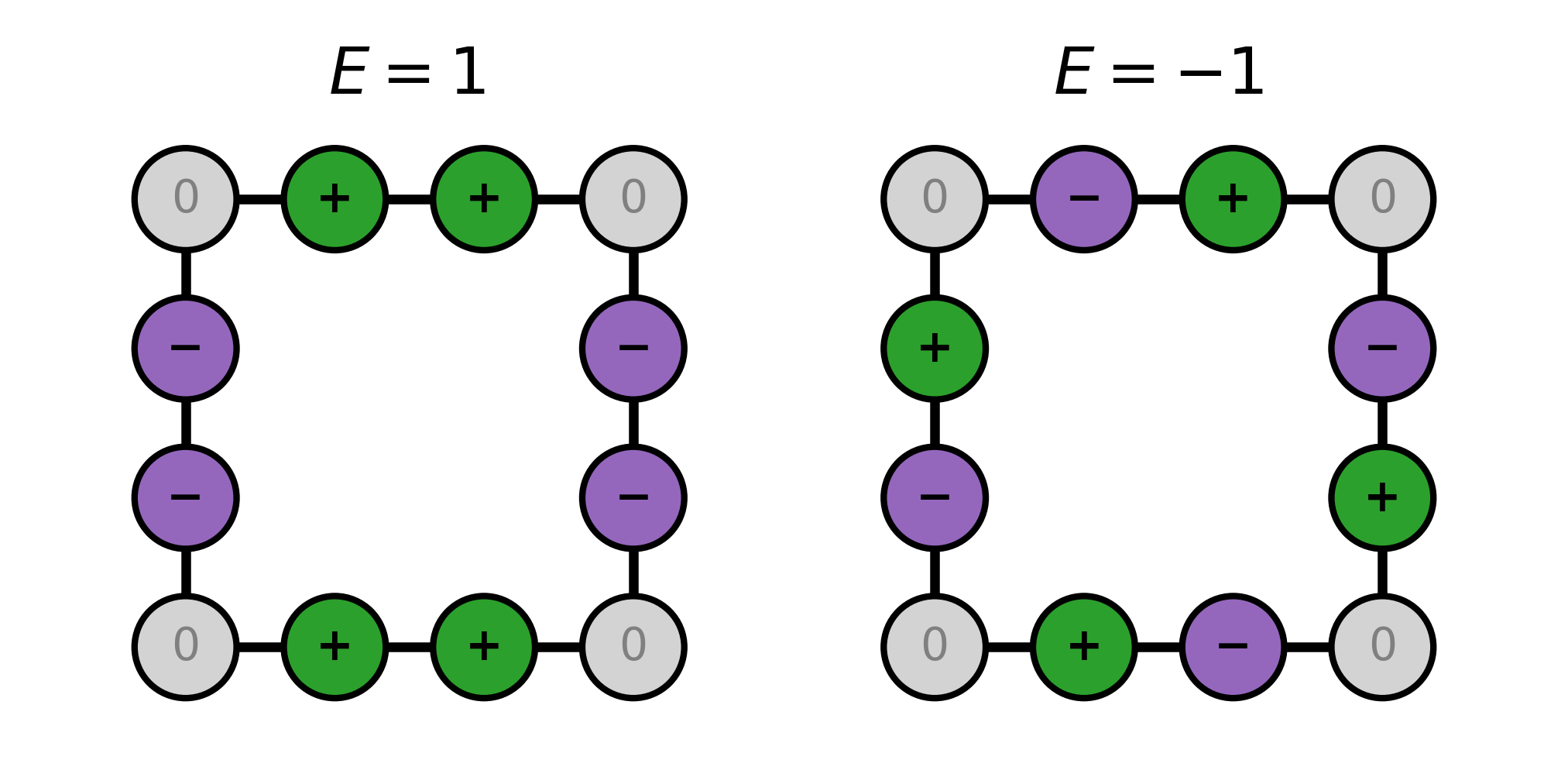}
\caption{The compact localized states that make up the two flat bands of the extended Lieb lattice, denoted with their respective energies.}
\label{Fig:2xLC_CLS_Sketch}
\end{figure}

The origin of these flat bands can be explained by constructing 
CLSs.
Consider one square on the extended Lieb-lattice. 
As shown in Fig.~\ref{Fig:2xLC_CLS_Sketch}, we can construct an arrangement of wavefunction amplitudes on the eights sites along the edges of this square 
that destructively interfere as they hop to the corners of the square. 
There are two distinct arrangements resulting in such destructive interference, one with energy $E=1$ [Fig.~\ref{Fig:2xLC_CLS_Sketch}(a)] and the other with $E=-1$ [Fig.~\ref{Fig:2xLC_CLS_Sketch}(b)]. 

\subsection{Exact Solution of 2xLieb-Cayley tree}\label{sec:2xLC_ExactSolution}
Drawing from the extended Lieb lattice, we introduce two additional ``Lieb nodes'' ($\textrm{L}1$ and $\textrm{L}2$) on each edge of the Cayley tree. 
For simplicity, we place these additional sites equidistantly along the edge, and we assume the same hopping amplitude $t_1 = 1$ between all NN sites.
Specifically, consider a Cayley tree with connectivity $K$ and with $M_\textrm{C}$ layers; then, by adding the Lieb nodes we triple the number of layers. 
In Fig.~\ref{Fig:2xLC_Sketch} we show a possible 2xLieb-Cayley tree. 
In graph theory, such construction is called a \emph{$2$-subdivision graph}; in the present case, the 2xLieb-Cayley tree `$S_2(\textrm{Cay})$' is the $2$-subdivision graph of the Cayley tree `$\textrm{Cay}$'.

The total number of states of this tree is given as 
\begin{equation}
\label{eqn:2xLieb-Cayley-counting}
\begin{aligned}
    &\phantom{=} N_\textrm{total} =  \\ 
    & = 1 + (K+1)\times \left(\sum_{l_\textrm{C}}^{M_\textrm{C}} K^{l_\textrm{C}-1} + \sum_{l_{\textrm{L}1}}^{M_{\textrm{L}1}} K^{l_{\textrm{L}1}-1} + \sum_{l_{\textrm{L}2}}^{M_{\textrm{L}2}} K^{l_{\textrm{L}2}-1} \right) \\
    & = 1 + \frac{K+1}{K-1}(K^{M_\textrm{C}} + K^{M_{\textrm{L}1}} + K^{M_{\textrm{L}2}} - 3).
\end{aligned}
\end{equation}
Here $M = M_\textrm{C} + M_{\textrm{L}1} + M_{\textrm{L}2}$, with $M_\textrm{C}$ the number of Cayley layers, $M_{\textrm{L}1}$ the number of first Lieb layers and $M_{\textrm{L}2}$ the number of second Lieb layers. In Fig.~\ref{Fig:2xLC_Sketch} we labeled a node on each of the corresponding layers with $\textrm{L}_1$, $\textrm{L}_2$ and $\textrm{C}$ respectively. 
We find that there are three different cases for the relationship between the numbers of layers:
\begin{itemize}
    \item Case~0: $M_{\textrm{L}1} = M_{\textrm{L}2} = M_\textrm{C}$, the tree ends on a Cayley layer.
    \item Case~1: $M_{\textrm{L}1} = M_{\textrm{L}2} + 1 = M_\textrm{C} + 1$, the tree ends on a first Lieb layer.
    \item Case~2: $M_{\textrm{L}1} = M_{\textrm{L}2} = M_\textrm{C} + 1$, the tree ends on a second Lieb layer
\end{itemize}
The enumeration of these cases follows the modulus 3 of the total layer number $M$. For example, a tree with 9 layers has 
$M = 0 \;\textrm{(mod $3$)}$, i.e., it is a ``Case~0'' tree.

\begin{figure}[t!]
\centering
\includegraphics[width=\linewidth]{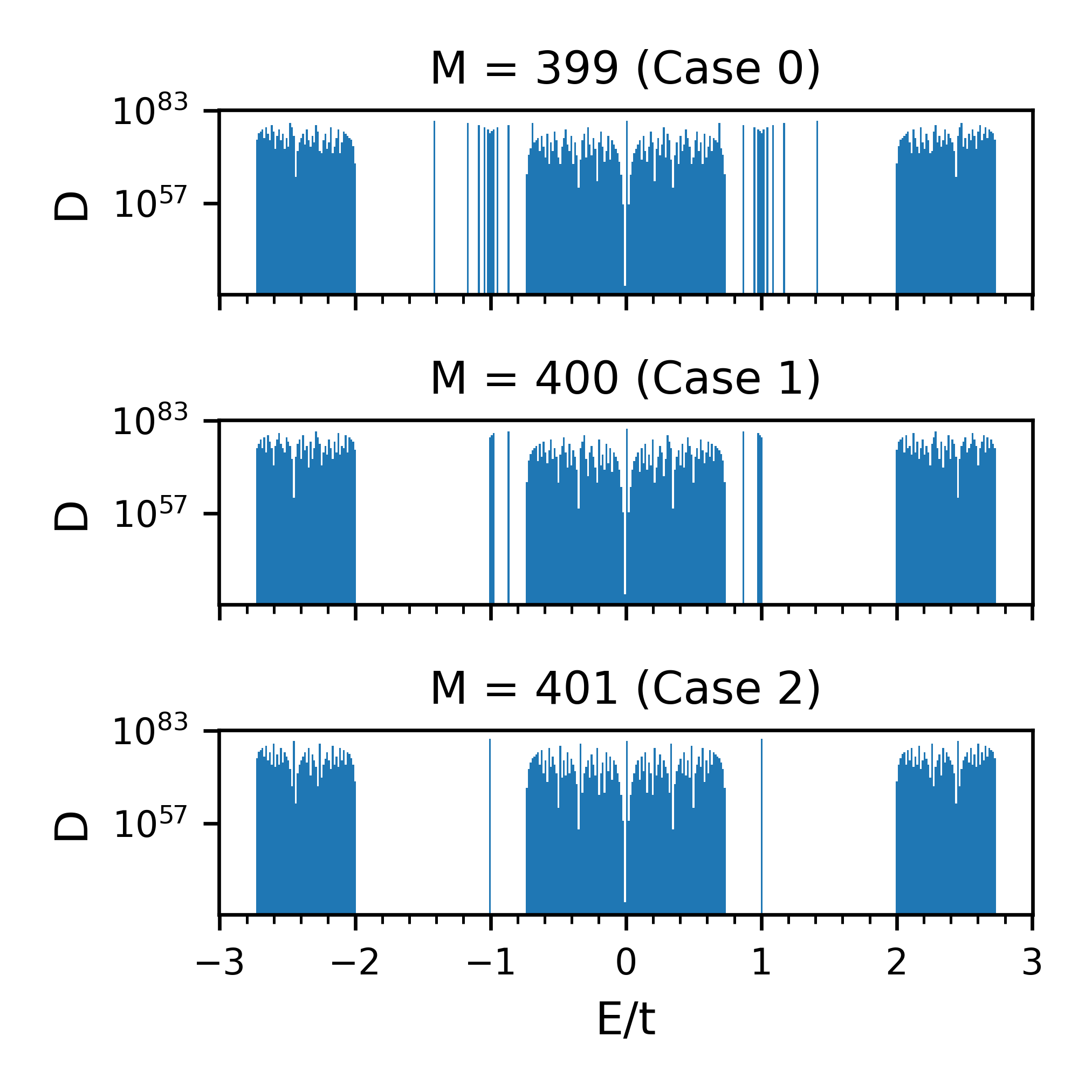}
\caption{Spectrum 
of a 2xLieb-Cayley tree for three choices of the number of shells $M$, each corresponding to a different termination condition, and with connectivity $K=4$. 
All spectra exhibit three energy bands as well as in-gap states at $E=\pm1$. 
Case~0 shows in-gap states that are symmetrically arranged near 
$E=\pm 1$, Case~1 shows a slight dispersion around $E=\pm1$, while Case~2 exhibits an exact flat band at both $E=1$ and $E=-1$.
In addition, an exact flat band at $E=0$ is visible in the middle of the central energy band for each boundary termination.}
\label{Fig:DoubleLiebCayley_Spectra}
\end{figure}

To extend the approach of symmetry-adapted basis states to the 2xLieb-Cayley tree, we proceed by the same principle as described for the Lieb-Cayley tree in Sec.~\ref{Sec:Lieb_2_exact}.
By adding two additional sets of (non-)symmetric states which form the relevant linear combinations across first and second Lieb layers, we find a complete and orthogonal set of states (explicitly listed in Appendix~\ref{Sec:App_2xLC}). 
Because Lieb nodes have coordination number $K_{\textrm{L}_i} = 1$ with $i \in [1,2]$, the symmetry-adapted hopping amplitude in the corresponding symmetry sectors will simply be $1$. 
The Hamiltonians of the symmetry sectors are therefore found to be
\begin{subequations}
\begin{equation}
    \mathcal{H}_{\textrm{sym.}} = \begin{pmatrix}
        0 & \sqrt{K+1} & 0 & 0 & 0 & 0 & \cdots \\
        \sqrt{K+1} & 0 & 1 & 0 & 0 & 0 & \cdots \\
        0 & 1 & 0 & 1 & 0 & 0 &  \cdots \\
        0 & 0 & 1 & 0 & \sqrt{K} & 0 & \cdots \\
        0 & 0 & 0 & \sqrt{K} & 0 & 1 &  \cdots \\
        0 & 0 & 0 & 0 & 1 & 0 & \cdots \\
        \vdots & \vdots & \vdots & \vdots & \vdots & \vdots & \ddots
        
    \end{pmatrix}
\end{equation}
and
\begin{equation}
\label{Eq:2xLiebCayley_Nonsymm_Hamiltonian}
    \mathcal{H}_{\textrm{nonsym.}}^\alpha = \begin{pmatrix}
        0 & 1 & 0 & 0 & 0 & 0 & \cdots \\
        1 & 0 & 1 & 0 & 0 & 0 & \cdots \\
        0 & 1 & 0 & \sqrt{K} & 0 & 0 &  \cdots \\
        0 & 0 & \sqrt{K} & 0 & 1 & 0 & \cdots \\
        0 & 0 & 0 & 1 & 0 & 1 &  \cdots \\
        0 & 0 & 0 & 0 & 1 & 0 & \cdots \\
        \vdots & \vdots & \vdots & \vdots & \vdots & \vdots & \ddots
    \end{pmatrix}
\end{equation}
\end{subequations}
We solve these according to the procedure presented in Sec.~\ref{Sec:Methods_B}, thereby obtaining the exact spectrum. 
In Fig.~\ref{Fig:DoubleLiebCayley_Spectra} we present the spectrum for three choices of $M$. 
We observe 
that only in trees with the number of layers $M = 2 \;\textrm{(mod $3$)}$ 
do the in-gap states form exact flat bands. 
Before we clarify 
the origin of the in-gap states, we comment on the implications dictated by the rank-nullity theorem.

\begin{figure}[t!]
\centering
\includegraphics[width=0.8\linewidth]{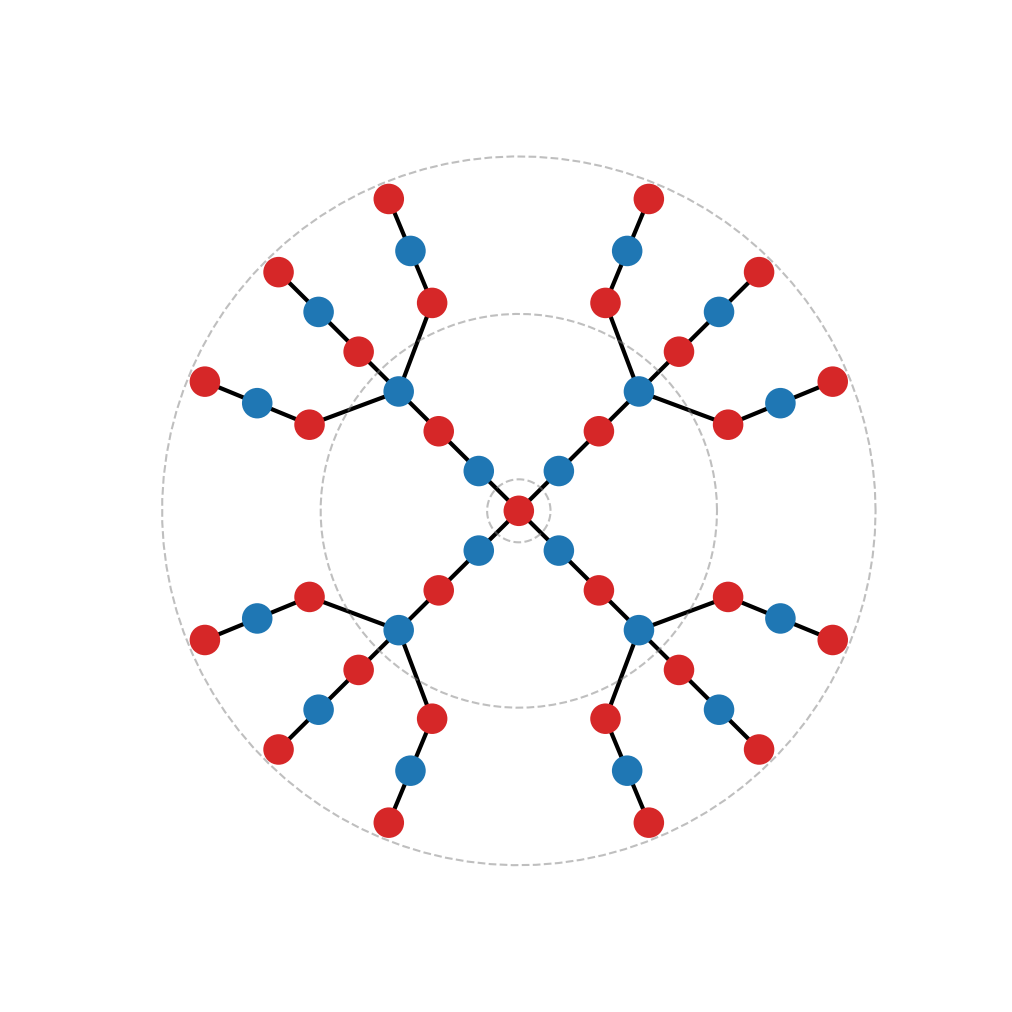}
\caption{A 2xLieb-Cayley tree with $K=3$ and $M=6$. Assigning each node to one of two sublattices, we find a pattern that predicts the number of exact $E=0$ states hidden in the bulk. 
The nodes are colored in red vs.~blue to denote their respective sublattice. 
The gray dashed lines demarcate 
the triple shells adopted in the presented counting of the sublattice imbalance.}
\label{Fig:2xLC_RankNullity}
\end{figure}

While not intuitive, one can still impose two sublattices on a 2xLieb-Cayley tree, such that the site imbalance between these two sublattices provides 
a lower bound on the zero-energy states of the system. 
Specifically, $N_{\text{FBS}} \geq \abs{N_A - N_B}$, with
the two sublattices $A$ and $B$ for the 2xLieb-Cayley tree 
shown in Fig.~\ref{Fig:2xLC_RankNullity}. 
To count the imbalance of the entire lattice, one needs to consider triple shells as marked with gray-dashed circles in Fig.~\ref{Fig:2xLC_RankNullity}. 
Each of these shells, assuming that it is complete, contributes 
$(-1)^m \times(K+1)\times  (2K^m - K^m) = (-1)^m \times(K+1)\times  K^m$ to $N_A - N_B$, 
where $m$ is the index of the triple shell. 
One can then write the full sum over all triple shells and the last incomplete layer as
\begin{equation}
\begin{aligned}
        N_A - N_B 
        = 1 - (K+1)\sum_{m = 0}^{\lfloor M/3 \rfloor} (-1)^m K^m \\
        + \begin{cases}
        (K+1)(-K)^{\lceil M/3 \rceil} & \textrm{if $M = 1\;\textrm{(mod $3$)}$} \\ 
        0 & \text{otherwise.}\end{cases}
\end{aligned}
\end{equation}
Using the geometric sum, this can be reformulated with 
\begin{equation}
    \sum_{m=0}^{A} (-K)^m= \frac{1 + (-1)^{A} K^{A +1}}{(K+1)}
\end{equation}
in the following short form:
\begin{equation}
        N_A - N_B 
        = \begin{cases}
        (-1)^{\lceil M/3 \rceil} K^{\lceil M/3 \rceil} & \textrm{if $M = 1\;\textrm{(mod $3$)}$} \\ 
        (-1)^{\lfloor M/3 \rfloor} K^{\lfloor M/3 \rfloor} & \text{otherwise.}\end{cases}
\end{equation}
For the bound $N_\textrm{FBS}$ on zero-energy states, the sign of $N_A - N_B$ 
does not matter; therefore $N_\textrm{FBS} \geq K^{[M/3]}$ where with `$[M/3]$' we mean either the ceiling or the floor of $M/3$. 
Through numerical calculation, we verify that this bound is saturated for each of the three choices of the boundary termination.
This flat band is recognizable in the spectra in Fig.~\ref{Fig:DoubleLiebCayley_Spectra} despite its occurrence inside an energy band.
The predicted degeneracy of the $E=0$ flat band is robust against perturbations that keep the Hamiltonian block-off-diagonal in the sublattice basis. This, in particular, applies to a bond disorder as previously considered for the Lieb-Cayley tree in Fig.~\ref{Fig:LiebCayley_BondDisorder}.

\subsection{Topological States}\label{sec:2xLC_topology}
In this subsection, we discuss the behavior of the in-gap states around $E=\pm1$, which is observed to depend 
on the termination condition of the tree. 
We show that exact eigenvalues at $E=\pm1$ impose exponential localization at the boundary with very specific termination conditions for shell-non-symmetric sectors. 
We further show that the 2xLieb-Cayley tree can be related to the trimer SSH (or `SSH3') model, which implies that these in-gap states have a topological origin.
\begin{figure}[t!]
\centering
\includegraphics[width=\linewidth]{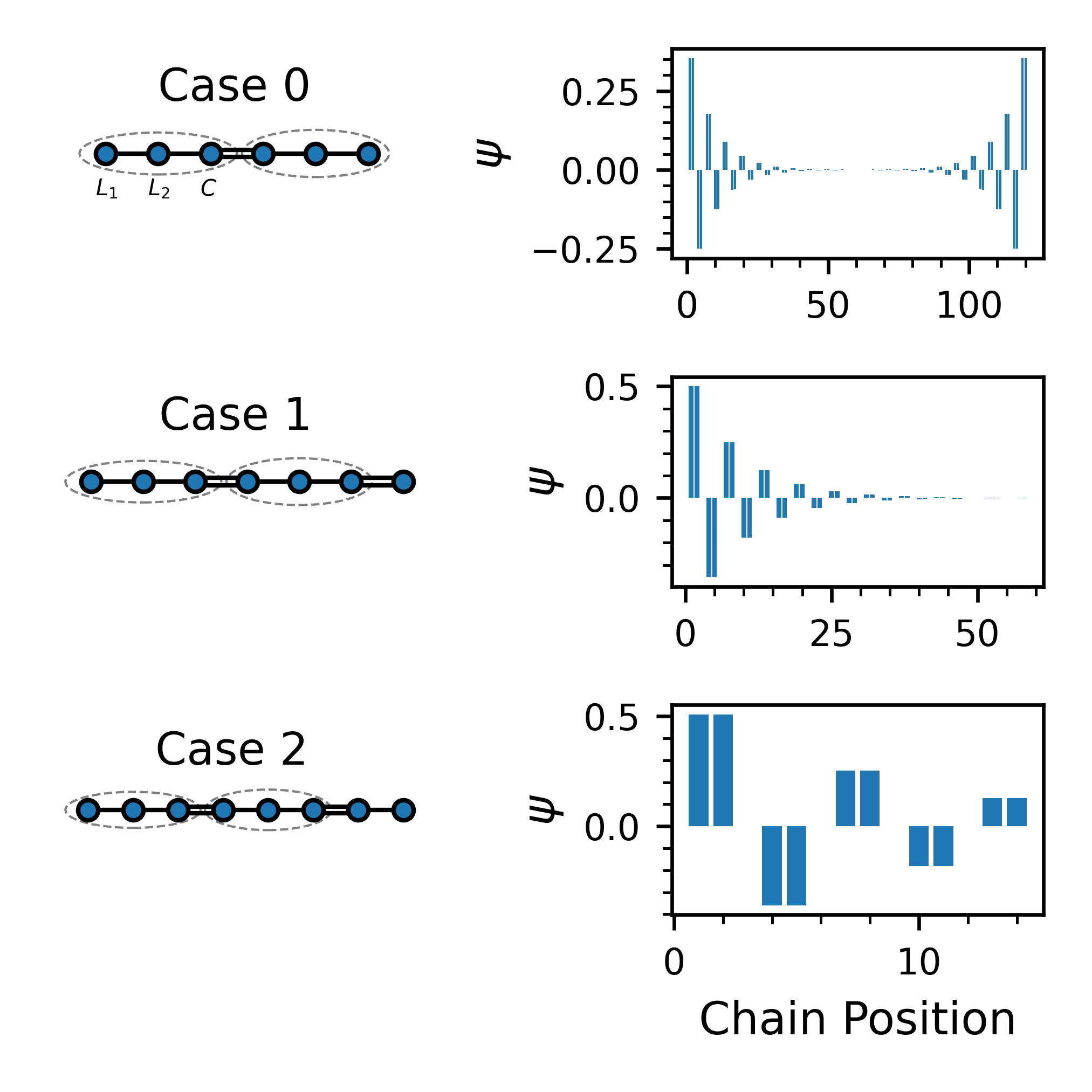}
\caption{On the left side we showcase examples of 1D chains described by the shell-non-symmetric sector of the 2xLieb-Cayley tree. Weak couplings are denoted with single bonds, while the strong couplings bonds are represented using double bonds. 
The unit cell with three atomic sites is indicated using a gray dashed oval. 
On the right side we show three examples of eigenstates hosted on these chains with their energies being close to or exactly at $E=1$, with precision $\epsilon = 10^{-3}$. 
Only in Case~2 is it possible to choose arbitrarily small chain lengths and still maintain an exact energy of $E=1$. 
Observe also that 
Case~0 hosts edge modes on both boundaries, allowing them to hybridize in finite chains; in contrast, Case~1 and Case~2 
host edge modes only at the beginning of the chain.}
\label{Fig:DoubleLiebCayley_Chains}
\end{figure}
To understand the origin and the energy distribution of the in-gap states shown in Fig.~\ref{Fig:DoubleLiebCayley_Spectra}, we need to discuss the properties of the shell-non-symmetric sector described by the Hamiltonian in Eq.~(\ref{Eq:2xLiebCayley_Nonsymm_Hamiltonian}), which hosts these in-gap states. 
This Hamiltonian describes an effective 1D chain with two weak couplings followed by a strong coupling. 
We show below that only the termination condition of Case~2 allows for a solution with an exact energy $E=\pm1$, while the two other cases approach this energy exponentially fast with the length $N$ of the chain. 
For all three cases, these solutions are exponentially localized at the boundary of the chain. 

In Fig.~\ref{Fig:DoubleLiebCayley_Chains}, we illustrate the three cases of the chain described by Eq.~(\ref{Eq:2xLiebCayley_Nonsymm_Hamiltonian}) depending on the termination condition, as well as an example of an eigenstate with $E=1 \pm 10^{-3}$ for each choice of the termination. 
Note that for Case~0 and Case~1 the chain needs to be sufficiently long to find an eigenstate with energy close enough to $E=1$. 
That this must be the case can be seen by considering the bulk recurrence relation of this 1D chain and the allowed termination conditions given $E=\pm1$. Using the Schrödinger equation $H|\psi \rangle = E \ket{\psi}$, we can write the bulk recurrence as:
\begin{equation}
    t_{n-1} \psi_{n-1} + t_n \psi_{n+1} =E\psi_n
\end{equation}
Where the couplings are
\begin{equation}
    t_n = \begin{cases}
        \sqrt{K} \quad \text{if } n = 0 \text{(mod 3)} \\
        1 \quad \quad \, \text{otherwise}
    \end{cases}
\end{equation}
For ease of notation, we will proceed with the following argument for $E=1$,the case $E=-1$ being analogous. 
With open boundaries the end equations are
\begin{equation}
    t_1\psi_2 = \psi_1, \quad t_{N-1} \psi_{N-1} = \psi_N,
\end{equation}
where $N$ is the length of the chain. Starting from the left-end condition $\psi_2 = \psi_1$, the bulk recurrence forces the pattern
\begin{equation}
\label{eqn:analytic-trimer-states}
    \psi_{3m} = 0, \quad \psi_{3m+1} = A(-\frac{1}{\sqrt{K}})^m, \quad \psi_{3m+2} = A(-\frac{1}{\sqrt{K}})^m.
\end{equation}
Here we assume $m\in\{0,1,2,\ldots\}$ 
and the free amplitude $A = \psi_1$ is set by normalization. 
This amplitude pattern confirms analytically that the edge modes decay exponentially. 
Next, we enforce the right-end equation $t_{N-1}\psi_{N-1} = \psi_N$. 
We see that only in the case $N=3m+2$, where $t_{N-1} =1$, can the bulk pattern satisfy $\psi_{3m + 1} = \psi_{3m + 2}$ and therefore have a nontrivial eigenstate at $E=1$. 
In contrast, for $N= 3m$ and $N = 3m + 1$, we need to set $A=0$ to satisfy the right-end equation, thereby arriving at a trivial solution. 
Therefore, we find that an eigenstate with $E=1$ with open boundaries exists only if $N = 2\; \textrm{(mod $3$)}$ 
(Case~2). 
However, as the length of the chain increases, the amplitudes $\psi_{3m+1}$ and $\psi_{3m + 2}$ determined by Eq.~(\ref{eqn:analytic-trimer-states}) get exponentially closer to zero and the energies of the corresponding eigenstates of Case~0 and Case~1, will converge closer to $E=1$ as well. 

To answer the question of whether or not these edge states can be considered topological, we need to investigate the bulk Bloch Hamiltonian of the shell-non-symmetric sectors. 
We define the unit cell according to the convention shown for the chains on the left side of 
Fig.~\ref{Fig:DoubleLiebCayley_Chains} and impose periodic boundary conditions. 
Adopting the basis $\textrm{L}_1,\textrm{L}_2,\textrm{C}$, the bulk Hamiltonian is then found to be
\begin{equation}\label{Eq:SSH3_H}
    H(k) =
    \begin{pmatrix}
        0 & 1 & \sqrt{K} e^{-ik} \\
        1 & 0 & 1 \\
        \sqrt{K} e^{ik} & 1 & 0
    \end{pmatrix}.
\end{equation}
This three-band tight-binding model, known as the SSH3 chain~\cite{martinez_alvarez_edge_2019,anastasiadis_bulk-edge_2022,Ghuneim:2024}, has attracted interest because it hosts robust edge modes even in the absence of conventional symmetry-protected topological phases. Its bulk Hamiltonian \eqref{Eq:SSH3_H} respects inversion symmetry, $P H(k) P^{-1} = H(-k)$,  which enforces a quantized Zak phase (i.e., the Berry phase accumulated over the Brillouin zone). 
Inversion symmetry also relates solutions at momenta $k$ and $-k$, and in the position space it implies 
that edge states appear in pairs on opposite ends of the chain. 
For a finite chain, inversion symmetry is preserved only when the number of unit cells is a multiple of three ($N \bmod 3 = 0$, the ``Case~0'' termination), in which case the quantized Zak phase predicts two edge modes that may hybridize for short chains (see top row of Fig.~\ref{Fig:DoubleLiebCayley_Spectra}).

When $N \bmod 3 \neq 0$, inversion symmetry is broken. Nevertheless, robust edge states at $E = \pm 1$ emerge, localized exclusively on a single boundary. These ``chiral edge states''~\cite{martinez_alvarez_edge_2019} are not captured by the standard Zak phase. Instead, their existence can be characterized by the normalized sublattice Zak phase (NSZP)~\cite{anastasiadis_bulk-edge_2022}. To define it, one expresses the eigenstates of the finite chain in terms of the Bloch solutions,
\begin{equation}
    \ket{\psi_\lambda(k)} = \Lambda \sum_{j=1}^M e^{ikj}\, 
    \ket{j} \otimes
    \begin{pmatrix}
        a_\lambda^{\textrm{L}_1}(k) e^{-i \theta_\lambda^{\textrm{L}_1}(k)} \\
        a_\lambda^{\textrm{L}_2}(k) e^{-i \theta_\lambda^{\textrm{L}_2}(k)} \\
        a_\lambda^\textrm{C}(k)
    \end{pmatrix},
\end{equation}
with band index $\lambda$ and normalization $\Lambda$. 
A gauge choice $\theta_\lambda^\textrm{C}=0$ fixes the $\textrm{C}$-sublattice phase. The NSZP is then
\begin{equation}
    Z_{\textrm{L}_1,\textrm{C}}^\lambda
    = \frac{i}{2}\oint dk\,\bra{\tilde{u}_\lambda(k)}\partial_k \tilde{u}_\lambda(k) \rangle
    = \int_0^\pi \! dk \,\frac{\partial \theta_\lambda^{\textrm{L}_1}}{\partial k}.
\end{equation}
Here, the first subscript indicates the projected sublattice, while the second one marks the chosen gauge. 
Intuitively, the NSZP measures the winding of the complex phase of a chosen sublattice component as $k$ runs over the Brillouin zone. 
Each $\pi$ of winding corresponds to one Bloch solution becoming inconsistent with the open–boundary quantization, and the ``missing'' bulk state reappears as an edge-localized mode. In this way, the difference of NSZPs directly counts the number of edge states contributed by a given band.

For a given termination, the open-boundary conditions fix which sublattice phase controls the counting of 
Bloch solutions with real wavenumber in band $\lambda$; the ``missing" ones correspond to solutions with complex wavenumbers and therefore to edge states. 
For $3N$ sites (Case~0) the quantization condition reads $\theta^\lambda_{\textrm{L}_1}(k)=(N+1)k-n\pi$; for $3N\!+\!1$ (Case~1) it is $\theta^\lambda_{\textrm{L}_2}(k)=(N+1)k-n\pi$; and for $3N\!+\!2$ (Case~2) the relevant gauge is $\theta^\lambda_\textrm{C}\!=\!0$ with a vanishing projected phase on $\textrm{C}$. 
In each case, the corresponding normalized sublattice Zak phase
\begin{equation}
    Z^\lambda_{\alpha,\;\textrm{C}}=\int_0^\pi \partial_k\theta^\lambda_\alpha(k)\,dk\in\pi\mathbb{Z}
\end{equation}
is quantized and jumps by $\pm\pi$ only when when the bands touch, facilitating a topological phase transition.
Since $Z^\lambda_{\alpha,\;\textrm{C}}$ depends on the gauge, we form the gauge-invariant \emph{difference} with a reference Hamiltonian that has \emph{no edge states} for the same termination:
\begin{equation}
{\;\;\#\,\text{edge states from band }\lambda\;=\;\frac{Z^\lambda_{\alpha,\textrm{C}}-Z^{\lambda,\text{ref}}_{\alpha,\textrm{C}}}{\pi}\;.\;\;}
\end{equation}
For Case~1, this predicts two edge states from the middle band,
while in Case~2 the gauge freedom in $\theta_\lambda^C$ implies no change relative to the reference Hamiltonian, i.e., the appearance of edge states is independent of the phase of the model. 
It is technically possible to count the actual number of edge states that occur in Case~2 using the presented approach. 
For brevity, in the next section we instead use the rank-nullity theorem to show that Case~2 hosts two edge states.

\subsection{Adapted Rank-Nullity Theorem}\label{sec:2xLC_RankNullity}
In this subsection, we show how one can rewrite the eigenvalue problem of the 2xLieb-Cayley tree to recover a form that proves that the rank-nullity theorem protects the exact flat band at $E=\pm1$ for the Case~2 termination. 
For this purpose, we order the position basis according to its sublattices ($\textrm{C},\textrm{L}_1,\textrm{L}_2$), thereby rewriting the eigenvalue problem for the eigenvalues $E=\pm1 \equiv \sigma$ as
\begin{equation}
\label{eqn:rank-nullity-pm1}
    \begin{pmatrix}
        0_\textrm{C} & X & Z^\dagger \\
        X^\dagger & 0_{\textrm{L}_1} & Y \\
        Z & Y^\dagger & 0_{\textrm{L}_2}
    \end{pmatrix}
    \begin{pmatrix}
        c \\ l_1 \\ l_2
    \end{pmatrix} = \sigma \begin{pmatrix}
        c \\ l_1 \\ l_2
    \end{pmatrix}.
\end{equation}
Here, $c/l_1/l_2$ are vectors of the wave function amplitudes on sites belonging to the sublattices $\textrm{C}/\textrm{L}_1/\textrm{L}_2$, and $0_{\textrm{C}/\textrm{L}_1/\textrm{L}_2}$ denotes the $N_{\textrm{C}/\textrm{L}_1/\textrm{L}_2} \times N_{\textrm{C}/\textrm{L}_1/\textrm{L}_2}$ zero matrix, with $N_{\textrm{C}/\textrm{L}_1/\textrm{L}_2}$ the number of Cayley / first Lieb / second Lieb nodes. 
Furthermore, $X$ is a $N_\textrm{C} \times N_{\textrm{L}_1}$ matrix containing the hopping amplitudes connecting the sublattices $C$ and $\textrm{L}_1$, $Y$ is a $N_{\textrm{L}_1} \times N_{\textrm{L}_2}$ matrix containing the hopping amplitudes which connect sites on the two different Lieb sublattices, and $Z$ is a $N_{\textrm{L}_2}\times N_\textrm{C}$ matrix connecting the sublattices $\textrm{L}_2$ and $\textrm{C}$ (with $X^\dagger,Y^\dagger,Z^\dagger$ their Hermitian conjugates).

We proceed with solving for $l_2$ through the 
matrix equation
\begin{subequations}
\begin{equation}
    \sigma l_2 = Z c + Y^\dagger l_1, \quad \textrm{therefore} 
    \quad l_2 = \sigma( Z c + Y^\dagger l_1).
\end{equation}
We can insert the obtained expression for $l_2$ 
in the remaining lines of Eq.~(\ref{eqn:rank-nullity-pm1}) to arrive at
\begin{gather}
    \sigma c = X l_1 + Z^\dagger \sigma \big(Z c + Y^\dagger l_1 \big), \\
    \sigma l_1 = X^\dagger c + Y \sigma \big( Z c + Y^\dagger l_1 \big).
\end{gather}
\end{subequations}
This pair of equations can, in turn, 
be rewritten as 
\begin{equation}
    \begin{pmatrix}
        Z^\dagger Z - \mathbb{1} & \sigma X + Z^\dagger Y^\dagger \\
        \sigma X^\dagger + YZ & Y Y^\dagger - \mathbb{1} 
    \end{pmatrix}
    \begin{pmatrix}
        c \\ l_1
    \end{pmatrix}
    = 0     \begin{pmatrix}
        c \\ l_1
    \end{pmatrix}.
\end{equation}
Provided that $Z$ and $Y$ are partial isometries, i.e., $Z^\dagger Z = \mathbb{1}_{N_\textrm{C}}$ and $YY^\dagger = \mathbb{1}_{N_{\textrm{L}_1}}$ (which requires $N_{\textrm{L}_2} \geq N_\textrm{C}$, $N_{\textrm{L}_2} \geq N_{\textrm{L}_1}$, and that the NN hopping amplitudes are normalized to $t_1 = 1$), 
the reduced matrix is chiral and the above Hamiltonian simplifies to
\begin{equation}
    \begin{pmatrix}
        0_\textrm{C} & \sigma X + Z^\dagger Y^\dagger \\
        \sigma X^\dagger + YZ & 0_{\textrm{L}_1}
    \end{pmatrix}
    \begin{pmatrix}
        c \\ l_1
    \end{pmatrix}
    = 0     \begin{pmatrix}
        c \\ l_1
    \end{pmatrix}.
\end{equation}
This is an equation of the form of Eq.~(\ref{Eq:RankNullity_BlockMatrix}) and we can apply the rank-nullity theorem according to the same procedure as outlined in Sec.~\ref{Sec:Lieb_1_euclidean}. 

Using this line of reasoning, we find the following bound on the number of flat band states for both $E=\pm1$:
\begin{equation}
    N_{\text{FBS}} \geq |N_{\textrm{L}_1} - N_\textrm{C}|.
\end{equation}
Note that this bound is only valid if we meet the partial isometry conditions stated above, which is not the case for termination Case~1. 
In contrast, the isometry conditions holds for 
Case~2 and Case~0, allowing us to formulate a bound on $N_\textrm{FBS}$ which depends on the imbalance between the first Lieb nodes and the Cayley nodes. 
This bound can be expressed using the sum over layers of the respective nodes as
\begin{equation}
    \abs{N_{\textrm{L}_1} - N_\textrm{C}}  = \abs{\frac{K+1}{K-1} (K^{{M_{\textrm{L}_1}}} - K^{M_\textrm{C}}) - 1 }.
\end{equation}
We therefore find the following bounds depending on the termination case of the tree.
\begin{equation}
    |N_{\textrm{L}_1} - N_\textrm{C}| = \begin{cases}
        1 \qquad \qquad \qquad  \text{ for Case~0} \\
        (K+1) 
        \times K^{M_\textrm{C}} \quad \text{ for Case~2}
    \end{cases}
\end{equation}
The number of flat-band states at $E=\pm 1$ derived 
for Case~2 corresponds exactly to the number of non-symmetric sectors (accounting for the different non-trivial roots of unity as well as for the multiple seed nodes within the Cayley layers). 
In other words, for Case~2 every shell-non-symmetric sector contributes exactly one $E=\pm1 $ state pair. 

Let us also point out that the isometry condition is lost for all three cases if one introduces bond disorder. 
Therefore, in sharp contrast to the odd-$M$ case of the Lieb-Cayley tree, where the exact zero-energy flat band is robust against bond-disorder (right column of Fig.~\ref{Fig:LiebCayley_BondDisorder}), such robustness against disorder is absent for the flat bands at $E=\pm 1$ of the 2xLieb-Cayley tree.

We conclude this section by restating our key findings. We have shown that the 2xLieb-Cayley tree hosts three flat bands: one at $E=0$ (covered by an energy band) and two at $E=\pm1$ (inside energy gaps). 
The origin of the flat bands at $E=\pm1$ can be traced to the topological invariant of the underlying SSH3 model. 
The states of these flat bands are exponentially localized at the inner boundary of the respective non-shell-symmetric sector and therefore inside the tree, with the Case 0 termination exhibiting additional edge states at these energies which are localized on the outer boundary of the tree. 
We characterized the exactness, the localization, and the robustness of these flat bands by studying their bulk recurrence relation.
In addition, we studied their protection through the rank-nullity theorem, explaining why the flat band is exact only for the Case~2 termination. 

\section{Husimi decoration}\label{Sec:HusimiDecoration}
In this section we discuss kagome-like decoration of the Cayley tree, where nodes with a common parent node are connected via an edge. 
This is similar to what has been called a ``Husimi tree''~\cite{Harary:1953} or a ``cactus graph'', which is why we call this tree a ``Husimi-Cayley'' tree. 
We show that the approach of symmetry-adapted basis states works for this tree as well, allowing us to find its 
exact spectrum. 
We find that the spectrum exhibits three features of interest: 
one originating in CLSs and the remaining two related to 
properties of line graphs. 

Our discussion of Husimi-Cayley trees is structured as follows.
We begin in Sec.~\ref{sec:Husimi_Euclidean} with a brief review of the kagome lattice, which exhibits a flat band and CLSs with energies $E=-2$ and that serves as a Euclidean analog to the Husimi-Cayley tree. 
In Sec.~\ref{sec:Husimi_ExactSolutions} we present the block-diagonalization of the Husimi-Cayley tree and showcase the exact spectrum derived from it. 
Finally, we discuss selected aspects of this spectrum in Sec.~\ref{sec:Husimi_LineGraph} and relate them to the line-graph nature of the Husimi-Cayley tree.

\subsection{Kagome Lattice}\label{sec:Husimi_Euclidean}

\begin{figure}[t!]
\centering
\includegraphics[width=\linewidth]{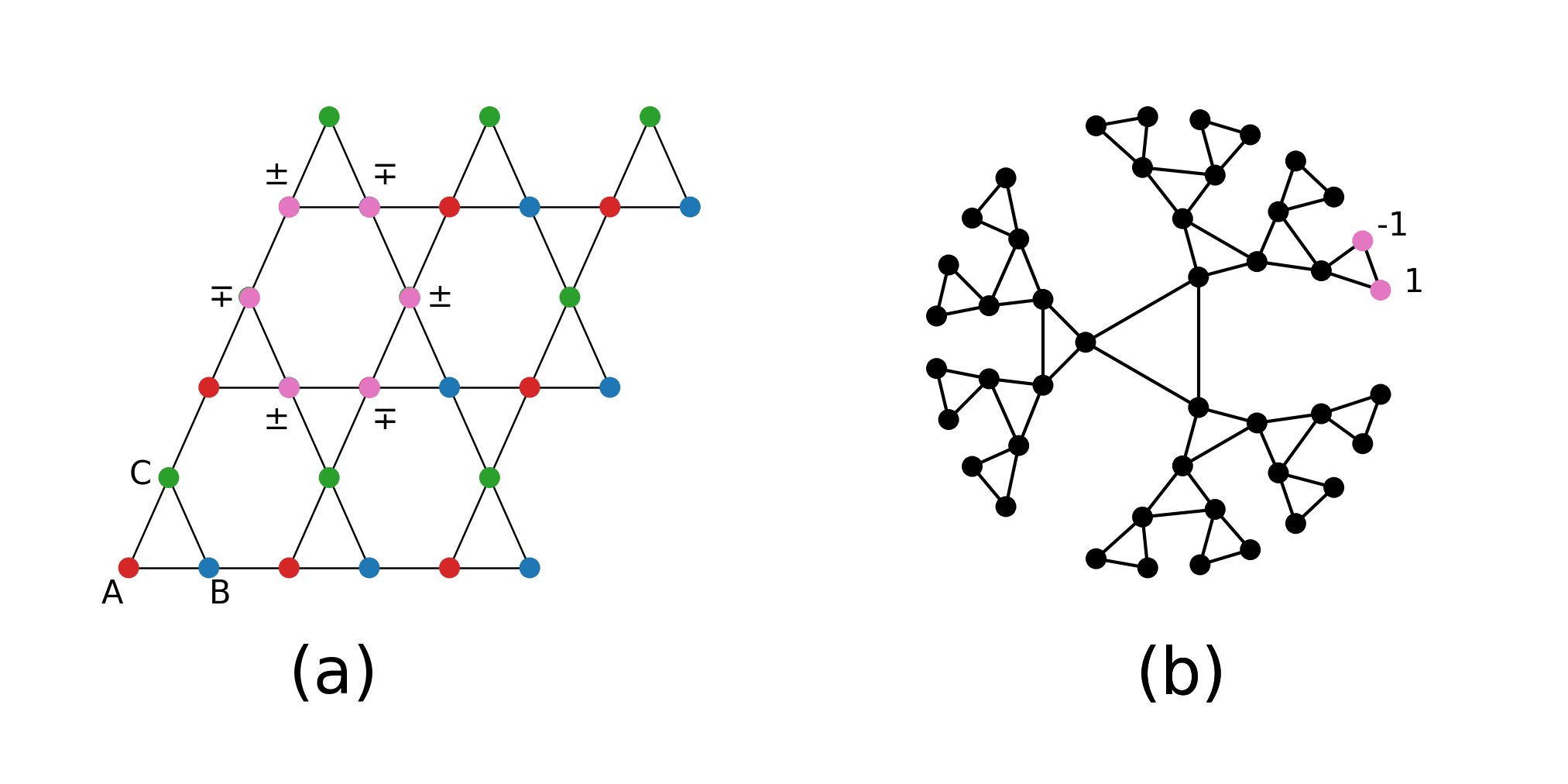}
\caption{
(a) Kagome lattice with the three sites of the unit cells $A$ in red, $B$ in blue, and $C$ in green. 
The lattice hosts CLSs on the hexagonal plaquettes; they are constructed by arranging oscillating amplitudes $\pm \psi$ on the sites forming the hexagon, as shown in pink.  
(b) Husimi-Cayley tree with branching factor 
$K=2$ and $M=4$ layers. 
Contrary to the simple Cayley tree (recall Fig.~\ref{Fig:NNN_CayleyTree_Sketch}), the Husimi-Cayley tree does not contain a root node; instead, it has a fully connected layer at the center. A CLS, as found on the boundary of the tree, is colored in pink.}
\label{Fig:Husimi_Sketch}
\end{figure}

The motivation for the Husimi tree decoration derives itself from the famous kagome lattice. 
The kagome lattice is a two-dimensional lattice composed of corner-sharing triangles. 
It is based on a triangular Bravais lattice with a three-point basis. 
The lattice vectors can be chosen 
as
\begin{subequations}
\begin{equation}
\label{eqn:triangular-Bravais}
    \bm{a}_1 = a(1, 0), \quad \bm{a}_2 = a\left( \tfrac{1}{2}, \tfrac{\sqrt{3}}{2} \right),
\end{equation}
\noindent where $a$ is the lattice constant. Each unit cell contains three inequivalent sites, which we denote as $A$, $B$, and $C$, as shown in Fig.~\ref{Fig:Husimi_Sketch}. These sites are located at the fractional coordinates
\begin{align}
    \bm{r}_A &= (0, 0), \quad 
    \bm{r}_B = \frac{1}{2} \bm{a}_1, \quad 
    \bm{r}_C = \frac{1}{2} \bm{a}_2.
\end{align}
\end{subequations}
The tight-binding Hamiltonian on the kagome lattice with only NN hoppings is written as
\begin{equation}
    H = t \sum_{\langle i,j\rangle }\ket{i}\bra{j} + \text{h.c.},
\end{equation}
where $t$ is the NN hopping amplitude and $\langle i,j\rangle $ labels pairs of NN sites on the kagome lattice. 
The Bloch Hamiltonian reads
\begin{equation}
\label{eqn:Bloch-kagome}
    H(\bm{k}) = 2t \begin{pmatrix}
    0 & \cos(k_1) & \cos(k_2) \\
   \cos(k_1) & 0 & \cos(k_3) \\
    \cos(k_2) & \cos(k_3) & 0
    \end{pmatrix},
\end{equation}
where we introduced $k_n = \bm{k}\cdot \bm{a_n}$ and $\bm{a_3} = \bm{a_1} - \bm{a_2}$. 
Upon diagonalizing $H(\bm{k})$, one finds that a completely flat band. 
Setting for concreteness the NN hopping amplitude to $t=1$,
the eigenvalues are~\cite{guo_topological_2009,tang_high-temperature_2011}:
\begin{subequations}
\begin{align}
    E_0(\bm{k}) &= -2 
    , \quad \text{(flat band)} \\
    E_{\pm}(\bm{k}) &= 1 
    \pm 
    \sqrt{4f(\bm{k}) - 3},
\end{align}
where 
\begin{equation}
    f(\bm{k}) = \cos^2(k_1) + \cos^2(k_2) + \cos^2(k_3).
\end{equation}
\end{subequations}
The flat band arises due to destructive interference and geometrical frustration. 
Namely, one can construct a CLS on a single hexagon by assigning alternating wave function amplitudes $\pm \psi$ to the vertices of the hexagon. 
This arrangement ensures that the amplitudes cancel out after tunneling to the adjacent 
sites. 
These CLSs form a macroscopic set of degenerate states which are exact eigenstates of $H$, forming a dispersionless band at energy $E=-2$ inside the momentum space.

It is also possible to explain the appearance of the flat band from a purely graph-theoretical perspective; specifically, by interpreting the kagome lattice as the line graph of another graph. 
We will not delve deeper into this argument here; instead, we revisit spectral properties of line graphs in Sec.~\ref{sec:Husimi_LineGraph} when discussing the derived energy spectra of Husimi-Cayley trees. 
Let us also remark that although the flat band of the kagome lattice is topologically trivial (it has vanishing Berry curvature), it can acquire non-trivial Chern number upon including staggered~\cite{Ohgushi:2000} or uniform~\cite{tang_high-temperature_2011} magnetic fluxes.

\subsection{Exact Solution for Husimi-Cayley tree}\label{sec:Husimi_ExactSolutions}
To imitate the kagome tiling on a Cayley tree, we introduce an additional edge between all nodes originating from the same parent. 
To ensure homogeneity, we further remove the central node $0$ (and the $K+1$ edges connected to it), thus ensuring that the tree locally looks the same from every bulk site.
The resulting structure for a small number $M$ of layers is shown for $K=2$ in Fig.~\ref{Fig:Husimi_Sketch}(a) and for $K=3$ in Fig.~\ref{Fig:Husimi_HigherK_Sketch}.

The resulting Husimi-Cayley tree has the same branching factor $K$ as the the parent Cayley tree. Assuming $M$ layers, it contains a total of 
\begin{equation} 
    N_\textrm{total} 
    = (K+1) \times \sum_{l=1}^{M}K^{l-1} = \frac{K+1}{K-1}(K^M -1)
\end{equation}
nodes, which is one less than the result in Eq.~(\ref{eqn:Cayley-number-of-sites}) for the number of nodes in the parent Cayley tree.
Observe also that the parent node $\alpha$ and its $K$ children form a fully connected set of nodes, which in the case $K=3$ produces the tetrahedra visible in Fig.~\ref{Fig:Husimi_HigherK_Sketch}. 
After adopting the symmetry-adapted basis states, 
the full connectivity among the child nodes generates 
an effective on-site potential for the effective 1D block Hamiltonians. 
(For a derivation, we refer the reader to Appendix~\ref{Sec:App_Husimi}.) 
Therefore, the Hamiltonians that describe the dynamics of the symmetry sectors of the tree are:
\begin{figure}[t!]
\centering
\includegraphics[width=0.8\linewidth]{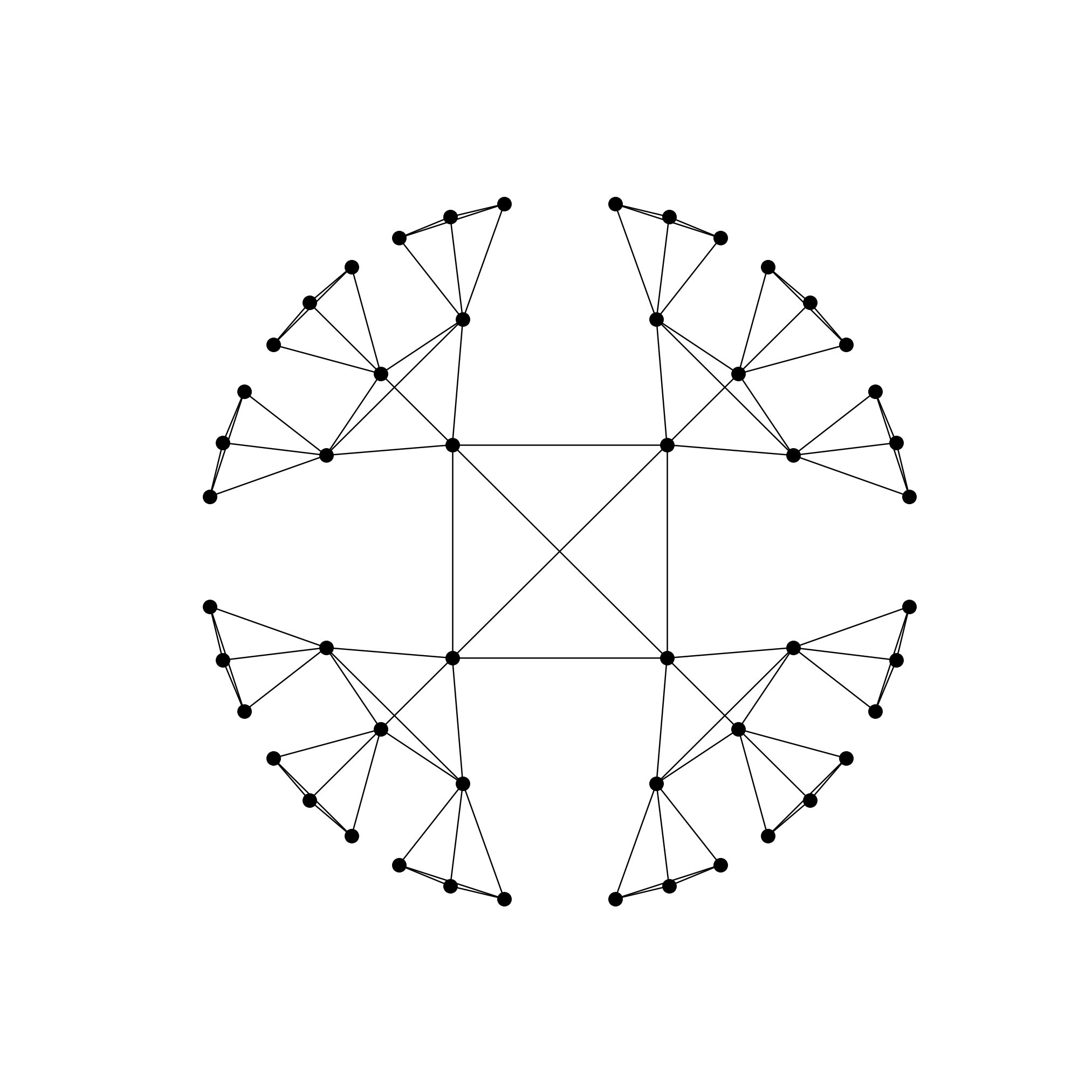}
\caption{A Husimi-Cayley tree with branching factor $K=3$ and $M=3$ layers. 
Due to the full connectivity of a parent node with all of its child nodes, the Husimi-Cayley tree looks like a collection of corner-sharing $K$-dimensional simplices (e.g., of $3$-dimensional tetrahedra for the plotted case).}
\label{Fig:Husimi_HigherK_Sketch}
\end{figure}
\begin{subequations}
\begin{equation}\label{Eq:Husimi_FullySymm}
        \mathcal{H}_{\textrm{sym.}} = \begin{pmatrix}
        K & \sqrt{K} & 0 & \cdots \\
        \sqrt{K} & K-1 & \sqrt{K} & \cdots\\
        0 & \sqrt{K} & K-1 & \cdots \\
        \vdots & \vdots & \vdots &  \ddots
    \end{pmatrix},
\end{equation}
\begin{equation}\label{Eq:Husimi_NonSymm}
        \mathcal{H}_{\textrm{nonsym.}}^\alpha = \begin{pmatrix}
        -1 & \sqrt{K} & 0 & \cdots \\
        \sqrt{K} & K-1 & \sqrt{K} & \cdots\\
        0 & \sqrt{K} & K-1 & \cdots \\
        \vdots & \vdots & \vdots &  \ddots
    \end{pmatrix}.
\end{equation}
\end{subequations}
These Hamiltonians describe one-dimensional chains with a single energy band and with an on-site potential. 
We do not expect any topological features for this system due to its single-band nature. 
However, upon solving for the spectrum of the tree, we do identify certain non-trivial aspects. 
In particular, the first site of the effective 1D models constitutes a defect due to its different on-site potential, meaning it can potentially capture a bound state.

\begin{figure}[t!]
\centering
\includegraphics[width=\linewidth]{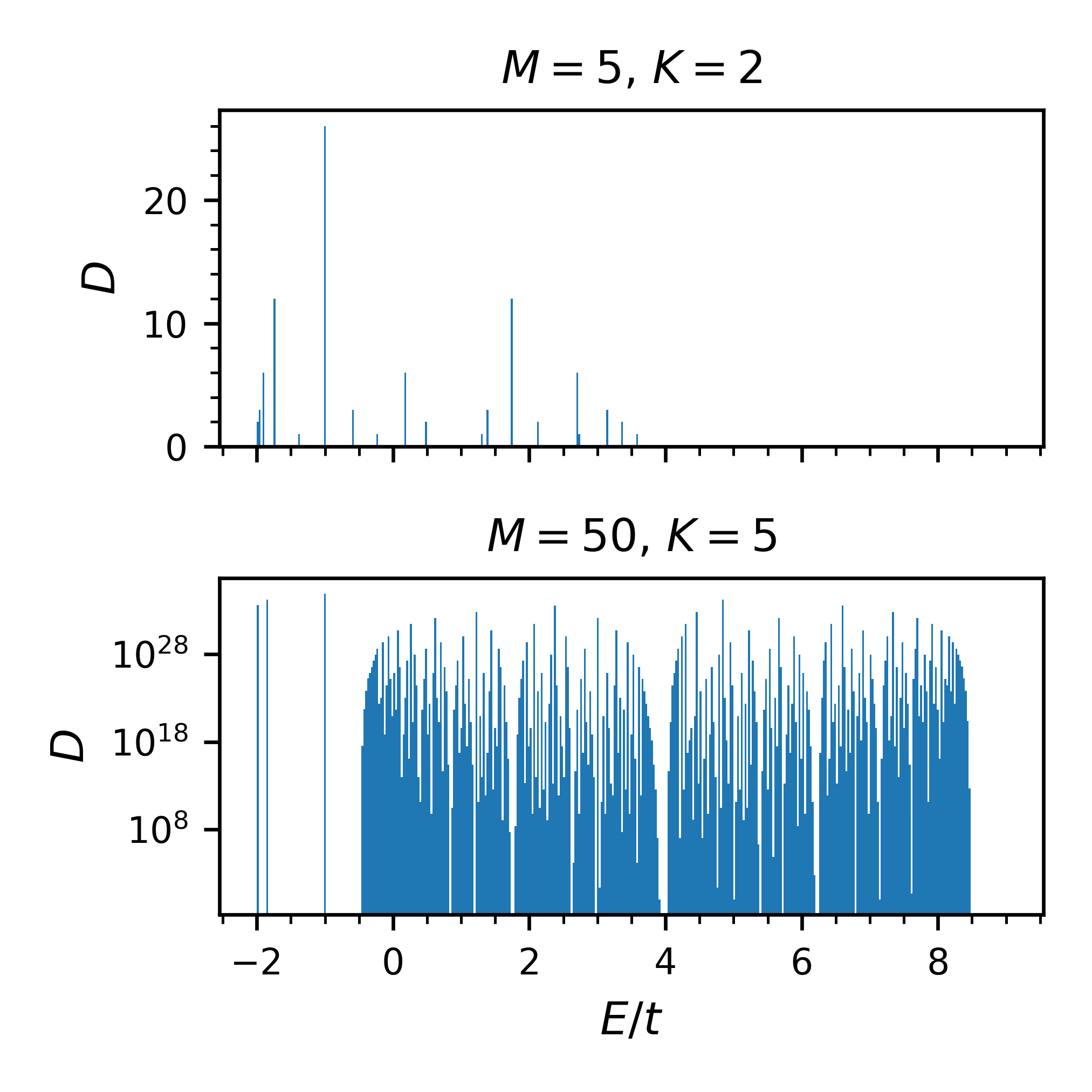}
\caption{
Spectra of a small and a large Husimi-Cayley tree with different branching factor $K$. 
Three features emerge: (1)~an exact flat band at $E=1$ whose origin lies in CLSs at the boundary of the tree, 
(2)~an accumulation of states near $E=-2$, and 
(3)~an energy band that is symmetrically arranged around $K-1$. 
The latter two observations hint at line-graph properties governing the spectrum of the Husimi-Cayley tree.}
\label{Fig:Husimi_Spectra}
\end{figure}

We show in Fig.~\ref{Fig:Husimi_Spectra} the spectra of two Husimi-Cayley trees with different values of $M$ and $K$. 
First, note the occurrence of a highly degenerate flat band at $E=-1$. 
This feature originates from CLSs at the boundary of the tree. 
Specifically, we can form such a CLS by taking the fully connected set of leaf nodes that originate from the same parent node $|\alpha\rangle$ and weighing them such that they interfere destructively as they hop to $|\alpha\rangle$. 
This construction is equivalent to the non-symmetric sector $\mathcal{H}_{\textrm{nonsym.}}^\alpha$ with the parent node in layer $l(\alpha) = M-1$. 
The CLS is built in the same way as in Eq.~(\ref{Eq:CLS_LiebCayley}) for the Lieb-Cayley tree, except for two differences: First, one here no longer needs to worry about two layers types (all sites of the Husimi-Cayley tree have the same number of neighbors), and second, the full connectivity among the child nodes shifts the energy of the CLS to $E=-1$. 
The total number of such CLS is 
\begin{equation}
\label{eqn:Husimi-Cayley-CLS-count}
    N_{\text{CLS}} = (K+1)\times K^{M -2} \times(K-1),
\end{equation}
according to the same derivation as presented in Eq.~(\ref{eqn:Lieb-Cayley-CLS-count}) of 
Sec.~\ref{Sec:Lieb_2_exact}. We find that this prediction agrees with the number of states found at $E=-1$ for branching factors $K \geq 4$ when the energy $E=-1$ lies inside the bulk energy gap; in contrast, the energy $E=-1$ lies inside the bulk energy band for branching factors $K\in\{2,3\}$, implying the presence of additional states at that energy beyond the count in Eq.~(\ref{eqn:Husimi-Cayley-CLS-count}). 

\subsection{Line-Graph Properties}\label{sec:Husimi_LineGraph}
In this subsection, we clarify the origin of two further features apparent in the spectrum of the Husimi-Cayley tree. 
Specifically, we want to elucidate the accumulation of states at $E=-2$ and the symmetric arrangement of the bulk band around the energy $E=K-1$. 
We formulate explanation of these two aspects in terms of graph-theoretical notion. 
For this reason, we here equivalently refer to nodes of a lattices as vertices of the corresponding graph.

In graph theory, the Husimi-Cayley tree is described as 
the line graph ``$L(\textrm{Cay})$'' of the Cayley tree ``$\textrm{Cay}$''. 
The line graph of a graph $G$ is the graph $L(G)$ that (1)~takes 
the edges of $G$ as its nodes, and where (2)~two nodes of $L(G)$ are connected if and only if the corresponding edges in $G$ 
are incident (i.e., meeting the same vertex) in $G$. 
A convenient definition of the line graph $L(G)$ uses the notion of the incidence matrix ``$B(G)$''~\cite{Kollar2020CMP}:
a matrix whose rows (columns) are indexed by the vertices (edges) of $G$, such that the element $B_{ve}(G)=1$ is non-zero 
if and only if the edge $e$ connects (i.e., ``is incident'') to vertex $v$. 
If $G$ has $\mathcal{V}$ vertices and $\mathcal{E}$ edges, 
then $B(G)$ is a matrix of size $\mathcal{V}\times \mathcal{E}$. 

The inner product of any two distinct rows of $B(G)$ is equal to the number of edges joining the corresponding vertices. 
For this reason, $B^{T}B$ is sometimes called the ``edge-edge matrix'', since any entry $(B^TB)_{e,e'}$ is equal to the number of vertices shared between edges $e$ and $e'$. 
In addition, the diagonal elements are fixed at $(B^TB)_{e,e} = 2$, since any edge shares two vertices (namely, both of its end-points) with itself.
It then follows that 
\begin{equation}
\label{eqn:line-graph-adjacency}
    A(L(G)) = B^TB - 2 \mathbb{I}_\mathcal{E}
\end{equation}
where $A(L(G))$ denotes the adjacency matrix (i.e., the NN-hopping Hamiltonian) of the line graph of $G$. 
Since matrices of the form $B^TB$ are 
positive semi-definite, it follows from Eq.~(\ref{eqn:line-graph-adjacency}) 
that the spectrum of a line graph is bounded by $E \geq -2$. 
This prediction aligns with the spectra shown in Fig.~\ref{Fig:Husimi_Spectra}.  

One can further show \cite{godsil_algebraic_2001} that
\begin{equation}
    BB^T = \Delta(G) + A(G)
\end{equation}
where $A(G)$ is the adjacency matrix of $G$, and $\Delta(G)$, called the degree matrix of $G$, is a diagonal matrix with the entry $\Delta(G)_{v,v}$ equal to the degree of the node $v$. 
From the last two equations, one can prove~\cite{godsil_algebraic_2001} that the characteristic polynomial $\phi(L,x)$ of the line graph $L(G)$ can be expressed as 
\begin{equation}\label{Eq:Husimi_CharacPolyonmial_nonRegular}
    \phi(L,x) = (x{+}2)^{\mathcal{E}-\mathcal{V}}
    \det[(x{+}2) \mathbb{I}_\mathcal{V} {-} \Delta(G) {-} A(G)].
\end{equation}
Equation~(\ref{Eq:Husimi_CharacPolyonmial_nonRegular}) suggests 
the emergence of a flat band at $E=-2$ with the degeneracy 
given by the difference between the number $\mathcal{E}$ of edges and the number $\mathcal{V}$ of vertices in graph $G$. [The degeneracy may be higher if $\det[(x{+}2) \mathbb{I}_\mathcal{V} {-} \Delta(G) {-} A(G)]$ contains an additional factor $(x+2)$].
Furthermore, if the underlying graph $G$ is $q$-regular, with $q$ the degree of \emph{all} its nodes, one can express $\Delta(G) = q \mathbb{I}$ and arrive~at~\cite{godsil_algebraic_2001}
\begin{equation}
\label{Eq:Husimi_CharacPolyonmial_kRegular}
    \phi(L,x) = (x+2)^{\mathcal{E}-\mathcal{V}}
    \phi(G,x -(q-2)).
\end{equation}
Equation~(\ref{Eq:Husimi_CharacPolyonmial_kRegular}) predicts that the line graph $L(G)$ of a $q$-regular graph $G$ has 
the same spectrum as $G$, except (1)~shifted by $q-2$, and (2)~with the additional eigenvalues at $E=-2$ as described earlier for general line graphs.

\begin{figure}[t!]
\centering
\includegraphics[width=\linewidth]{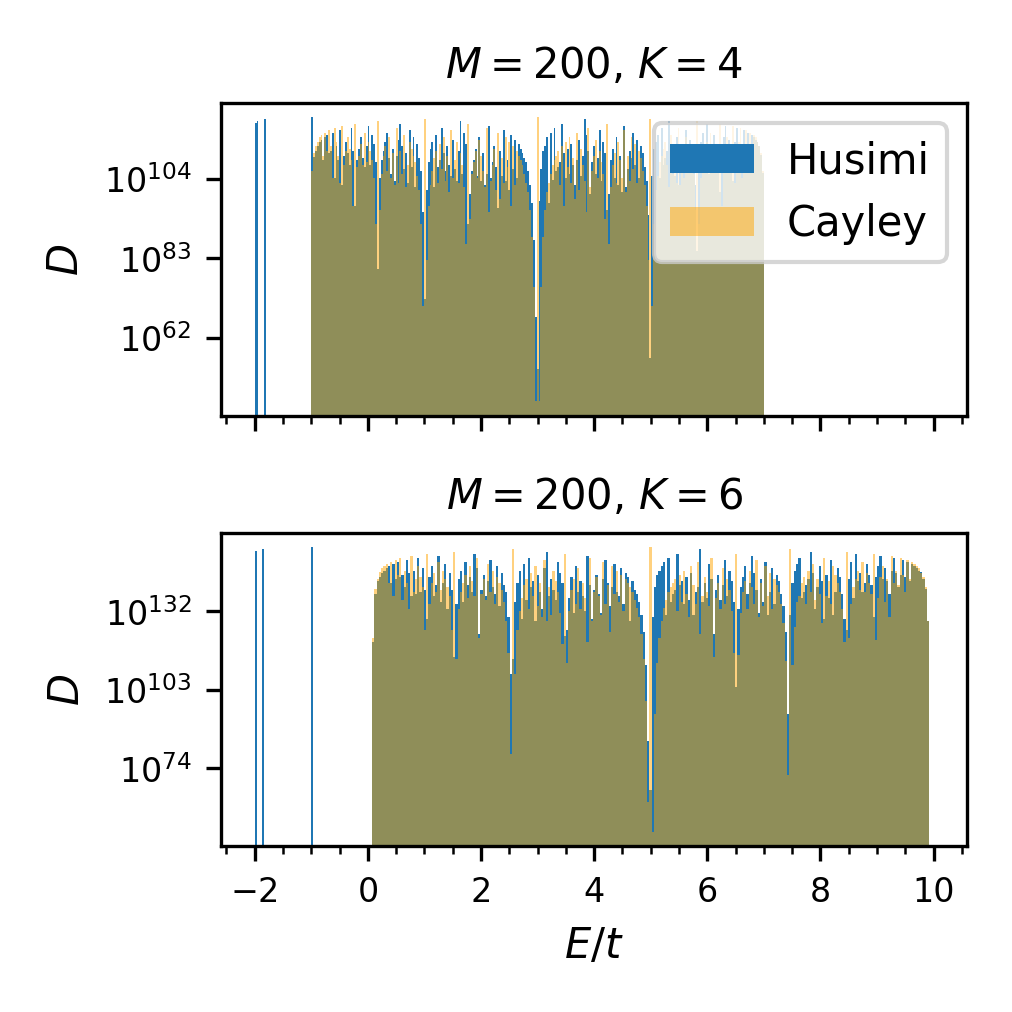}
\caption{
The spectra of two Husimi-Cayley trees with different branching factor $K$, each compared against the spectrum of a simple Cayley tree of the same $K$ and $M$ shifted by energy $\Delta E = K-1$. 
We observe that there is a mismatch between the spectra, most manifestly in the highly degenerate states of the shifted Cayley spectrum at $E = K-1$, which is absent in the Husimi-Cayley spectrum}.
\label{Fig:Husimi_ShiftedCayleyComparison}
\end{figure}

Let us discuss what the above general theory for line graphs implies for the spectrum of the Husimi-Cayley tree.
First, 
we compare the spectrum of the Husimi-Cayley tree with the spectrum of a Cayley tree shifted by $q - 2 \equiv  K-1$ as shown in Fig.~\ref{Fig:Husimi_ShiftedCayleyComparison}.
We observe that there is a small but noticeable mismatch between the two spectra. 
This is because the Cayley tree is not a $q$-regular graph: a finite fraction of the nodes are leaf nodes that 
have a lower degree than the inner nodes. 
Nevertheless, the spectrum obeys the general shift by 
$K-1$. 
For an infinite Husimi-Cayley tree, i.e., in the absence of boundaries, the tree becomes $q$-regular and it is expected to obey Eq.~(\ref{Eq:Husimi_CharacPolyonmial_kRegular}) exactly.

Second, we consider the degeneracy of the flat band at $E=-2$ predicted by Eq.~(\ref{Eq:Husimi_CharacPolyonmial_nonRegular}). 
The difference between the edge count and the vertex count 
of the Cayley tree is easily found to be $\mathcal{E}-\mathcal{V} = -1$. 
This result can be seen inductively, considering how branching from a parent node to its $K$ children always introduces $K$ new edges, leaving the root node as the only node without a corresponding edge. 
The negative exponent is brought to $\mathcal{E}-\mathcal{V}+c_0 = 0$ after accounting for $c_0 = 1$ factors $(x+2)$ originating from the determinant part of Eq.~(\ref{Eq:Husimi_CharacPolyonmial_nonRegular}).
To understand this, recall that the right-hand side of Eq.~(\ref{Eq:Husimi_CharacPolyonmial_nonRegular}) contains $\det[(x+2)\mathbb{I}_\mathcal{V}-B B^T]$. 
If the incidence matrix $B$ does not have the full rank $\mathcal{V}$, then this determinant contains factor $(x+2)^{\mathcal{V}-\rank(B)}$. 
It can be proved~\cite{godsil_algebraic_2001} that $\mathcal{V}-\rank(B) = c_0$, where $c_0$ is the number of bipartite connected components of graph $G$.
Since the Cayley tree has a single connected component, which is also bipartite, we have $c_0 = 1$, resulting in $\mathcal{E}-\mathcal{V}+c_0 = 0$ eigenstates with exact energy $E=-2$.

Nonetheless, we observe an accumulation of a large number of states near $E=-2$. 
This can be explained by studying the large $M$ limit of the eigenvalues of the non-symmetric sector $\mathcal{H}_{\textrm{nonsym.}}^\alpha$.
After shifting Eq.~(\ref{Eq:Husimi_NonSymm}) by constant energy $-K+1$~to 
\begin{equation}
\tilde{\mathcal{H}}_{\textrm{nonsym.}}^\alpha = \mathcal{H}_{\textrm{nonsym.}}^\alpha - (K-1)\mathbb{I},
\end{equation}
the tridiagonal nature of the obtained matrix allows us to use the Laplace expansion to express its determinant recurrently~as 
\begin{equation}
    \text{det}(\tilde{\mathcal{H}}_{\textrm{nonsym.}}^\alpha 
    - \lambda \mathbb{I}) \equiv d_{m}(\lambda) = - \lambda d_{m-1}(\lambda) - Kd_{m-2}(\lambda)
\end{equation}
for $m \geq 3$, where $m$ is the linear dimension of the Hamiltonian.
The initial conditions for the determinant of short-chain shell-non-symmetric Hamiltonians are $d_1(\lambda) = - \lambda - K$ and $d_2(\lambda)=\lambda(K+\lambda)-K$. 
This recurrence problem can be solved analytically as
\begin{equation}
\begin{aligned}
    d_m(\lambda) &=\frac{1}{2^{m+1}\Delta} \left[ (-2K -\lambda + \Delta)(-\lambda + \Delta)^{m} \right.  \\
    &\phantom{=}\qquad \qquad \left.+ (2K + \lambda + \Delta)(-\lambda - \Delta)^m\right],
\end{aligned}
\end{equation}
where we introduced $\Delta = \sqrt{-4K + \lambda^2}$ to achieve brevity.

To find a measure for how an eigenvalue of $\mathcal{H}_{\textrm{nonsym.}}^\alpha$ approaches $E=-2$ (which corresponds to an eigenvalue of $\tilde{\mathcal{H}}_{\textrm{nonsym.}}^\alpha$ approaching $-K-1$) as we increase the length of the Hamiltonian sector,
we substitute $\lambda = -(K+1)+\delta $ in the expression for $d_M(\lambda)$. 
This allows us to perform a Taylor expansion around $\delta = 0$ to first order and to solve the linearized characteristic polynomial for $\delta$. 
After these steps, we arrive at 
\begin{equation}
\label{eqn:husimi-convergence-result}
    \delta = \frac{(K-1)^2}{K^{\tilde{M}+1} + \tilde{M} - K(\tilde{M}+1)},
\end{equation}
where $\tilde{M} = M - l(\alpha)$ is the length of the non-symmetric sector with seed $\alpha$. 
The result in Eq.~(\ref{eqn:husimi-convergence-result})
provides an analytical approximation for the distance between the eigenvalues of the non-symmetric sector 
and $E=-2$. 
We find that as the length $\tilde{M}$ increases, one eigenvalue approaches $E=-2$ exponentially, thereby confirming the accumulation of states at $E=-2$ as observed in the spectral plots. 
The occurrence of an eigenstate near energy $E=-2$, which lies outside of the single energy band of the effective 1D chain, follows from the different on-site potential at the outermost site in Eq.~(\ref{Eq:Husimi_NonSymm}). 
This potential acts as an impurity that a binds a localized state near the chain boundary, which at the level of the Husimi-Cayley tree corresponds to a localized state in the~bulk.

In conclusion, we have shown that the approach of symmetry-adapted basis states 
works for the Husimi decoration of the Cayley tree and that the resultant symmetry sectors help us 
elucidate the behavior of the spectrum. 
In particular, we described the CLSs which generate 
the flat band at $E=-1$, and we provided 
an explanation for the accumulation of states near $E=-2$ due to an eigenvalues of all the non-symmetric sectors approaching the bound exponentially. 
Finally, we anticipate that the overall shift of the spectrum by $K-1$ is related to the fact that the Husimi-Cayley tree is approximately the line graph of the Cayley tree (except for the sites at the boundary), with this statement becoming exact in the limit of the infinite Husimi-Cayley tree.

\section{Clique decoration}\label{Sec:CliqueDecoration}
In this section, we discuss the clique decoration of the Cayley tree where each node is replaced by a clique of nodes. 
This decoration is reminiscent of the Lieb-Cayley tree, except for the inclusion of addition edges connecting Lieb nodes that share the same parent node, and with the root node at the center removed; alternatively, it corresponds to the Husimi-Cayley tree with additional edges added between the corner-sharing simplices. 
This dual interpretation motivated the name ``Husimi-Lieb'' for the same decoration in Ref.~\citenum{duss_exploration_2025}.
We adapt the approach of symmetry-adapted basis states to clique-Cayley trees, finding their exact spectrum which exhibits in-gap states.
We find no indication that these states are topological edge states; however, we show that some of them can be understood as 
perturbed SSH edge~states. 

Our treatment of the clique-Cayley tree is organized as follows.
In Sec.~\ref{sec:clique_euclidean} we introduce the Euclidean analog of the clique-decorated tree, which we call the star lattice. 
It hosts two flat bands at energies $E=0$ and $E=-2$, both of which can be explained through the construction of CLSs. 
In Sec.~\ref{sec:clique_exact_solution} we present the block Hamiltonians of the clique-Cayley trees that result from the symmetry-adapted basis states. 
We find that the blocks describe a one-dimensional chain with two sites per unit cell (thus, with two energy bands) and 
with a staggered on-site potential. 
In addition, the on-site potential exhibits a defect on the first site of the chain, reminiscent of the defect at the boundary also found in the block Hamiltonians for the Husimi-Cayley tree.
The spectrum exhibits an accumulation of a large number of in-gap states near energies $E=0$ and $E=-2$, although these differ from the star lattice in that they are not exactly flat. 
Finally, in Sec.~\ref{sec:clique_spectrum_properties} we study the bulk model of the block Hamiltonians and find that it does not exhibit any topological 
phases. 
Nonetheless, one can understand the appearance of the in-gap edge states by studying the perturbation of edge states in the SSH model by a staggered on-site potential. 
Considering further the altered on-site potential at the beginning of the 1D chain results in a cancellation of the energy shift imposed by the staggered on-site potential, causing some of the edge states to occur near $E=0$ just as for the SSH model.

\subsection{Star Lattice}\label{sec:clique_euclidean}

The motivation for the clique decoration derives from the so-called ``star'' lattice (also called the ``expanded kagome'' lattice). 
In graph theory, the star lattice is obtained as the line graph of the subdivision graph of the hexagonal graph. 
This lattice shares certain similarities with both 
the kagome lattice and the honeycomb lattice, and it has been shown to exhibit topological quantum phase transitions when imposing additional interactions \cite{Yao:2007,chen_topological_2012,fan_two-dimensional_2023,rojas_geometrically_2019}. 

\begin{figure}[t!]
\centering
\includegraphics[width=\linewidth]{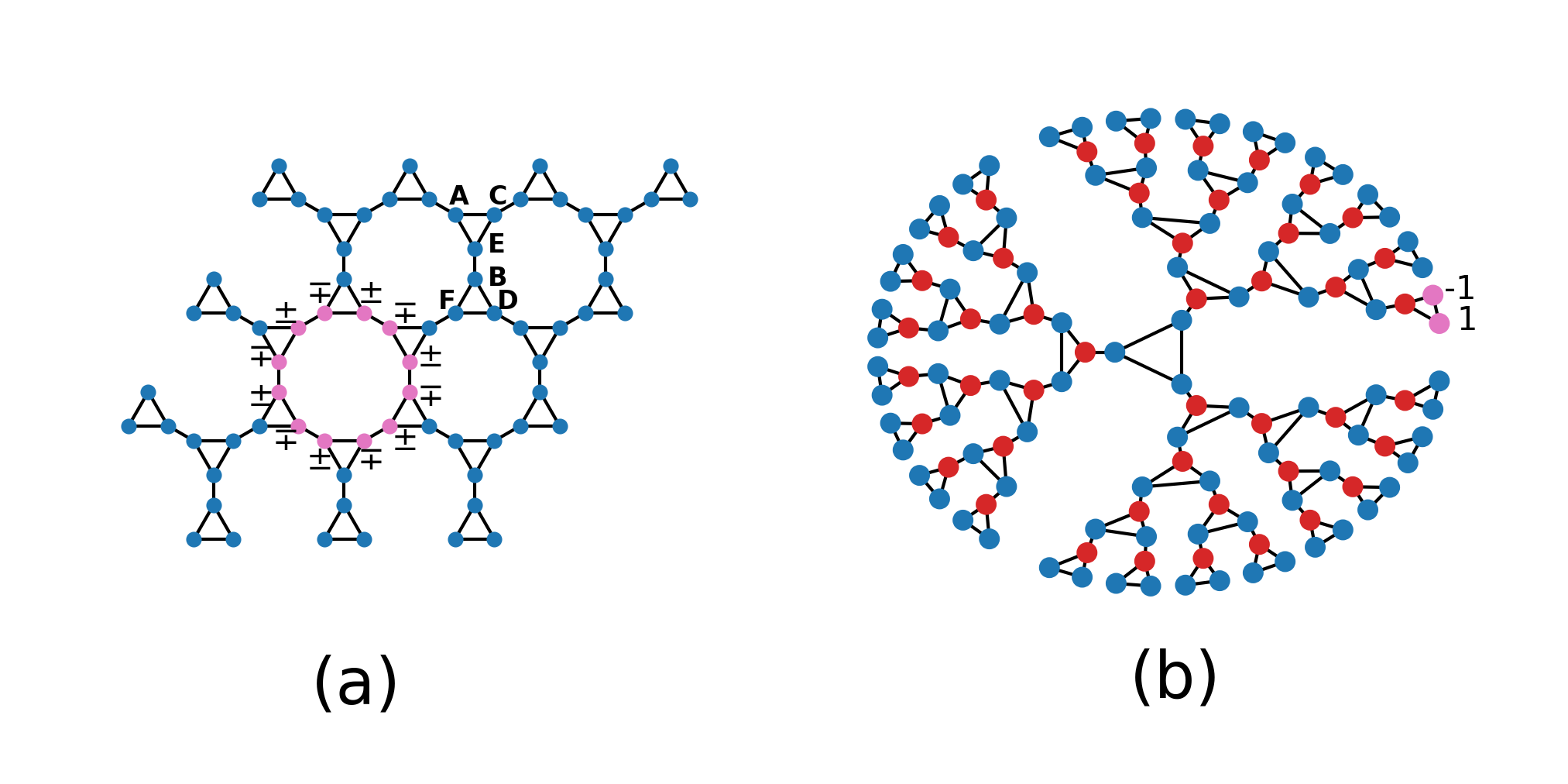}
\caption{
(a)~The star lattice is the line graph of the subdivision graph of the honeycomb lattice. 
The unit cell contains the six sites, labeled $\{A,B,\ldots,F\}$, and exhibits two 
flat energy bands due to CLSs. One example of CLS has been added in pink, which corresponds to alternating amplitude on a dodecagon plaquette.
(b)~Clique-Cayley tree with branching factor $K=2$ and $M=9$ layers. 
This decoration is similar to the Lieb-Cayley tree (with the original Cayley nodes colored red and Lieb nodes shown in blue), with additional edges between the Lieb nodes that share the same parent and with the root node at the center removed. Here we also find CLSs at the boundary of the tree, an example illustrated in pink.}
\label{Fig:Clique_LatticeSketch}
\end{figure}

The star lattice has the same triangular Bravais lattice as the kagome lattice, specified earlier in Eq.~(\ref{eqn:triangular-Bravais}).
Each unit cell contains six inequivalent sites, which we denote as $A,B,\ldots,F$, as shown in Fig.~\ref{Fig:Clique_LatticeSketch}. 
Assuming tight-binding model with NN hopping amplitude $t$, the diagonalization reveals six energy bands~\cite{fan_two-dimensional_2023}
\begin{subequations}
\begin{gather}
    E_{1,2}(\mathbf{k}) = \frac{1}{2}t \big(1 \pm \sqrt{13 + 4f(k)} \big) \\
    E_{3,4}(\mathbf{k}) = \frac{1}{2}t \big(1 \pm \sqrt{13 - 4f(k)} \big) \\
    E_{5,6}(\mathbf{k}) = -t \pm t
\end{gather}
\end{subequations}
where $f(\mathbf{k}) = \sqrt{3 + 2\cos(k_1) + 2 \cos (k_2) + 2\cos(k_3)}$ 
with $k_{1,2,3}$ defined below Eq.~(\ref{eqn:Bloch-kagome}).
For simplicity, we set $t=1$ in the following.

The spectrum exhibits two flat bands, one of which occurs at $E=-2$ as expected due to the line graph nature of the lattice, and the other one at $E=0$. 
Both flat bands can be understood by considering CLSs on the graph. 
Let ${|i\rangle}_{i=1}^{12}$ be the twelve sites arranged clockwise at the vertices of a dodecagon on the lattice. 
Then, by placing equal-sized positive and negative amplitudes in an alternating pattern, we find
\begin{subequations}
\begin{equation}
    H |\text{CLS}_1\rangle = H \Big(\frac{1}{\sqrt{12}} \sum_{i=1}^{12} (-1)^{i} |i\rangle \Big) = -2 |\text{CLS}_1\rangle 
\end{equation}
and
\begin{equation}
    H |\text{CLS}_2\rangle = H \Big(\frac{1}{\sqrt{12}} \sum_{i=1}^{6} (-1)^{i } \Big[|2i\rangle + |2i+1\rangle\Big]\Big) = 0.
\end{equation}
\end{subequations}
This corresponds exactly to the observed flat bands. 

\subsection{Exact Solution of the Clique Tree}\label{sec:clique_exact_solution}
In this subsection, we introduce the clique-decorated Cayley trees and present the symmetry sectors that govern their dynamics. 
Utilizing properly adjusted symmetry-adapted basis states (see Appendix~\ref{Sec:App_Clique}), we obtain a block Hamiltonian that corresponds to one-dimensional SSH chains with staggered on-site potentials.
The simple form of the Hamiltonians allows us to derive the complete spectrum of the clique-Cayley trees and to discuss the CLSs that live on their boundary. 

We want to place ($K+1$)-gon cliques (i.e., a collection of $K+1$ vertices in which each pair is connected by an edge) on the Cayley tree, which for $K=2$ imitates the tiling of the star lattice. 
The latter can be understood as a hexagonal tiling of triangles connected by ``bridge edges''. 
To construct such a decoration, we take the Lieb-Cayley tree of Sec.~\ref{Sec:Lieb_2_exact}, connect all Lieb nodes originating from the same parent Cayley node, and finally we remove the root node at the center.
The resulting tree, which we call the clique-Cayley tree 
is shown in Fig.~\ref{Fig:Clique_LatticeSketch}. 
It is the line graph of the Lieb-Cayley tree. 

\begin{figure}[t!]
\centering
\includegraphics[width=\linewidth]{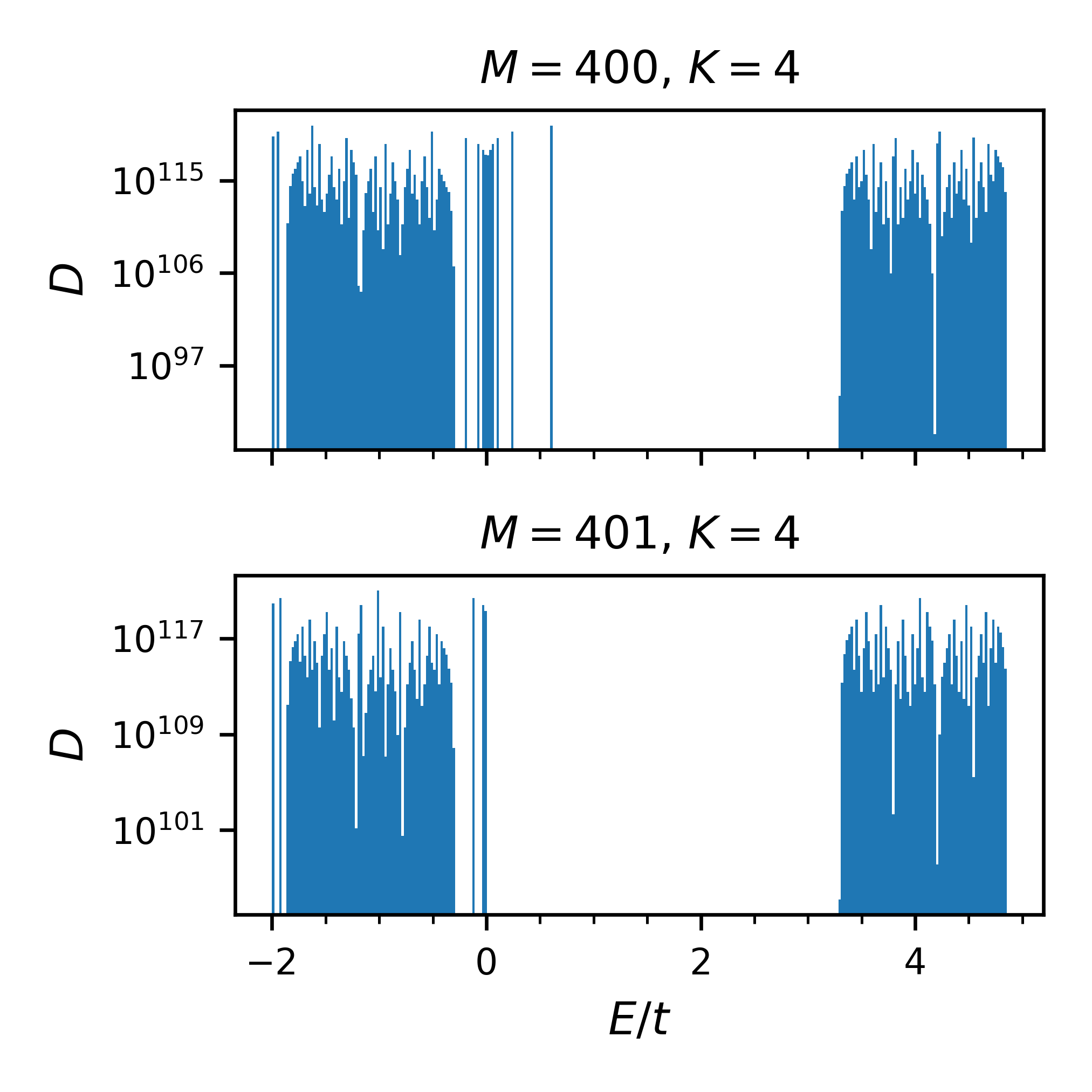}
\caption{Spectra of clique-Cayley trees with the same branching factor $K=4$ and for different termination conditions. 
In both cases we find the occurrence of in-gap states. 
For $M$ even, we find a set of states that are symmetrically arranged around $E=0$, further states near $E=-2$, as well as an isolated flat band at $E\approx 0.62$. 
In addition, a flat band with high degeneracy occurs at $E\approx -1.62$ inside the bulk energy band.
For $M$ odd, we find states accumulating at $E=0$ without a symmetric pattern, 
a set of states near $E=-2$ similar to the case of $M$ even, as well as a highly degenerate set of states at $E=-1$ inside the bulk~band. 
}
\label{Fig:Clique_Spectra}
\end{figure}

The Clique tree has the same number of nodes as the Lieb-Cayley tree minus the central node $0$, i.e., 
\begin{equation}
\label{eq:total_number_of_states_LC_lattice}
    \begin{aligned}
    N_\textrm{total} 
    & =  (K+1) \times \sum_{l_\textrm{C} = 1}^{M_\textrm{C}} K^{l_\textrm{C} - 1} + (K+1) \times \sum_{l_\textrm{L} = 1}^{M_\textrm{L}} K^{l_\textrm{L} - 1} \\
    & = \frac{(K+1)}{(K-1)} (K^{M_\textrm{C}} + K^{M_\textrm{L}} -2)
  \end{aligned}  
\end{equation}
Using the symmetry-adapted basis states, the Hilbert space decouples into sectors analogous 
to those of Sec.~\ref{Sec:Lieb_2_exact}, except that the derived block Hamiltonians 
include an on-site potential on the Lieb layers; namely,
\begin{subequations}
\begin{equation}
        \mathcal{H}_{\textrm{sym.}} = \begin{pmatrix}
        K & 1 & 0 & 0 & 0 &\cdots \\
        1 & 0 & \sqrt{K} & 0 & 0 & \cdots\\
        0 & \sqrt{K} & K-1 & 1 & 0 & \cdots \\
        0 & 0 & 1 & 0 & \sqrt{K} & \cdots \\
        0 & 0 & 0 & \sqrt{K} & K-1 & \cdots\\
        \vdots & \vdots & \vdots & \vdots  & \vdots  & \ddots
    \end{pmatrix},
\end{equation}
and
\begin{equation}\label{Eq:clique_nonSymm_Sector_H}
        \mathcal{H}_{\textrm{nonsym.}}^\alpha = \begin{pmatrix}
        -1 & 1 & 0 & 0 & 0 &\cdots \\
        1 & 0 & \sqrt{K} & 0 & 0 & \cdots\\
        0 & \sqrt{K} & K-1 & 1 & 0 & \cdots \\
        0 & 0 & 1 & 0 & \sqrt{K} & \cdots \\
        0 & 0 & 0 & \sqrt{K} & K-1 & \cdots\\
        \vdots & \vdots & \vdots & \vdots  & \vdots  & \ddots
    \end{pmatrix}.
\end{equation}
\end{subequations}
These Hamiltonians describe a 1D chain with two bands and with a staggered on-site potential. 
As we show in the next section, the on-site potential breaks chiral symmetry and inversion symmetry of the SSH Hamiltonian that emerged in the analogous description of the Lieb-Cayley tree. 
However, the tree does exhibit in-gap states, as visible in Fig.~\ref{Fig:Clique_Spectra}. 

In analogy with the earlier sections, the first place to look for the origin of these in-gap states is the boundary of the tree. 
The position states on the boundary of the tree again contribute a highly degenerate set of eigenstates. 
For $M$ odd, the tree terminates on the Lieb layer, and we find the occurrence of a flat band due to CLSs with energy $E=-1$.
These states are equivalent to the CLSs at the same energy as found earlier for the Husimi Cayley tree, discussed in Sec.~\ref{sec:Husimi_ExactSolutions}. 
For $M$ even, the smallest available symmetry sector has $\dim(\mathcal{H}_{\textrm{nonsym.}}^\alpha) =2$, 
and therefore hosts two states.
Their energies are $E=\frac{1}{2}(\pm \sqrt{5} - 1)$, which explains the isolated set of states at $E\approx 0.62$. 
However, these considerations fail to explain the states accumulating around $E=0$ and $E=-2$ in both termination cases. 
We clarify their origin in the next section.

\subsection{Discussion of Spectrum Properties}\label{sec:clique_spectrum_properties}

In this subsection, we first discuss the bulk 
of the finite 1D chain models described by Eq.~(\ref{Eq:clique_nonSymm_Sector_H}) to clarify 
the origin of the additional in-gap states of the model. 
After excluding a topological origin, we will move on to show the impact of the defect at the beginning of the 1D chain on the eigenenergies of the system. 

For the purpose of investigating the bulk model, we can ignore 
the defect of the on-site potential at the beginning of the chain. 
The resulting 1D chain has two sites per unit cell with a staggered hopping amplitude, equivalent to the SSH model, as well as an additional staggered on-site potential.
Assuming periodic boundary conditions, we find the following momentum-space Hamiltonian
\begin{subequations}
\begin{equation}
\label{Eq:HL_bulk_momentumspace_hamiltonian}
    H (k) = \begin{pmatrix}
K-1 & 1 +\sqrt{K} e^{-ik}\\
1 +\sqrt{K} e^{ik} & 0 
\end{pmatrix}.
\end{equation}
Rewriting in Pauli matrices, we find
\begin{equation}
\label{eqn:clique-Cayley-2D-bulk-Ham}
    H(k) = \frac{K-1}{2}\sigma_0 + d_x(k)\sigma_x + d_y(k)\sigma_y + d_z \sigma_z,
\end{equation}
\end{subequations}
with
\begin{equation}
    d_x = 1 +\sqrt{K}\cos (k), \quad d_y = \sqrt{K} \sin (k), \quad d_z = \frac{K-1}{2}.
\end{equation}
The energy bands are
\begin{equation}
    E_{\pm}(k) = \frac{K-1}{2} \pm \sqrt{d_z^2 + |1 + \sqrt{K}e^{ik}|^2}.
\end{equation}
Since all three Pauli matrices $\sigma_{x,y,z}$ are used in Eq.~(\ref{eqn:clique-Cayley-2D-bulk-Ham}), the chiral and the inversion symmetry are both broken. 
The absence of inversion symmetry implies that we do not expect there to be a quantized Zak phase in any termination case, i.e., the in-gap states are not of topological origin. 
On the other hand, the absence of the chiral symmetry is manifested by the fact that the near-zero-energy states are not pinned to stay there: upon introducing a tuning parameter in the clique-Cayley tree (namely, assuming different hopping implitudes within the ``cliques'' than on the ``bridges'' between them), it is possible to move the in-gap states away~$E=0$ (not shown in the Figures). 

However, it is noteworthy that the clique tree has approximate flat bands at $E=-2$ and $E=0$, paralleling its Euclidean analog. 
We anticipate that the precise values of these eigenvalues can be obtained through a derivation analogous do the one presented for edge states accumulating near $E=-2$ for the Husimi-Cayley tree. 
The doubled unit cell size implies that the recurrence relation for $\det(\mathcal{H}_{\textrm{nonsym.}}^\alpha - \lambda \mathbb{I}) \equiv d_{m}(\lambda)$ obtained via the Laplace expansion has to be expressed in a matrix form; specifically, one can show that
\begin{equation}
\left(\begin{array}{c}
d_{2n+2} \\
d_{2n+1}
\end{array}\right) = \left(\begin{array}{cc}
-\lambda(K - 1 - \lambda) - 1   & \lambda K \\
K-1-\lambda     & - K
\end{array}\right)\left(
\begin{array}{c}
d_{2n} \\
d_{2n-1}
\end{array}\right).
\end{equation}
This recurrence can, in principle, be solved analytically by finding biorthogonal eigenstates of the $2\times 2$ matrix, and the sought Hamiltonian eigenvalues are then obtained by expanding the derived expressions for $d_m(\lambda)$ to linear order around $\lambda = 0$ resp. $\lambda = -2$.
However, as the resulting expressions are rather long and not particularly illuminating, we do not follow this path further.

Regarding the states at $E=0$, their approximately symmetric distribution around zero energy for the $M$-even case is reminiscent of the symmetric arrangement around the same energy found earlier for even-$M$ Lieb-Cayley trees in Fig.~\ref{Fig:LiebCayley_EvenVsOdd}. 
Indeed, one can understand the Bloch Hamiltonian in Eq.~(\ref{Eq:HL_bulk_momentumspace_hamiltonian}) as the SSH model with an additional on-site potential on odd sites in the position basis. 
This motivates us to study the perturbation of the edge states in a finite SSH chain under the introduction of a sublattice on-site potential. 
For the purpose of this argument, let $J = K-1$ and let $|\delta| \propto C|\frac{t}{t'}|^L$ be the energy shift of the hybridized edge states, where $L$ is the length of the chain, $t=1$, $t'=\sqrt{K}$ and $C$ a constant of order unity.

We project to the subspace spanned by the two edge states in the finite SSH model $\{ \ket{L}, \ket{R}\}$, where $\ket{L}$ is the state exponentially localized on the left edge, which 
has support on sublattice $A$ only, while $\ket{R}$ is exponentially localized on the right edge and has support on sublattice $B$ only. 
We find the effective Hamiltonian to be
\begin{equation}
    H_{\text{eff}} = \begin{pmatrix}
        J & \delta \\
        \delta & 0
    \end{pmatrix}.
\end{equation}
In this matrix, the diagonal terms correspond to the on-site potentials  $V$ in the bulk; namely, 
$\bra{L}V\ket{L} \approx J$ and $\bra{R}V \ket{R} \approx 0$.
We neglect mixing with the bulk states, since they are far in energy from the edge states for the SSH model. 
We can solve this effective Hamiltonian for its eigenvalues to find
\begin{subequations}
\begin{equation}
    E_{\pm}(J) = \frac{J \pm \sqrt{J^2 + 4\delta^2 }}{2}.
\end{equation}
For large $J$, we can approximate 
\begin{equation}
\label{eqn:approx-clique-edge}
    E_+ (J) \approx J + \frac{\delta^2}{J}, \quad E_- (J) \approx - \frac{\delta^2}{J}.
\end{equation}
\end{subequations}
This implies that one of the states remains close to $E=0$ while the other moves with $J$. 
However, studying Fig.~\ref{Fig:Clique_Spectra}, we observe symmetrically arranged states, implying that both energies remain close to $E=0$ regardless of $J$. 

To explain the mismatch above, we need to include the defect at the beginning of the clique chain. 
We again place a staggered on-site potential $J$ on all sublattice $A$ sites except the first one, labeled $|1\rangle$, which receives a potential $J_{\ket{1}}$.  
\begin{equation}
    V = J\sum_{j\in A, j \neq 1} \ket{j}\bra{j} + J_{\ket{1}} \ket{1}\bra{1}
\end{equation} 
Now because the left edge state is mainly localized on site $|1\rangle$ we find that $J_L = \bra{L}H_{\text{eff}}\ket{L}$ 
will be an interpolation between the defect potential $J_{\ket{1}}$ and the sublattice potential $J$, where the interpolation depends on the extension of the edge state, which in turn depends on the relationship between $t$ and $t'$. 
Assuming the spatial dependence of the edge-state wave function taking the exponential form 
\begin{equation}
\langle 2n - 1|L\rangle \equiv 
\psi_L(2n-1) = A(-r)^{n-1}
\end{equation}
with $r=|\frac{t}{t'}| = \frac{1}{\sqrt{K}}$, one finds that for large number of sites $N$, the normalization factor converges to $A=\sqrt{1-r^2}$. Calculating $J_L$ explicitly
\begin{equation}
\begin{aligned}
        J_L &= \bra{L} V \ket{L} = J_{\ket{1}}(1-r^2) + J\sum_{n \geq 2} (1-r^2)r^{2(n-1)} \\
        &= J_{\ket{1}} + (J - J_{\ket{1}})r^2.
\end{aligned}
\end{equation}
If we now insert $J_{\ket{1}} = -1$ and $J=K-1$ and $r^2 = 1/K$, we find $J_L = 0$. 
In other words, the relationship between $r$, $J$ and the defect $J_{\ket{1}}$ on the clique tree leads to an exact cancellation of the effective on-site potential in the edge-state subspace. 
This assumes that $N$ is sufficiently large for the leading terms to dominate in the various derivations performed above. 
For the $N$ odd case, where there is no integer number of unit cells and only a single edge state on the left $|L\rangle$, one finds that the energy shift of the edge state depends solely on $J_L$. 
This then explains why we find an accumulation of states at $E=0$ in the $M$ odd case as the tree gets larger. 

\begin{figure}[t!]
\centering
\includegraphics[width=\linewidth]{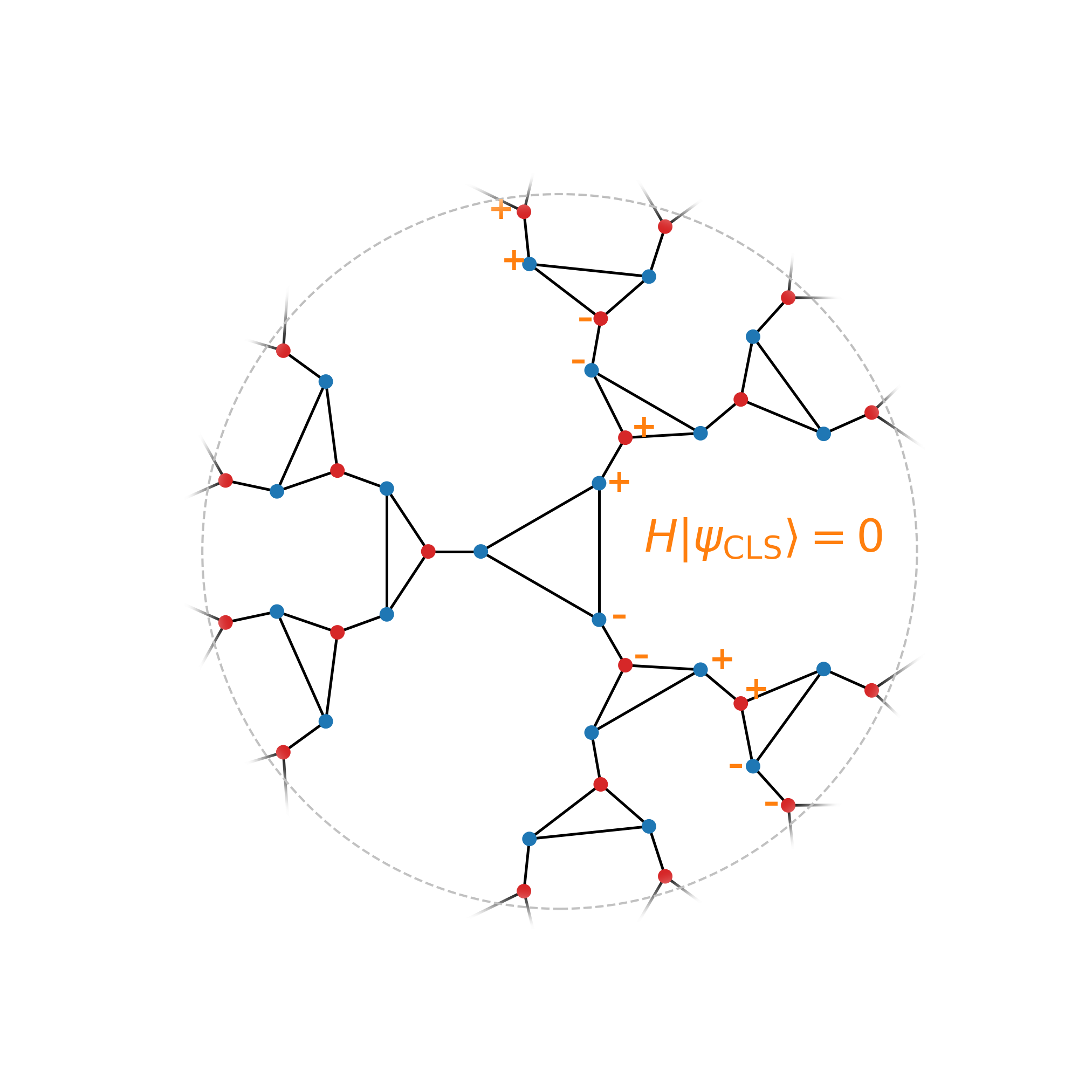}
\caption{The clique-decorated tree in its infinite $M = \infty$ (Bethe) limit hosts a large set of CLSs at the exact energies $E=0$ and $E=-2$. Every such state corresponds to an infinite chain of alternating amplitudes $\pm 1$ that thread \emph{ad infinitum} across the whole tree, as indicated in this figure.
Every decorated tree with only NN hoppings, 
whose Euclidean analog hosts a flat band of CLSs, will host such infinite string states 
with their energies matching exactly the energy of the CLSs of the related 
Euclidean lattice.
}
\label{Fig:Clique_InfiniteCLS}
\end{figure}

In conclusion, as the tree grows we expect edge states to get closer to $E=-2$ and $E=0$. 
Furthermore, in the limit of infinite system size, the tree will host exact flat bands at the aforementioned energies. 
This is because it is possible to find infinite string states that stretch throughout the infinite tree and whose energies are exactly 
at $E=-2 $ and $E=0$. 
We show an example of such an infinite string state with $E=0$ in Fig.~\ref{Fig:Clique_InfiniteCLS}. 
This is not a coincidence and it can be shown that an infinite tree will always ``inherit'' the flat bands of its Euclidean analog, as we show in the next section.

\section{Flat bands on infinite trees}\label{Sec:CoveringGraphs}

In this section, we explain the existence of flat bands on infinite versions of the decorated trees considered in previous sections from the perspective of covering graphs.
Later, we comment on how the result transfers to the case of finite decorated trees.
A graph $\mathcal{L}$ is called a covering graph of a graph $L$ if there exists a covering map $f: \mathcal{L}\rightarrow L$ i.e., a surjective map that is a local isomorphism \cite{godsil_algebraic_2001}. 
In the canonical definition, the neighborhoods are taken to be the adjacent sites to a given vertex $v$, and the local isomorphism means the bijection between edges starting from the vertex $v\in \mathcal{L}$ and the projection $f(v)\in L$. 
However, for our aims, we adopt a slightly more restrictive definition of local isomorphism. 
Namely, our `neighborhoods' $U(v)$ consist of three ingredients: (1)~an initial vertex $v$, together with (2)~all its adjacent sites $\{u_i\}_{i=1}^{\textrm{nn}(v)}$ [where $\textrm{nn}(v)$ is the number of nearest neighbors of $v$], as well as (3)~all the edges connecting any pair of sites within the set $S_v=\{v,u_1,\ldots,u_{\textrm{nn}(v)}\}$. 
By adopting this definition, we ensure that triangles (as well as their collections) meeting at one vertex, such as those arising in the Husimi and the clique decoration, are preserved by the covering map.

We should also note that among all possible lifts $\mathcal{L}$ of graph $L$ (assumed to be connected), there is a unique `maximal' one called the \emph{universal covering graph} $\mathscr{L}$, which
(\emph{i})~is connected, (\emph{ii})~does not contain any loops (other than those contained within the neighborhoods $U_v$), and (\emph{iii})~is the covering graph of all coverings $\mathcal{L}$. The absence of loops hints that universal covering graphs allow us to elevate Euclidean lattices to Bethe lattices, as we elaborate later.

Across the considered Euclidean models that motivated the presented Cayley tree decorations, we focused on CLSs with non-trivial support on one-dimensional cycles. 
In the following discussion, we show that these CLSs can be lifted to eigenstates on covering graphs with the same eigenvalue. 
More precisely, we introduce the notion of \emph{string states} that have support on a finite or infinite chain on a graph. 
Then the Euclidean CLSs considered in the previous sections [e.g., those in Fig.~\ref{Fig:LiebLattice_LiebCayleyTree}(a) and in Fig.~\ref{Fig:2xLC_CLS_Sketch}]
are examples of finite string states. 
We claim that string states on the graph $L$ are lifted to string states on the covering graph $\mathcal{L}$ with the covering map defined above. 

Before proceeding with the argumentation, we remark that for a given graph $L$ and its covering graph $\mathcal{L}$, one can lift the Hamiltonian $H$ defined on $L$ to the Hamiltonian $\mathcal{H}$ defined on $\mathcal{L}$. 
This is possible to do because the considered nearest-neighbor Hamiltonians $H$ match the adjacency matrix of the lattice graph $L$. 
The lifting is then made by assigning the local hoppings in $\mathcal{H}$ according to the hoppings in $H$ via the local isomorphisms induced by the covering map $f$. 
More precisely, the matrix elements of Hamiltonian $\mathcal{H}$ can be defined as
\begin{gather}
\begin{cases}
\mathcal{H}_{v' u'}{=}H_{vu},\, \mathrm{if\,} \mathrm{e}(v,u)\in L\,\, \mathrm{and}\,\, \mathrm{e}(v',u')\in \mathcal{L}\\
\mathcal{H}_{v' u'}{=}0,\,\,\mathrm{otherwise},
\end{cases}
\end{gather}
where $v'{\in} f^{-1}(v)$, $u'{\in} f^{-1}(u)$, with  $f^{-1}$ being the full preimage of the covering map $f$, and $\mathrm{e}(u,v)$ means an edge connecting the vertices $v$ and $u$.

The Hamiltonians on the infinite versions of decorated trees studied in the previous sections are examples of such lifts of the Hamiltonians defined on the Euclidean side. For example, in the case of decorations of the square lattice (i.e., the Lieb and the double-Lieb lattice) one should first observe that a regular tree graph with coordination number $q=4$ is the universal covering graph of the square lattice. 
Since the decorations are obtained by adding vertices on the edges, i.e. a local change which contributes more features to the local neighborhoods, applying the decoration to both the square lattice and the corresponding tree (both assumed to be infinite, i.e., never terminating at a boundary), one preserves the covering structure. 
Similar covering correspondence can be derived for the Husimi and for the the clique decoration. 
Starting with the kagome lattice [Fig.~\ref{Fig:Husimi_Sketch}(a)] or the star lattice [Fig.~\ref{Fig:Clique_LatticeSketch}(a)], and noting that each triangle (together with all its vertices and edges) belongs to the neighborhood $U_v$ of some vertex $v$, we establish the infinite $q=3$ Husimi-Cayley (clique-Cayley) tree as the universal cover of the infinite kagome (star) lattice.

Now, we are set to discuss the lifting of string states.
Let $\phi$ be a finite string state with energy $\varepsilon$ defined on the lattice $L$ with Hamiltonian $H$ (for simplicity, we consider normalizable string states having finite support):
\begin{equation}
    H\phi=\varepsilon \phi.
\end{equation}
Then, there exists a lift of $\phi$, which we call $\varphi$, such that $\varphi$ is a string state of $\mathcal{H}$ with the same energy $\varepsilon$:
\begin{equation}\label{eq:lift_CLS}\mathcal{H}\varphi=\varepsilon \varphi.
\end{equation}
In other words, a string state on a lattice that supports compact localized states can be lifted to a string state on a covering graph with the same energy. 
This property is the direct consequence of the local nature of the wave-function amplitude of the string state, which can be shown by a direct construction as follows.

First, we should understand how cycles are lifted to the covering graph. The covering property implies that a cycle $C$ is lifted to a \emph{set} of cycles, which we denote 
\begin{equation}
\label{eqn:lifting-cycles}
    f^{-1}(C)=\{\mathcal{C}_i\},
\end{equation} 
where the cycles $\mathcal{C}_i$ can also have infinite lengths. 
In particular, each $\mathcal{C}_i$ has infinite length in the universal cover $\mathscr{L}$.
The statement in Eq.~(\ref{eqn:lifting-cycles}) 
can be shown by indexing the vertices in $C$ as $C=\{v_1,v_2,\ldots,v_l\}$ and taking a closed path through $C$ starting from $v_1$ and ending at the same vertex. Let us choose a particular lift $v'_1\in f^{-1}(v_1)$ of vertex $v_1$ as the starting point of the path. Then, the path is lifted uniquely to the covering graph $\mathcal{L}$ via consequent applications of local isomorphisms. Therefore, we have only two options: either the listed path starting from $v'_1$ returns to $v'_1$, or not. 
The first option corresponds to the case when the lift $\mathcal{C}_i$ is finite, while the second one corresponds to an infinite cycle $\mathcal{C}_i$.
By considering all possible choices of $v'_1\in f^{-1}(v_1)$ we find all lifts of $C$~in~$f^{-1}(C)$.

Having defined the lifts of cycles, we move to describing the lift of a finite string state $\phi$ supported on a cycle $C$. 
We choose a cycle $\mathcal{C}_i$ from $f^{-1}(C)$ and denote the amplitude of the state $\phi$ at the vertex $v$ as $\phi_v$. 
Using this notation, we can define the non-normalized lift $\varphi^{(i)}$ of $\phi$ by the following formula:
\begin{equation}
\label{eqn:lifted-state-on-one-cycle}
\begin{cases}
\varphi^{(i)}_{v'}=\phi_{v}, \quad v\in C, \quad v'\in f^{-1}(v)\,\, \mathrm{and}\,\, v'\in \mathcal{C}_i\\
\varphi^{(i)}_{v'}=0, \quad \mathrm{otherwise},
\end{cases}
\end{equation}
where $f^{-1}$ means the full preimage of the covering map $f$. 
Informally speaking, for all vertices with the same neighborhood, we assign the same amplitudes of the wave function as in the covered graph $L$. Since for each neighborhood the energy is defined only by the local amplitudes, the energy stays the same when passing from the graph $L$ to the covering graph $\mathcal{L}$. 
Therefore, $\varphi^{(i)}$ is an eigenstate of $\mathcal{H}$ localized to a single cycle $\mathcal{C}_i \in \mathcal{L}$, which has the same energy $\varepsilon$ as the CLS state $\varphi$ has with respect to the Hamiltonian $H$. 
Note that it is important that we have adopted the more restricted version of the covering map $f$, as defined in the beginning of the section, since this definition preserves the destructive interference on sites that are adjacent to multiple sites of the initial CLS (such destructive interference would be lost if the covering graph were constructed using the canonical definition of covering map, which can remove \emph{all} loops). 

In the case of finite $\mathcal{C}$, we can normalize $\varphi^{(i)}$:
\begin{equation}
    \varphi=\frac{\varphi^{(i)}}{||\varphi^{(i)}||}.
\end{equation}
The resulting state $\varphi$ is the lift of $\phi$ satisfying Eq.~(\ref{eq:lift_CLS}). 
owever, we should note that in the case of infinite trees, the lift of a finite Euclidean string state is an infinite string state (as illustrated in Fig.~\ref{Fig:Clique_InfiniteCLS}); hence, the lift is non-normalizable and one can only use Eq.~\eqref{eqn:lifted-state-on-one-cycle}. In other words, the lift $\varphi^{(i)}$ does not belong to the space $\ell^2(\mathcal{L})$ of square-integrable functions~on~$\mathcal{L}$.

As discussed in the previous sections, finite trees typically
do not support states that exactly reproduce the flat bands generated by Euclidean CLS states. Nevertheless, the associated energy levels converge exponentially toward the flat bands of the corresponding Euclidean lattices as the tree size increases. This confirms that the flat bands of Cayley trees are related to infinite string states of infinite trees. More precisely, one can still consider a finite string state on a Cayley tree that, within the bulk, is indistinguishable from an infinite string chain. The isomorphism breaks only at the endpoints, which coincide with the boundary of the Cayley tree. 
However, the larger the length of such a string state, the closer the state is to an eigenstate with exact flat band energy. 
This statement can be formulated more rigorously as follows. 
Let us denote such a normalized string state with length $L$ as $\psi_L$ and the corresponding flat-band energy (taken to be the \emph{exact} flat band energy of the Euclidean parent model) as $\varepsilon$. 
Since the mismatch between $\psi_L$ and an eigenstate with energy $\varepsilon$ occurs only at the endpoints, we can conclude that an appropriately chosen norm is a decreasing function of $L$:
\begin{equation}\label{eq:string_state_norm}
||(\mathcal{H}-\varepsilon)\psi_L||\le\frac{c}{\sqrt{L}}.
\end{equation}
Here, $c$ is a constant that depends on $\varepsilon$, hoppings, and on the components of the corresponding Euclidean CLS. We also used the fact that the components of $\psi_L$ arise from the lift of Euclidean CLS, and therefore depend on $L$ only through the normalization. By applying spectral decomposition, one can further show that the energy $\varepsilon$ lies \emph{at most} $c/\sqrt{L}$ away from the true eigenspectrum of the tree Hamiltonian. Although in practice, a much faster convergence is observed, as manifested, for example, by the perturbative result for the Husimi tree in Eq.~\eqref{eqn:husimi-convergence-result}.
We should also note, that in some cases, such as for Lieb–Cayley trees with Lieb boundary [Fig.~\ref{Fig:LiebCayley_RankNullity}(a)], the existence of an edge does not break compatibility with the exact flat-band energy, and exponentially localized bulk states can be obtained as a linear combination of finite string states lying within the support of the corresponding non-symmetric sector. 

\section{Conclusion and outlooks}\label{Sec:Conclusion}

In this work, we have extended the concept of flat energy bands to the setting of decorated Cayley trees. 
By introducing tree analogs of several well-known Euclidean lattices, namely of the Lieb, double Lieb, kagome, and star lattices, we demonstrated that exact flat bands or nearly flat bands can persist when passing from the Euclidean to the tree geometry. 
In the infinite limit, these decorated Bethe lattices inherit the exact flat bands of their Euclidean counterparts with the same energy, which can be explained by the covering property of the decorated trees. 
For finite Cayley trees, the energies corresponding to Euclidean flat bands can either stay exactly the same or converge exponentially towards these values as the number of generations in the tree increases.

Our analysis shows that not only can  flat energy bands arise in the tree analogs of the specified Euclidean lattices, but that the transition from Euclidean lattices to the tree geometry can further enrich the phenomenology of the flat bands. 
For example, the flat bands can become topological and protected by chiral symmetry or its generalized versions, as in the case of the Lieb and the double-Lieb decorations. 
In particular, we uncovered an exact mathematical correspondence between the $E=0$ flat band states of the Lieb-Cayley tree and topological edge states of Su-Schrieffer-Heeger chains.
In other cases, the flat bands on the decorated trees can be mapped onto states localized on edge defects in one-dimensional chains, as in the case of the Husimi and the clique decorations. Remarkably, when the states on the decorated trees acquire a topological interpretation, they need not be
localized at the boundary of the tree; instead, they can be exponentially localized at a site inside the \emph{bulk}, in a sharp contrast with the conventional bulk-boundary correspondence known from Euclidean lattices.

Our construction provides a unified framework for understanding how topological and flat band phenomena manifest on homogeneous non-Euclidean, loopless structures. 
The analysis based on symmetry-adapted sectors and their mapping to one-dimensional topological chains establishes an exact correspondence between the number of topological zero modes and the prediction of the rank–nullity theorem. 
Moreover, we showed that such topological bulk states persist even for moderate system sizes, suggesting their potential realizability in experimental platforms, such as photonic or circuit networks, that have already been used to emulate some of the tree systems ~\cite{Junsong:2025,Weststrom:2023} and hyperbolic lattices~\cite{Kollar:2019,Lenggenhager:2022,Boettcher:2020,Dey:2024}.
We expect flat bands, as theoretically described by our study, to be experimentally detectable in such photonic and circuit platforms by identifying energies (frequencies) with large local density of states (LDOS) but vanishing transmission between distant nodes. 
Here, large LDOS signals a narrow energy band, while the absence of long-range transmission indicates the band is dispersionless: both clear fingerprints of an approximately flat band.
The LDOS is accessed by exciting a single node and measuring the on-site spectral response.
In electric circuits, this is achieved by measuring the impedance $Z_{ii}(\omega)$ at site $i$ to ground~\cite{Imhof:2018,Helbig:2020,Lenggenhager:2022}.
In waveguides, we envision the LDOS to be approximately captured by integrating transmission from site $i$ to its immediate neighbors (within the support of the anticipated compact localized states).
Vanishing group velocity is then validated by suppressed two-site impedance $Z_{ij}(\omega)$ resp.~by suppressed transmission to distant nodes at the same frequency. (The latter is, essentially, the diagnostic used by Ref.~\citenum{Kollar:2019} to detect flat bands in a circuit QED realization of the hyperbolic-kagome circuit.)

Looking ahead, several directions appear promising. One concerns extending our analysis to decorated Cayley trees with inhomogeneous hoppings, similar to Ref.~\cite{singh_arboreal_2024}, which may reveal new mechanisms of topological protection. Another avenue involves exploring topological states on genuine hyperbolic lattices, where higher-dimensional topology may lead to modified notions of the bulk-boundary correspondence.
It is also interesting to ponder whether non-trivial band topology can be manifested through states localized inside the bulk also in lattices with loops provided that one maintains the expander property (i.e., negative curvature).
This avenue becomes especially interesting if one takes into account the recent discovery of bulk topological states on quasicrystals \cite{Johnstone:2022,Nielsen:2025} and the connection between quasicrystals and hyperbolic lattices \cite{Flicker:2020,Kulp:2025}. More broadly, our results point toward a richer landscape of flat-band and topological phenomena in non-Euclidean systems, where geometry alone can generate and stabilize unconventional quantum states.

\section*{Data availability}
The code used to generate the data presented in this work is publicly available in the data repository~\citenum{duss_cayleytree_2026}.

\begin{acknowledgments}
We would like to thank Shu Hamanaka for motivating us to investigate decorated Cayley trees and to Mykhailo Pavliuk for valuable discussions about the presented exact solutions.
W.P.D.~acknowledges support through a MaNEP internship grant.
A.I.~and T.B.~were supported by the Starting Grant No.~211310 by the Swiss National Science Foundation (SNSF). 
A.I. acknowledges support from the UZH Postdoc Grant No.~FK-24-104.
\end{acknowledgments}

\appendix

\section{Basis states in Lieb decoration}\label{Sec:App_LiebCayley}

In this section, we present the construction of the symmetry-adapted basis for the Lieb-Cayley tree and prove its completeness. 
Let the position states be $|l,j_{\textrm{L}/\textrm{C}},m \rangle$ with  $j_\textrm{L} = 1,\cdots,K^{l-1}$ and $j_\textrm{C} = 1,\cdots,K^{l-2}$ in the $l$-th ($l=1,\cdots,M$) generation of the branch $m$ ($m = 1,2,\cdots,K+1$).
In addition, there is the root node at the center of the tree, labeled $\ket{0}$.
Constructing the basis of the shell-symmetric sector $\mathcal{H}_{\textrm{sym.}}$, we obtain the following structure for the symmetric states:
\begin{equation}
\label{eqn:Lieb-Cayley-sym-basis-states}
    \begin{aligned}
        & |l)^\textrm{L} = \frac{1}{\sqrt{(K+1)K^{l-1}}} \sum_{m=1}^{K+1}\sum_{j_\textrm{L}=1}^{K^{l-1}} |l,j_\textrm{L},m\rangle \quad \text{for $l$ odd}, \\
        & |l)^\textrm{C} = \frac{1}{\sqrt{(K+1)K^{l-2}}} \sum_{m=1}^{K+1}\sum_{j_\textrm{C}=1}^{K^{l-2}} |l,j_\textrm{C},m\rangle  \quad \text{for $l$ even.}
    \end{aligned}
\end{equation}
In the above, we have adopted $|\cdots \rangle$ to denote the position basis and $| \cdots )$ for the symmetry-adapted basis. 

We next apply the Hamiltonian to these states to see how they transform. 
For example, in a model with $K = 2$:
\begin{eqnarray}
        \mathcal{H} |2)^\textrm{C} \!& =& \! \frac{1}{\sqrt{3}} \!\! \sum_{\langle i,j \rangle} \!\! (|i\rangle \langle j| \!+\! |j\rangle \langle i |) (|2,\!1_\textrm{C},\!1\rangle \!+\! |2,\!1_\textrm{C},\!2\rangle \!+\! |2,\!1_\textrm{C},\!3\rangle ) \nonumber \\
        \! & =& \! \frac{1}{\sqrt{3}}(|1,1_\textrm{L},1\rangle + |1,1_\textrm{L},2\rangle + |1,1_\textrm{L},3\rangle ) \nonumber \\
        & \phantom{=}& + \frac{1}{\sqrt{3}} (|3,1_\textrm{L},1\rangle + |3,2_\textrm{L},1\rangle + |3,1_\textrm{L},2\rangle \nonumber  \\
        & \phantom{=}&+ |3,2_\textrm{L},2\rangle + |3,1_\textrm{L},3\rangle + |3,2_\textrm{L},3\rangle ) \nonumber \\
        \!& =&\! \frac{1}{\sqrt{3}} \sqrt{3} |1)^\textrm{L} + \frac{1}{\sqrt{3}} \sqrt{6} |3)^\textrm{L} \nonumber  \\
        \!& =& \! |1)^\textrm{L} + \sqrt{2} |3)^\textrm{L}
    \end{eqnarray}
Generalizing from here, we obtain a recursion relation for the symmetric eigenstates of the form 
\begin{eqnarray}
    E \psi_0 &=& \sqrt{K+1} \psi_{1}^\textrm{L}, \\
    E \psi_{l}^\textrm{L} &=& \sqrt{K} \psi_{l-1}^\textrm{C} + 1 \psi_{l+1}^\textrm{C}, \\
    E \psi_{l}^\textrm{C} &=& 1 \psi_{l-1}^\textrm{L} + \sqrt{K} \psi_{l+1}^\textrm{L},
\end{eqnarray}
where $\psi^{\textrm{L}/\textrm{C}}_{l}$ 
are the wavefunction components in the symmetry-adapted basis. Counting the number of shell-symmetric states, we find that there are $M$ states. 
Adding the $|0\rangle = |0)$ state, we find the the total number of symmetric states as
\begin{equation}
    N_S =M + 1 = (M_\textrm{C} + M_\textrm{L}) + 1.
\end{equation}

We proceed to construct the basis of the non-symmetric sectors $\mathcal{H}_{\textrm{nonsym.}}^\alpha$. 
As has been shown in Sec.~\ref{Sec:Methods_B}, construction of non-symmetric basis states 
on a Cayley tree requires choosing a node $\alpha$ of the tree and constructing the state from there. 
It turns out that it suffices to restrict $\alpha$ to nodes on layers $l_\textrm{C} \in M_\textrm{C}$ of the Lieb-Cayley tree, as these are the only nodes at which we can permute sub-branches. 
Specifically, consider the $K \geq 2$ sub-branches rooted at this node. 
Then $\alpha$ will be on some even $l$-layer with the number of remaining layers being $M-l$. Let $r = 1,2,\cdots,M-l$ and let $\omega$ be a non-trivial $K$-th root of unity. 
A non-symmetric basis state $|l,r,\omega)_\alpha$ is then constructed by weighting the position basis $|r,k,n\rangle_\alpha$ (with $k = 1,\cdots,K^{\lceil r/2 \rceil -1}$) in the $r$-th generation with the powers of $\omega$, such that all states in the same sub-branch have the same weight,~i.e,
\begin{equation}
\label{eqn:lieb-cayley-nonsym-def}
    |l_\textrm{C},r,\omega)_\alpha := \frac{1}{K^{\lceil r/2 \rceil}} \sum^K_{n = 1} \omega^n \sum_{k=1}^{K^{\lceil r/2 \rceil}} |r,k,n\rangle_\alpha.
\end{equation}
The states $|r,k,n\rangle_\alpha$ on the right-hand side lie in a Cayley layer for $r$ even and in a Lieb layer for $r$ odd. 

Below we provide some example states for $K$ = 2, where the only non-trivial root of unity is $\omega = -1$.
\begin{eqnarray}
    |l_\textrm{C},r = 1,\omega = -1)_\alpha &=&  (|1,1,1\rangle - |1,1,2\rangle)/\sqrt{2}, \\
        |l_\textrm{C},r = 2,\omega = -1)_\alpha &=&  (|2,1,1\rangle - |2,1,2\rangle)/\sqrt{2}, \\
        |l_\textrm{C},r = 3,\omega = -1)_\alpha &=& (|3,1,1\rangle + |3,2,1\rangle \\
        &\phantom{=}& \quad - |3,1,2\rangle -|3,2,2\rangle)/\sqrt{4}.\;\;\;\;\;
\end{eqnarray}
We can now apply the Hamiltonian to these states, e.g.:
\begin{equation}
    \begin{aligned}
        & \mathcal{H} |l_\textrm{C},r= 2,\omega = -1)_\alpha \\
        & = \mathcal{H} (|2,1,1\rangle - |2,1,2\rangle )/\sqrt{2} \\
        & =  (|3,1,1\rangle + |3,2,1\rangle - |3,1,2\rangle \\
        & \phantom{=} - |3,2,2\rangle)/\sqrt{2} + \frac{1}{\sqrt{2}} (|1,1,1\rangle - |1,1,2\rangle )/\sqrt{2} \\
        & = \sqrt{2} |l_\textrm{C},3,-1)_\alpha + 1 |l_\textrm{C},1,-1)_\alpha. 
    \end{aligned}
\end{equation}
Generalizing to the full eigenvalue equation $\mathcal{H} |\Psi\rangle = E |\Psi\rangle$ one finds the recursion relations
\begin{eqnarray}
    E \phi_{l_\textrm{C},r,\omega}^\alpha &=& \sqrt{K} \phi_{l_\textrm{C},r-1,\omega}^\alpha + \phi_{l_\textrm{C},r+1,\omega}^\alpha \quad \text{for $r$ odd,} \;\;\;\;\;\;\; \\
    E \phi_{l_\textrm{C},r,\omega}^\alpha &=&  \phi_{l_\textrm{C},r-1,\omega}^\alpha  + \sqrt{K} \phi_{l_\textrm{C},r+1,\omega}^\alpha\quad \text{for $r$ even,} \;\;\;\;\;\;\;\;\;\; \\
    0 &=& \phi_{l_\textrm{C},0,\omega}^\alpha = \phi_{l_\textrm{C},M-l + 1,\omega}^\alpha.
\end{eqnarray}
Coefficients in these recursion relations define components of the non-symmetric Hamiltonian block $\mathcal{H}_\textrm{nonsym}^\alpha$.

The constructed symmetry-adapted basis states are orthogonal to each other. Furthermore, counting of these states reveals that they form a complete basis for the Lieb-Cayley tree. 
To count the non-symmetric basis states, note that their origin  
can be chosen from $(K+1) \times K^{l_\textrm{C} - 1}$ nodes of the $l_\textrm{C}$ Cayley-layer. Each of these nodes will produce $M-l = M_\textrm{C} + M_\textrm{L} - 2\times l_\textrm{C}$ non-symmetric states, where we used that the layer-number $l$ will always be a Cayley-layer, which are the even numbered layers. Furthermore, we have $K-1$ nontrivial $K$-th roots of unity $\omega$. This gives us the following expression for the number of non-symmetric states
\begin{equation}
    N_{B_{l\geq 1}} \!=\! \sum_{l_\textrm{C} = 1}^{M_\textrm{C}} (K\!-\!1)\times(K\!+\!1) \times K^{l_\textrm{C} - 1} (M_\textrm{C} + M_\textrm{L} - 2l_\textrm{C})
\end{equation}
Additionally, we get a set of $M$ non-symmetric states originating from the central node, as seen with the simple Cayley tree in Sec.~\ref{Sec:Methods_B}. These have degeneracy $K$ due to $K$ non-trivial $(K+1)$-th roots of unity. They are characterized by Hamiltonian $\mathcal{H}_\textrm{nonsym.}^0$ that takes the same general form as the Hamiltonian $\mathcal{H}_\textrm{nonsym.}^\alpha$ considered earlier. The number of such states totals 
\begin{equation}
    N_{B_{l=0}} = K\times M.
\end{equation}

We next determine the total number of states. 
First, we reformulate the number of non-symmetric states with root $\alpha \neq 0$ into a closed form as
\begin{eqnarray}
        N_{B_{l\geq1}} & =& \sum_{l_\textrm{C} = 1}^{M_\textrm{C}} (K-1)\times(K+1) \times K^{l_\textrm{C} - 1} (M_\textrm{C} + M_\textrm{L} - 2l_\textrm{C}) \nonumber \\
        & =& \frac{K+1}{K-1} \big{(} M_\textrm{L}(K-1)(K^{M_\textrm{C}} - 1) \\
        & \phantom{=}& + 2(K^{M_\textrm{C}}-1) - M_\textrm{C} (K-1)(K^{M_\textrm{C}} + 1) \big{)}.  \nonumber
    \end{eqnarray}
In the above, we adopted the geometric sum twice, and we further used the following sum:
\begin{equation}
\begin{aligned}
     & \sum_{l_\textrm{C} = 1}^{M_\textrm{C}} (K-1)\times(K+1) \times 2l_\textrm{C} \times  K^{l_\textrm{C} - 1} \\ 
     & = \frac{K+1}{K-1} \left(2M_\textrm{C} K^{M_\textrm{C} + 1} - 2(M_\textrm{C} + 1)K^{M_\textrm{C}} + 2\right)
\end{aligned}
\end{equation}
We sum all symmetry-adapted basis states together, finding
\begin{eqnarray}
        && \hspace{-2mm} N_S + N_{B_{l=0}} + N_{B_{l\geq 1}} 
         \\
        & =& (K+1)(M_\textrm{C} + M_\textrm{L}) + 1 + N_{B_{\geq 1}} \nonumber  \\ 
        & =& \frac{K+1}{K-1} \!\big( (K\!-\!1)(M_\textrm{C} \!+\! M_\textrm{L})  + (K\!-\!1) + M_\textrm{L}(K\!-\!1)(K^{M_\textrm{C}} \!-\!1) \nonumber \\
        & \phantom{=}& + 2(K^{M_\textrm{C}}-1) - M_\textrm{C} (K-1)(K^{M_\textrm{C}} + 1)
        \big) \nonumber \\
        & =& \frac{- 3 + K + (K+1)K^{M_\textrm{C}}\times (-2 - (K-1)M_\textrm{L} + (K-1)M_\textrm{C})}{K -1}. \nonumber 
\end{eqnarray}

We proceed to show that $N -  N_{B_{l=0}} - N_{B_{l\geq 1}} - N_S = 0$, where the total number of sites $N$ on the Lieb-Cayley tree 
is given by Eq.~(\ref{eq:total_number_of_states_lieb_lattice}). 
Inserting results in
\begin{eqnarray}
        && \hspace{-2mm} N - N_S -  N_{B_{l=0}} - N_{B_{l\geq 1}}  \\
        &=& 1 + \frac{K+1}{K-1}(K^{M_\textrm{L}} + K^{M_\textrm{C}} -2) \nonumber \\
        & \phantom{=}& - \frac{- 3 + K + (K\!+\!1)K^{M_\textrm{C}}\times (-2 - (K\!-\!1)M_\textrm{L} + (K\!-\!1)M_\textrm{C}}{K \!-\!1} \nonumber \\ 
        & =& \frac{K+1}{K-1} \left(K^{M_\textrm{C}} - K^{M_\textrm{L}} + (K\!-\!1)K^{M_\textrm{C}} M_\textrm{L} - (K\!-\!1)K^{M_\textrm{C}} M_\textrm{C}\right). \nonumber 
    \end{eqnarray}
For the above expression to yield zero, we need to have this last bracket equal to zero, i.e.,
\begin{equation}
    (K^{M_\textrm{C}} - K^{M_\textrm{L}} + (K-1)K^{M_\textrm{C}} M_\textrm{L} - (K-1)K^{M_\textrm{C}} M_\textrm{C}) \overset{!}{=} 0.
\end{equation}
To prove the above, it is important to note that there are only two cases for the relationship between $M_\textrm{C}$ and $M_\textrm{L}$ in a Lieb-Cayley tree with the prescribed radius. 
Either (1)~$M_\textrm{L} = M_\textrm{C}$ or (2)~$M_\textrm{L} = M_\textrm{C} + 1$,  i.e. either the Lieb-Cayley tree ends on a (1)~Cayley-layer or on a (2)~Lieb-layer. 
It is then easy to verify 
that in both cases the above expression equals zero. 
Considering finally that all the constructed basis states are orthogonal to each other, we conclude our proof of the completeness of the constructed basis.

\section{Basis states in double Lieb decoration}\label{Sec:App_2xLC}

We again introduce position states $|l,j_{\textrm{C}/\textrm{L}1/\textrm{L}2},m \rangle$ with $j_{\textrm{L}1} = 1,\cdots,K^{l-1}$, $j_{\textrm{L}2} = 1,\cdots,K^{l-2}$ and $j_\textrm{C} = 1,\cdots,K^{l-3}$ in the $l$-th ($l=1,\cdots,M$) generation of the branch $m$ ($m = 1,2,\cdots,K+1$). 
In analogy with the previous case, we construct the symmetric states as
\begin{eqnarray}
        && |l)^{\textrm{L}1} \!=\! \frac{1}{\sqrt{(K\!+\!1)K^{l-1}}} \!\! \sum_{m=1}^{K+1} \! \sum_{j_{\textrm{L}1}=1}^{K^{l-1}} \!\! |l,j_{\textrm{L}1},m\rangle \quad \!\! \textrm{for $l$ mod 3 = 1}\;\;\;\;\;\;\; \\
        && |l)^{\textrm{L}2} \!=\! \frac{1}{\sqrt{(K\!+\!1)K^{l-2}}} \!\! \sum_{m=1}^{K+1} \! \sum_{j_{\textrm{L}2}=1}^{K^{l-2}} \!\! |l,j_{\textrm{L}2},m\rangle \quad \!\! \textrm{for $l$ mod 3 = 2}\;\;\;\;\;\;\; \\
        && |l)^\textrm{C} \!=\! \frac{1}{\sqrt{(K\!+\!1)K^{l-3}}} \!\! \sum_{m=1}^{K+1} \! \sum_{j_\textrm{C}=1}^{K^{l-3}} \!\! |l,j_\textrm{C},m\rangle \quad \!\! \textrm{for $l$ mod 3 = 0}\;\;\;\;\;\;\; 
    \end{eqnarray}
Applying the Hamiltonian to these states gives essentially the same behavior as in the previous Appendix~\ref{Sec:App_LiebCayley}. Skipping these examples due to their redundancy, we immediately show the energy recursion relations, which take the form
\begin{eqnarray}
    E \psi_0 &=& \sqrt{K+1} \psi_{1}^{\textrm{L}1}, \\
    E \psi_{l}^{\textrm{L}1} &=& \sqrt{K} \psi_{l-1}^{\textrm{C}} + 1 \psi_{l+1}^{\textrm{L}2},  \\
    E \psi_{l}^{\textrm{L}2} &=& 1 \psi_{l-1}^{\textrm{L}1} + 1 \psi_{l+1}^\textrm{C},  \\
    E \psi_{l}^{\textrm{C}} &=& 1 \psi_{l-1}^{\textrm{L}2} + \sqrt{K} \psi_{l+1}^{\textrm{L}1}.
\end{eqnarray}

We find the total number of symmetric states to be:
\begin{equation}
    N_S = M + 1 = (M_\textrm{C} + M_{\textrm{L}1} + M_{\textrm{L}2}) + 1
\end{equation}
By the exact same procedure as in Appendix~\ref{Sec:App_LiebCayley}, we construct the non-symmetric basis states (i.e., basis states of the non-symmetric Hamiltonian sectors). 
Again, these only start from a node $\alpha \in M_\textrm{C}$ in a Cayley layer.
A non-symmetric basis state $|l,r,\omega)_\alpha$ is constructed by weighting the position basis $|r,k,n\rangle_\alpha$ (where $k = 1,\cdots,K^{\lceil r/3 \rceil -1}$) in the $r$-th generation with the powers of $\omega$ (non-trivial $K$-th roots of unity), such that all states in the same branch have the same weight, i.e., 
\begin{equation}
    |l_\textrm{C},r,\omega)_\alpha := \frac{1}{K^{\lceil r/3 \rceil}} \sum^K_{n = 1} \omega^n \sum_{k=1}^{K^{\lceil r/3 \rceil}} |r,k,n\rangle_\alpha.
\end{equation}
Compared to the the simple Lieb-Cayley tree we now have 3 layers (instead of 2) with the same number of nodes. 

By the same procedure as applied to the Lieb-Cayley tree in the preceding section, one can show that the eigenvalue equation of the 2xLieb-Cayley tree results in the energy recursion relation of the form
\begin{eqnarray}    
    E \phi_{l_\textrm{C},r,\omega}^\alpha \!&=&\! \sqrt{K} \phi_{l_\textrm{C},r-1,\omega}^\alpha \!+\! \phi_{l_\textrm{C},r+1,\omega}^\alpha \quad \text{for $r$ mod 3 = 1,} \;\;\;\;\;\;\;\;\;\\
    E \phi_{l_\textrm{C},r,\omega}^\alpha \!&=&\! \phi_{l_\textrm{C},r-1,\omega}^\alpha \!+\! \phi_{l_\textrm{C},r+1,\omega}^\alpha \quad \text{for $r$ mod 3 = 2,} \;\;\;\;\;\;\;\;\;\\
    E \phi_{l_\textrm{C},r,\omega}^\alpha \!&=&\!  \phi_{l_\textrm{C},r-1,\omega}^\alpha  \!+\! \sqrt{K} \phi_{l_\textrm{C},r+1,\omega}^\alpha \quad \text{for $r$ mod 3 = 0,} \;\;\;\;\;\;\;\;\;\\
    0 \!&=&\! \phi_{l_\textrm{C},0,\omega}^\alpha \!=\! \phi_{l_\textrm{C},M-l + 1,\omega}^\alpha. \;\;\;\;\;\;\;\;\;
\end{eqnarray}

We next count the total number of non-symmetric states. 
The origin of our non-symmetric states can be chosen from $(K+1)\times K^{l_\textrm{C}-1}$ nodes of the $l_\textrm{C}$-th Cayley layer with $l_\textrm{C} \in M_\textrm{C}$. Each of these nodes will produce $M-l = M_{\textrm{L}1} + M_{\textrm{L}2} + M_\textrm{C} - 3l_\textrm{C}$ non-symmetric states. Here we used the fact that the layer $l$ on which the node is placed will always be a Cayley-layer with $3l_\textrm{C} = l$. 
We further have $K-1$ non-trivial roots of unity $\omega$, giving the following expression for the total number of non-symmetric states:
\begin{equation}
\begin{aligned}
    & N_{B_{l\geq1}} \\
    & = \sum_{l_\textrm{C} = 1}^{M_\textrm{C}} (K-1)\times(K+1) \times K^{l_\textrm{C} - 1} (M_\textrm{C} + M_{\textrm{L}1} + M_{\textrm{L}2} - 3l_\textrm{C}) \\
    & = \frac{K+1}{K-1} [(M_{\textrm{L}1}(K-1)(K^{M_\textrm{C}} - 1) + M_{\textrm{L}2}(K-1)(K^{M_\textrm{C}} - 1) \\
    &  \phantom{=}+ 3(K^{M_\textrm{C}}-1) - M_\textrm{C} (K-1)(2K^{M_\textrm{C}} + 1) ]
\end{aligned}
\end{equation}
Additionally, we have $N_{B_{l=0}} =K\times M$ non-symmetric states that originate from the central (i.e., root) site, with their dynamics captured by Hamiltonian $\mathcal{H}_\textrm{nonsym}^0$ that takes the same form as $\mathcal{H}_\textrm{nonsym}^\alpha$ shown in Eq.~(\ref{Eq:2xLiebCayley_Nonsymm_Hamiltonian}) of the main text. Counting all symmetry-adapted basis states and comparing this count against the number of sites in position bases, similar calculations as in the preceding section lead to
\begin{equation}
    \begin{aligned}
        & N_\textrm{total}
        - N_S - N_{B_{l=0}}- N_{B_{l\geq1}} \\
        & =  \frac{K+1}{K-1} [K^{M_{\textrm{L}1}} +K^{M_{\textrm{L}2}} - 2 K^{M_\textrm{C}} \\
        & \phantom{=} + (K-1)K^{M_\textrm{C}}(-M_{\textrm{L}1} - M_{\textrm{L}2} + 2M_\textrm{C})].
    \end{aligned}
\end{equation}
One can show that this expression evaluates to zero for all the possible choices of $M_\textrm{C}/M_{\textrm{L}1}/M_{\textrm{L}2}$, categorized as Case 0/1/2 below Eq.~(\ref{eqn:2xLieb-Cayley-counting}) of the main text.
For example, by inserting $M_{\textrm{L}1} = M_{\textrm{L}2} = M_\textrm{C} + 1$ into the above, we find:
\begin{eqnarray}
        && K^{M_{\textrm{L}1}} \!+\!K^{M_{\textrm{L}2}} \!-\! 2 K^{M_{\textrm{C}}} + (K\!-\!1)K^{M_{\textrm{C}}}(-M_{\textrm{L}1} \!-\! M_{\textrm{L}2} \!+\! 2M_\textrm{C}) \nonumber \\
        & = &K^{M_{\textrm{L}1}} + K^{M_{\textrm{L}1}} - 2K^{M_{\textrm{L}1}-1}\nonumber  \\
        & \phantom{=} &+ (K-1)K^{M_{\textrm{L}1}-1}(-M_{\textrm{L}1} - M_{\textrm{L}1} + 2M_{\textrm{L}1} -2)  \\
        & = &2K^{M_{\textrm{L}1}} - 2K^{M_{\textrm{L}1}} + 2K^{M_{\textrm{L}1}} - 2 K^{M_{\textrm{L}1}} \nonumber \\
        & = & 0 \nonumber 
    \end{eqnarray}
Since the constructed basis states are orthogonal to each other, we conclude that we have found a complete set of basis states.

\section{Basis states in Husimi decoration}\label{Sec:App_Husimi}

We define the position basis states for the Husimi-Cayley tree in the same way as they are defined for the Cayley tree, except for removing the central (root) node, i.e., we use the labels
$|l,j,m \rangle$ with  $j = 1,\cdots,K^{l}$ in the $l$-th ($l=1,\cdots,M$) generation of the branch $m$ ($m = 1,2,\cdots,K+1$). 
The definition of the symmetric basis states for Husimi-Cayley tree follows precisely the same formula as for the Cayley tree [Eq.~(\ref{eqn:Lieb-sym-basis-states}) of the main text], i.e., 
\begin{equation}
    \begin{aligned}
        & |l) = \frac{1}{\sqrt{(K+1)K^{l-1}}} \sum_{m=1}^{K+1}\sum_{j=1}^{K^{l-1}} |l,j,m\rangle .
    \end{aligned}
\end{equation}
If we apply the nearest-neighbor hopping Hamiltonian to the state above, we find a similar dynamics as for the Cayley tree, except for the appearance of an additional 
on-site potential. 
For example, for $K = 2$ we obtain
\begin{equation}
\begin{aligned}
    \mathcal{H} |1) & = \mathcal{H}\big( |1,1,1\rangle + |1,1,2\rangle + |1,1,3\rangle \big)/\sqrt{3} \\
    & = \Big( (|1,1,2\rangle + |1,1,3\rangle)\sqrt{3} \\
    & \phantom{=} + (|1,1,1\rangle + |1,1,3\rangle) + (|1,1,1\rangle + |1,1,2\rangle) \\
    &\phantom{=} + \big[ |2,1,1\rangle + |2,2,1\rangle + |2,1,2\rangle \\ 
    & \phantom{=} + |2,2,2\rangle + |2,1,3\rangle + |2,2,3\rangle \big] \Big)/\sqrt{3} \\
    & = 2|1) + \sqrt{2} |2).
\end{aligned}  
\end{equation}

We observe (in contrast with the simple Cayley tree) that the bond connecting the position states within the $l = 1$ layer generates a contribution on the right-hand side proportional to $|l)$.
The analysis above can be generalized to the following recursion relations for the amplitudes of symmetric eigenstates,
\begin{eqnarray}
    E \psi_1 &=& K \psi_{1}  + \sqrt{K} \psi_2, \\
    E \psi_{l,m} &=& \sqrt{K} \psi_{l-1} + (K-1) \psi_{l} + \sqrt{K} \psi_{l+1}, \label{eqn:Husimi-symmetric-second-eq} \\
    0 &=& \psi_{M+1}.
\end{eqnarray}
There is exactly one set of $M$ states captured by the above equations, which correspond to eigenstates of the Hamiltonian in Eq.~(\ref{Eq:Husimi_FullySymm}) of the main text.
The factor $(K-1)$ in Eq.~(\ref{eqn:Husimi-symmetric-second-eq}) derives from the fact that the fully-connected set of branching nodes can hop to more nodes on the same layer for higher $K$.

Non-symmetric states for the Husimi-Cayley tree are defined in exactly the same way as for the simple Cayley tree. 
Consider the $K$ branches rooted at a node $\alpha$.
Let $\alpha$ be on layer $l$ so that the number of remaining layers is $M-l$; we use $r = 1,2,\cdots,M-l$ for labeling the remaining layers and $\omega$ for non-trivial $K$-th roots of unity. 
A non-symmetric basis state $|l,r,\omega)_\alpha$ is then constructed by weighting the position basis $|r,k,n\rangle$ ($k = 1,\cdots,K^{r -1}$) in the $r$-th generation with the powers of $\omega$, such that all states in the same branch have the same weight, i.e.,
\begin{equation}
    |l,r,\omega)_\alpha := \frac{1}{\sqrt{K^{r}}} \sum^K_{n = 1} \omega^n \sum_{k=1}^{K^{r-1}} |r,k,n\rangle_\alpha,
\end{equation}
which exactly reproduces Eq.~(\ref{eqn:shell-non-sym-state-def-nonroot}) of the main text.

To illustrate the action of the Hamiltonian on shell-non-symmetric basis states,
we consider the case $K$ = 3 for $|l,1,\omega)_\alpha$ with $M-l > 1$. We find 
\begin{eqnarray}
    \mathcal{H} |l,1,\omega)_\alpha \!& = &\!\mathcal{H} \frac{1}{\sqrt{3}} \! \left( \omega_1 |1,1,1\rangle_\alpha \!+\! \omega_2 |1,1,2\rangle_\alpha \!+\! \omega_3 |1,1,3\rangle_\alpha \right) \nonumber \\
    \!& = &\![ (\omega_1 + \omega_2 + \omega_3)|\alpha) +  (\omega_2 + \omega_3) |1,1,1\rangle_\alpha \nonumber \\
    \!& \phantom{=} & \!+ (\omega_1 + \omega_3) |1,1,2\rangle_\alpha + (\omega_1 + \omega_2) |1,1,3\rangle_\alpha \nonumber \\
    \! & \phantom{=} & \!+ \omega_1 (|2,1,1\rangle_\alpha + |2,2,1\rangle_\alpha + |2,3,1\rangle_\alpha) \nonumber \\ 
    \! & \phantom{=} & \! + \omega_2(\dots) + \omega_3 (\dots)]/\sqrt{3} \nonumber \\
    \! & = &\! 0 -|l,1,\omega)_\alpha +  \sqrt{K}|l,2,\omega)_\alpha.
\end{eqnarray}
Here, $|\alpha)$ is the seed node. 
We further used the fact that the sum of the $K$-th roots of unity gives zero, i.e.,
\begin{equation}\label{Eq:Sum_of_Roots_of_Unity2}
    \sum_i^K \omega_i = 0, \quad \text{therefore} \quad \sum_{i, i \neq j}^K \omega_i = - \omega_j. 
\end{equation}

Note that, due to the fully-connected property of the branching nodes, we will always get $-1$ for the $|l,1,\omega)$ state. 
This is due to the ``mixing'' of the $r=1$ position states, which produces a combination of all position states except the original one for any choice of $K$ and $\omega$.
To be more explicit, consider only in-layer hopping at $r=1$,
\begin{equation}\label{Eq:HC_Roots_Unity_Mixing}
\begin{aligned}
     \mathcal{H}_{r=1} \left( \sum_{n=1}^{K} \omega^n  |1,k,n\rangle_\alpha \right) & =  \sum_{i=1}^{K}\left( \sum_{n=1, n \neq i}^{K} \omega^n  |1,k,i\rangle_\alpha \right) \\ 
     = \sum_{i=1}^{K}\left(-\omega^i  |1,k,i\rangle_\alpha \right) & = - \sum_{n=1}^{K} \omega^n  |1,k,n\rangle_\alpha
\end{aligned}
\end{equation}
The energy recursion relations are therefore 
\begin{eqnarray}
    E \phi_{l,1,\omega}^\alpha &=& \sqrt{K} \phi_{l,2,\omega}^\alpha  - \phi_{l,1,\omega}^\alpha \\
    E \phi_{l,r,\omega}^\alpha &=& \sqrt{K} \phi_{l,r-1,\omega}^\alpha  + \sqrt{K} \phi_{l,r+1,\omega}^\alpha  + (K-1) \phi_{l,r,\omega}^\alpha \;\;\;\;\;\;\;\;\;\\
    0 &=& \phi_{l,0,\omega}^\alpha = \phi_{l,M-l + 1,\omega}^\alpha.
\end{eqnarray}
The multiplicity of these states is the same as for the simple Cayley tree, i.e., 
\begin{eqnarray}
    N_{B_{l\geq1}} &=& \sum_{l=1}^{M-1} (K-1)(K+1) \times K^{l-1}(M-l) \\
    &=& N_\textrm{total}
    - (K+1)M
\end{eqnarray}

Moving on, we investigate the non-symmetric states with non-zero amplitudes on the first layer $l=1$. For consistency with prior notation, we formulate these states as if there exists a central root node at $l=0$ from which these states originate. Although this is not the case, the results are consistent with such an interpretation.
\begin{equation}
\label{eqn:nonsym-basis-Husimi}
        |r,\varpi)_{0} = \frac{1}{\sqrt{(K+1)K^{r-1}}}\sum_m^{K+1} \varpi^m \big( \sum_{j=1}^{K^{r-1}} |r,j,m\rangle \big)
\end{equation}
Here $\varpi$ is a non-trivial ($K+1$)-th root of unity. 
We arrive at the energy recursion relations 
\begin{eqnarray}
    E \phi_{r,\varpi}^0 &=& \sqrt{K} \phi_{r-1,\varpi}^0 + (K-1) \phi_{r,\varpi}^0 +  \sqrt{K} \phi_{r+1,\varpi}^0 \;\;\;\;\; \\
    0 &=& \psi_{M+1,\varpi} = \psi_0
\end{eqnarray}
The altered values of the roots of unity do not appear in the above equations explicitly, but they enhance the degeneracy to $K$,
such that there are 
$N_{B_{l=0}}=M\times K$ states in total. 

To conclude this section, we show that the constructed set of symmetry-adapted basis states is complete. 
For this purpose, one simply adds up the number of symmetric and non-symmetric states to verify that
\begin{eqnarray}
    N_\textrm{total}
    &=& N_{B_{l\geq 1}} + N_{B_{l=0}} + N_S \\
    &=& N_\textrm{total}
    - (K+1)M + K\times M + M,
\end{eqnarray}
which is mathematical consistent.
This result has been anticipated, considering that we simply removed a single node from the center of the simple Cayley tree while adopting essentially identical construction of symmetry-adapted basis states as in the case of the simple Cayley tree.

\section{Basis states in clique decoration}\label{Sec:App_Clique}

We pointed out in the main text that the clique-Cayley tree can be interpreted as the Lieb-Cayley tree with additional bonds connecting all Lieb nodes emanating from the same Cayley node, minus the node at the center of the Lieb-Cayley tree.
Therefore, let the position states be $|l,j_{\textrm{L}/\textrm{C}},m \rangle$ with  $j_\textrm{L} = 1,\cdots,K^{l-1}$ and $j_\textrm{C} = 1,\cdots,K^{l-2}$ in the $l$-th ($l=1,\cdots,M$) generation of the branch $m$ ($m = 1,2,\cdots,K+1$).
We arrive at the same symmetric states as formerly constructed for the Lieb-Cayley tree:
\begin{equation}
    \begin{aligned}
        & |l)^\textrm{L} = \frac{1}{\sqrt{(K+1)K^{l-1}}} \sum_{m=1}^{K+1}\sum_{j_\textrm{L}=1}^{K^{l-1}} |l,j_\textrm{L},m\rangle \quad \text{for $l$ odd}\\
        & |l)^\textrm{C} = \frac{1}{\sqrt{(K+1)K^{l-2}}} \sum_{m=1}^{K+1}\sum_{j_\textrm{C}=1}^{K^{l-2}} |l,j_\textrm{C},m\rangle \quad \text{for $l$ even}
    \end{aligned}
\end{equation}
In the above expressions, we use $|\cdots \rangle$ to denote the position basis and $| \cdots )$ for the symmetrized basis. 
While the shell-symmetric basis states are the same as for the Lieb-Cayley tree [cf.~Eq.~(\ref{eqn:Lieb-Cayley-sym-basis-states})], the dynamics of this symmetry sector is altered.
Applying the Hamiltonian to a state in a Lieb layer in a tree with $K=2$, we find
\begin{eqnarray}
        \mathcal{H} |1)^\textrm{L} & = & \mathcal{H}(|1,1_\textrm{L},1\rangle + |1,1_\textrm{L},2\rangle + |1,1_\textrm{L},3\rangle )/\sqrt{3} \nonumber \\
        & = & (|1,1_\textrm{L},2\rangle + |1,1_\textrm{L},3\rangle + |1,1_\textrm{L},1\rangle  \nonumber \\
        & \phantom{=} & + |1,1_\textrm{L},3\rangle ) + |1,1_\textrm{L},1\rangle + |1,1_\textrm{L},2\rangle )\sqrt{3} \\
        & \phantom{=} & + (|2,1_\textrm{C},1\rangle + |2,1_\textrm{C},2\rangle + |2,1_\textrm{C},3\rangle )/\sqrt{3} \nonumber \\
        & = & \frac{2}{\sqrt{3}} \sqrt{3} |1)^\textrm{L} + \frac{1}{\sqrt{3}} \sqrt{3} |2)^\textrm{C} \nonumber \\
        & = & 2|1)^\textrm{L} + 1|3)^\textrm{C} \nonumber 
\end{eqnarray}
We find again that bonds among the sites emanating from the same seed
induce effective on-site potentials.

The full energy recursion relation takes the form
\begin{eqnarray}
    E \psi_{1}^\textrm{L} &=& \psi_{2}^\textrm{C}  + K\psi_{1}^\textrm{L} \\
    E \psi_{l}^\textrm{L} &=& \sqrt{K} \psi_{l-1}^\textrm{C} +(K-1)\psi_{l}^\textrm{L} + \psi_{l+1}^\textrm{C} \\
    E \psi_{l}^\textrm{C} &=& \psi_{l-1}^\textrm{L} + \sqrt{K} \psi_{l+1}^\textrm{L}
\end{eqnarray}
where $\psi^{\textrm{L}/\textrm{C}}_{l}$ are the wave function components of a symmetry-adapted state. 
There are $N_S = M = (M_\textrm{C} + M_\textrm{L})$ symmetric states.
The non-symmetric states are again rooted on nodes $\alpha$ inside the Cayley-layers only. 
Consider the $K$ branches rooted at this node. Then $\alpha$ will be on some even $l$-layer ($l$ being the global layer index) with the number of remaining layers being $M-l$. 
Let $r = 1,2,\cdots,M-l$ and let $\omega$ be a non-trivial $K$-th root of unity. 

A non-symmetric basis state $|l,r,\omega)_\alpha$ is then constructed by weighting the position basis $|r,k,n\rangle_\alpha$ ($k = 1,\cdots,K^{\lceil r/2 \rceil -1}$) in the $r$-th generation with the powers of $\omega$, such that all states in the same branch have the same weight
\begin{equation}
    |l_\textrm{C},r,\omega)_\alpha := \frac{1}{K^{\lceil r/2 \rceil}} \sum^K_{n = 1} \omega^n \sum_{k=1}^{K^{\lceil r/2 \rceil}} |r,k,n\rangle_\alpha,
\end{equation}
which is exactly the same as Eq.~(\ref{eqn:lieb-cayley-nonsym-def}) for the Lieb-Cayley tree.
We can now apply the Hamiltonian to these states; for example,
\begin{eqnarray}
        & & \mathcal{H} |l_\textrm{C},r = 1,\omega = -1)_\alpha \nonumber \\
        & = & \mathcal{H} \frac{1}{\sqrt{2}} (|1,1,1\rangle_\alpha - |1,1,2\rangle_\alpha ) \nonumber \\
        & = &  \frac{1}{\sqrt{2}}(|2,1,1\rangle_\alpha \! -\!  |2,1,2\rangle_\alpha ) - \frac{1}{\sqrt{2}} (|1,1,1\rangle_\alpha \!- \!  |1,1,2\rangle_\alpha ) \nonumber \\
        & = & |l_\textrm{C},2,-1)_\alpha - |l_\textrm{C},1,-1)_\alpha.
    \end{eqnarray}
As with the Husimi-Cayley tree, we find an effective negative on-site potential at the start of the chain. 
Note that for higher $K$ we encounter the same features as previously discussed in Eq.~(\ref{Eq:HC_Roots_Unity_Mixing}), i.e., the roots of unity combine in a way to produce an effective $-1$ on-site potential. 

We have not yet commented on the non-symmetric states that originate from the central node (which is removed, i.e., non-existent, in the clique-Cayley tree). 
We label these basis states $|r,\varpi)_0$ in analogy with Eq.~(\ref{eqn:nonsym-basis-Husimi}).
When we act with the Hamiltonian on one of these states, we obtain
\begin{eqnarray}
        \mathcal{H} |1,\varpi)_0 & = &\frac{1}{\sqrt{3}} \mathcal{H}(\varpi^1|1,1,1\rangle + \varpi^2|1,1,2\rangle + \varpi^3|1,1,3\rangle ) \nonumber  \\
        & = &\frac{1}{\sqrt{3}}\big( (\varpi^2+\varpi^3)|1,1,1\rangle + (\varpi^1+\varpi^3)|1,1,2\rangle \nonumber \\
        & \phantom{=} &+ (\varpi^1 + \varpi^2)|1,1,3\rangle ) \big) \nonumber \\
        & \phantom{=} &+ \frac{1}{\sqrt{3}}(\varpi^1|2,1,1\rangle + \varpi^2|2,1,2\rangle + \varpi^3|2,1,3\rangle ) \nonumber \\
        & = &\frac{-1}{\sqrt{3}} \sqrt{3} |1,\varpi) + \frac{1}{\sqrt{3}} \sqrt{3} |2,\varpi) \nonumber \\
        & = & -1|1,\varpi)_0 + 1|2,\varpi)_0
    \end{eqnarray}
where $r \in [1,...,M]$.
We find that this symmetry sector is described by the same dynamics as the other non-symmetric sectors. 
Generalizing to the full eigenvalue equation $\mathcal{H} |\Psi\rangle = E |\Psi\rangle$, we find the recursion relations
\begin{eqnarray}
    E \phi_{l_{C},1,\omega}^\alpha &=& \phi_{l_{C},2,\omega}^\alpha  - \phi_{l_{C},1,\omega}^\alpha \\
    E \phi_{l_\textrm{C},r,\omega}^\alpha &=& \sqrt{K} \phi_{l_\textrm{C},r-1,\omega}^\alpha +  (K-1)\phi_{l_\textrm{C},r,\omega}^\alpha\nonumber \\
    &\phantom{=} & \qquad +\phi_{l_\textrm{C},r+1,\omega}^\alpha \quad \text{for r odd} \\
    E \phi_{l_\textrm{C},r,\omega}^\alpha &=&  \phi_{l_\textrm{C},r-1,\omega}  + \sqrt{K} \phi_{l_\textrm{C},r+1,\omega}\quad \text{for r even} \;\;\;\;\;\;\; \\
    0 &=& \phi_{l_\textrm{C},M-l + 1,\omega}^\alpha  \phi_{l_\textrm{C},0,\omega}^\alpha.
\end{eqnarray}
for non-symmetric sectors with seed $\alpha \neq \ket{0}$. The same equations are also found for the wave function amplitudes $\phi_{r,\varpi}^0$ for eigenstates of the non-symmetric sector with seed at the center.

Counting of the constructed symmetry-adapted basis states reveals that we have found a complete basis for the clique-Cayley tree. 
Specifically, the origin of the non-symmetric states rooted at $\alpha \neq \ket {0}$ can be chosen from $(K+1) \times K^{l_\textrm{C} - 1}$ nodes of the $l_\textrm{C}$ Cayley-layer. 
Each of these nodes will produce $M-l = M_\textrm{C} + M_\textrm{L} - 2\times l_\textrm{C}$ non-symmetric states, where we used that the layer-number $l$ will always be a Cayley-layer, which are the even numbered layers. 
Furthermore we have $K-1$ nontrivial roots of unity $\omega$. 
This gives us the following expression for the number of non-symmetric states with seed at $\alpha \neq \ket{0}$,
\begin{equation}
    N_{B_{l>1}} = \sum_{l_\textrm{C} = 1}^{M_\textrm{C}} (K-1)\times(K+1) \times K^{l_\textrm{C} - 1} (M_\textrm{C} + M_\textrm{L} - 2l_\textrm{C})
\end{equation}
For the non-symmetric states with seed at $\alpha = \ket{0}$, we have $M$ states for each of the $K$ nontrivial $(K+1)$-th roots of unity. 
In total, this gives
\begin{equation}
    N_{B_{l=0}} = K\times M
\end{equation}
We therefore find that 
\begin{equation}
    N_S + N_{B_{l=0}} = (K+1)\times M.
\end{equation}
This is exactly the same as for the Lieb-Cayley tree, minus the central node. One can therefore use the exact same derivation shown in Appendix~\ref{Sec:App_LiebCayley}, in the context of the Lieb-Cayley tree discussion, to show that we have found a complete set of states. 

\bibliography{bib}
\end{document}